\documentclass[10pt, two column, twoside]{IEEEtran}
\usepackage{color}
\usepackage[colorlinks, linkcolor=color1, anchorcolor=blue, citecolor=color1]{hyperref}

\usepackage{amsmath,graphicx,amsfonts,amsthm}
\usepackage{mathtools}
\usepackage{epstopdf}
\usepackage{bm}
\usepackage{mathrsfs}
\usepackage[ruled,linesnumbered,commentsnumbered]{algorithm2e}
\usepackage{multirow} 
\usepackage{xcolor}
\usepackage{color}
\usepackage{verbatim}
\usepackage{amssymb}
\usepackage{cite}
\usepackage{longtable}
\usepackage{supertabular}
\usepackage{booktabs} 
\definecolor{colorhkust}{RGB}{20,43,140}
\definecolor{colortsinghua}{RGB}{116,52,129}
\definecolor{color1}{RGB}{128,0,0}

\newcommand{\tabincell}[2]{\begin{tabular}{@{}#1@{}}#2\end{tabular}}

\newcommand{\mini}{\sf{minimize}}

\newcommand{\diagg}{\rm{diag}}

\newcommand{\R}{\mathbb{R}}  
\newcommand{\N}{\mathbb{N}}  
\newcommand{\C}{\mathbb{C}} 
\newcommand{\E}{\mathbb{E}}

\date{}

\begin{document}

\title{{Machine Learning for Large-Scale Optimization in 6G Wireless Networks}}
\author{Yandong~Shi,~\IEEEmembership{Member,~IEEE,}
         Lixiang Lian, \IEEEmembership{Member, IEEE},
        Yuanming Shi, \IEEEmembership{Senior Member, IEEE}, \\
        Zixin Wang, \IEEEmembership{Student Member, IEEE},
         Yong Zhou, \IEEEmembership{Senior Member, IEEE}, 
         Liqun Fu, \IEEEmembership{Senior Member, IEEE}, \\
         Lin Bai,  \IEEEmembership{Senior Member, IEEE}, 
        Jun Zhang, \IEEEmembership{Fellow, IEEE},
         and Wei Zhang, \IEEEmembership{Fellow, IEEE}
         \thanks{Yandong Shi is with the China Telecom Research Institute, Guangzhou 510660, China (e-mail: shiyd2@chinatelecom.cn).}
          \thanks{Lixiang Lian, Yuanming Shi, and Yong Zhou are with the School
of Information Science and Technology, ShanghaiTech University, Shanghai 201210, China (e-mail: lianlx@shanghaitech.edu.cn; shiym@shanghaitech.edu.cn; zhouyong@shanghaitech.edu.cn).}
 \thanks{Zixin Wang is with the School of Information Science and Technology, ShanghaiTech University, Shanghai 201210, China, also with the Shanghai Institute of Microsystem and Information Technology, Chinese Academy of Sciences, Shanghai 200050, China, and also with the University of Chinese Academy of Sciences, Beijing 100049, China (e-mail: wangzx2@shanghaitech.edu.cn).}
 \thanks{Liqun Fu is with the School of Informatics and the Key Laboratory
 	of Underwater Acoustic Communication and Marine Information Technology
 	(Ministry of Education), Xiamen University, Xiamen 361005, China (e-mail: liqun@xmu.edu.cn).}
\thanks {Lin Bai is with the School of Cyber Science and Technology, Beihang University, Beijing 100191, China (e-mail: l.bai@buaa.edu.cn).}
\thanks {Jun Zhang is with the Department of Electronic and Computer Engineering, Hong Kong University of Science and Technology, Hong Kong (e-mail: eejzhang@ust.hk).}
 \thanks{Wei Zhang is with the School of Electrical Engineering and Telecommunications, The University of New South Wales, Sydney, NSW 2052, Australia (e-mail: w.zhang@unsw.edu.au).}}
                           
\maketitle
\IEEEpeerreviewmaketitle


\maketitle

\begin{abstract}
The sixth generation (6G) wireless systems are envisioned to enable the paradigm shift from “connected things” to “connected intelligence”, featured by ultra high density, large-scale, dynamic heterogeneity, diversified functional requirements and machine learning capabilities, which leads to a growing need for highly efficient intelligent algorithms. The classic optimization-based algorithms usually require highly precise mathematical model of data links and suffer from poor performance with high computational cost in realistic 6G applications. Based on domain knowledge (e.g., optimization models and theoretical tools), machine learning (ML) stands out as a promising and viable methodology for many complex large-scale optimization problems in 6G, due to its superior performance, generalizability, computational efficiency and robustness. In this paper, we systematically review the most representative “learning to optimize" techniques in diverse domains of 6G wireless networks by identifying the inherent feature of the underlying optimization problem and investigating the specifically designed ML frameworks from the perspective of optimization. In particular, we will cover algorithm unrolling, learning to branch-and-bound, graph neural network for structured optimization, deep reinforcement learning for stochastic optimization, as well as end-to-end learning for semantic optimization, for solving challenging large-scale optimization problems arising from various important wireless applications. To enable ML implementation in distributed wireless networks across massive number of end devices, federated learning for distributed optimization will further be presented. Through the in-depth discussion, we shed light on the excellent performance of ML-based optimization algorithms with respect to the classical methods, and provide insightful guidance to develop advanced ML techniques in 6G networks. Neural network design, theoretical tools of different ML methods, implementation issues, as well as challenges and future research directions are also discussed to support the practical use of ML model in wireless applications.

\end{abstract}


\section{Introduction}\label{sec:1}
The sixth generation (6G) wireless systems have recently attracted considerable attention from both industry and academia, whose visions are towards ubiquitous 3D coverage (space-air-ground-sea integrated network) \cite{hongzhi22sags}, the intelligent and green networks \cite{bomin22green6g}, Internet of everything (IoE) \cite{kb196G}, etc. 
Compared with the previous generations, 6G can provide services with more stringent requirements, such as higher throughput, lower latency, higher reliability, denser connection, higher energy efficiency, as well as connected intelligence with machine learning capability \cite{kb196G}.  
{Driven by the new industrial and technological revolution, 6G can also support new services/applications beyond 5G such as immersive cloud extended reality (XR), holographic communications, sensory interconnection, digital twins and so on \cite{IMT6G}, which may demand new performance metrics to facilitate diversified and personalized user services}. 
The requirements of 6G system have made the fine-grained optimization of radio resources and effective learning of network-related information urgent necessities. 
Due to the large-scale, high density, heterogeneous qualities of services, and integrated multi-functional cross-layer design, the optimization problems in 6G can be extremely time-sensitive and complex, which pose great challenges for efficient optimization algorithm design. 
Machine learning (ML) has been recently leveraged as a disruptive technology to solve the challenging optimization problems in 6G, as well as support ubiquitous artificial intelligence (AI) services and IoE applications \cite{tataria20216g,Yuanming_CST20,IMT6G} including synaesthesia internet, digital twins, smart industry, smart agriculture, super traffic, precision medicine and blockchain economy. 
In this section, we first discuss the properties of optimization problems in 6G wireless networks and summarize the advantages and disadvantages of classic optimization-based methods. 
Then we introduce the motivation for ML-based optimization frameworks and summarize the existing design paradigms to solve different classes of optimization problems. 
Table \ref{ta:notation} summarizes the main notations used throughout this paper.

\subsection {Large-Scale Optimization for 6G}
The performance of 6G wireless networks can be enhanced by adopting various optimization algorithms, which solve the practical engineering problem through the mathematical tools. 
Constructing and solving optimization problems can effectively handle the technical issues in engineering and guide the performance-related policy development. 
The properties of optimization problems in 6G networks lay in the following three aspects:
(1) The objective functions of 6G optimization problems can be complicated to meet personalized services of heterogeneous networks and highly non-convex due to the enabling of integrated functions in 6G, such as joint sensing, communication, computing and control \cite{feng2021joint}.
Other factors such as diversified services 6G facilitates (e.g. distributed edge training and inference \cite{yuanming22edgeai}) and integrated cross-layer designs will further complicate the objective function. 
For example, the mutual information is adopted as the objective function in semantic communications \cite{beck2022semantic} to optimize the efficiency-accuracy trade-off, which is intractable. 
(2) Optimization variables and model parameters of 6G optimization problems can be of high dimension due to the massive devices, large-scale antennas in wireless networks and large amounts of data in various 6G technologies and services \cite{kb196G}. 
The feasible region of optimization variables can be highly stringent to satisfy practical network conditions under resource constraints and provide robust and reliable services for ubiquitous networking.
For example, the spectral-efficiency maximization problem in heterogeneous networks involves lots of non-convex constraints to support different transmission types (i.e., uplink and downlink transmission constraints) \cite{li2022heterogeneous}. 
(3) 
The optimization problems in 6G wireless networks usually involve real-time network-dependent parameters, such as the network structure, channel state information (CSI), traffic condition, etc.  
Therefore, the near-optimal performance of various optimization-based
algorithms (OAs) should be achieved in real time, which is a fundamental challenge.
  
Highly effective algorithms based on classical optimization theory have been extensively developed for various classes of optimization problems in 6G. 
For example, many iterative algorithms (e.g., approximate message passing (AMP), orthogonal matching pursuit (OMP) and alternating direction method of multipliers (ADMM)) are designed for signal recovery (e.g., signal detection \cite{mehrdad20MIMOdetect} and channel estimation \cite{liangtian21ADMMnet}). 
The semidefinite relaxation (SDR) and successive convex approximation (SCA) techniques are widely applied to solve non-convex optimization problems (e.g., non-convex quadratically constrained quadratic programming problems \cite{Yuanming_fang2021over}) in wireless communications.
Despite that some of these algorithms can achieve good performance through theoretical analysis and numerical simulations, classic optimization-based algorithms (COAs) in realistic 6G applications face many challenges. 

\subsubsection{Optimization Performance}
Due to the highly non-convexity, COAs can be intractable, suboptimal or heuristic without performance guarantees.  
Besides, the tuning of free-parameters in COAs relies on prior knowledge or model assumption, whose setting can greatly affect the achievable performance of COAs. 
For example, iterative soft thresholding algorithm (ISTA) for sparse recovering suffers from an inherent trade-off between estimation performance and convergence rate, which is controlled by the choice of regularization parameter \cite{shi2022twc4ma}.

\subsubsection{Computational Cost}
Most of the COAs are iterative in nature, which typically induces high computational cost to obtain optimal solutions. 
For example, for solving mixed combinatorial optimization problems, the complexity of branch-and-bound (BB) algorithm grows exponentially with the scale of problem \cite{yifei20lorm}. 
However, most signal processing techniques in 6G have stringent latency requirement. 
Furthermore, the COAs depend on the real-time environmental parameters (e.g., network topology, channel conditions). 
When the environment changes, the iterative COAs need to be executed repeatedly to accommodate to the dynamic environment, which is unfordable for time-sensitive applications. 

\subsubsection{Tractability of System Modeling}
The design of COAs highly depends on the availability and accuracy of system modeling. 
However, it is hard to precisely capture the network architecture, communication environment and transmission data links using mathematical models due to the time-varying, heterogeneity, complexity and nonlinearity in 6G wireless networks. The imperfect and mismatched system model can greatly deteriorate the performance of COAs when applied in practical systems.

Therefore, traditional COAs are computational inefficient and scale poorly for large-scale optimization problems in 6G systems. The reliance on perfect mathematical models and possible intractability of optimal solutions make the COAs entail serious performance gap between theoretical design and real-time application \cite{Yu2022RoleOD}.
Motivated by the disadvantages of COAs, ML has surged as a powerful technique to solve the challenging optimization problems in wireless networks. 

\begin{table}[htbp]
	\centering
	\caption{LIST OF ACRONYMS IN ALPHABETICAL ORDER}
	\label{ta:notation}
	\begin{tabular}{cc}
		\toprule
		Acronym & Explanation                                   \\ \toprule
		6G      & 6th Generation                                \\ \hline
		ADMM    & Alternating Direction Method of Multipliers   \\ \hline
		AE      & Auto-Encoder                                  \\ \hline
		AI      & Artificial Intelligence                      \\ \hline
		AMP     & Approximate Message Passing                  \\ \hline
    BB      &  Branch-and-Bound                            \\ \hline
		BS      & Base Station                                 \\ \hline
		CMCP    & Complex Modulus Constrained Problem         \\ \hline
		CNN     & Convolutional Neural Network                 \\ \hline
		COA     & Classic Optimization-Based Algorithm         \\ \hline
		CS     & Compressive Sensing                       \\ \hline
		CSI     & Channel State Information                     \\ \hline
		D2D     & Device to Device Communication                \\ \hline
		DL      & Deep Learning                                 \\ \hline
		DNN/NN     & Deep Neural Network/Neural Network                             \\ \hline
		DQN/DQL     & Deep Q-Network/Deep Q-Learning                               \\ \hline
		DRL     & Deep Reinforcement Learning                   \\ \hline
		FL      & Federated Learning                            \\ \hline
		GCN     & Graph Convolutional Network                   \\ \hline
		GNN     & Graph Neural Network                    \\ \hline
		IB      & Information Bottleneck                        \\ \hline
		IoT     & Internet of Things                            \\ \hline
		ISTA    & Iterative Soft Thresholding Algorithm         \\ \hline
   JADCE    & Joint Activity Detection and Channel Estimation       \\ \hline
		JSCC    & Joint Source-Channel Coding                   \\ \hline
		LBB     & Learning to Branch-and-Bound                  \\ \hline
		MAS/MADRL     & Multi-Agent System/Multi-Agent DRL        \\ \hline
		MDP     & Markov Decision Process                   \\ \hline
		MIMO    & Multi-Input Multi-Output                      \\ \hline
		MINLP    & Mixed Integer Nonlinear Problem                   \\ \hline
		MIP    & Mixed Integer Programming                 \\ \hline
		ML      & Machine Learning                              \\ \hline
		MLP     & Multi-Layer Perceptron                        \\ \hline
		MOA     & Machine Learning-Based Optimization Algorithm \\ \hline
		MSE     & Mean Square Error                             \\ \hline
    OA      & Optimization-Based Algorithm                  \\ \hline
    OMP     & Orthogonal Matching Pursuit                  \\ \hline
		QoS     & Quality of Service                            \\ \hline
		RL      & Reinforcement Learning                   \\ \hline
		RNN     & Recurrent Neural Network                      \\ \hline
		RIS     & Reconfigurable Intelligent Surfaces                    \\ \hline
 		SNR     & Signal to Noise Ratio                         \\ \hline
		SDR     & Semidefinite Relaxation                         \\ \hline
		SCA     & Successive Convex Approximation               \\ \hline
		VAE     & Variational Auto-Encoder                   \\ \hline
		WMMSE   & Weighted Minimum Mean-Square Error           \\ \bottomrule
	\end{tabular}
\end{table}

\subsection {Machine Learning in 6G} 
The goal of ML-based optimization algorithm (MOA) design is to achieve near-optimal performance with high computational efficiency for challenging large-scale optimization problems in 6G wireless networks, enabling a paradigm shift from classic optimization theory-based approaches to employing more promising deep learning (DL) architectures \cite{kb196G}. 
ML for large-scale optimization features the following advantages \cite{amos2022tutorial,shlezinger2022model}.

\subsubsection{\textit{Superior Performance}} 
MOAs entail near-optimal or superior learning performance compared with COAs due to the data-driven feature as well as the sophisticated design of neural networks (NNs) and learning strategies.  
For example, algorithm unrolling methods enjoy a superior performance for signal detection \cite{hengtao2020MIMOdetection}, channel estimation \cite{xiuhong21GM-LAMP} and precoding design \cite{Hu2020unrolling} compared with their corresponding COA counterparts and other traditional algorithms. 
Graph neural network (GNN) enjoys better performance for resource allocation problems with fewer iterations compared with traditional weighted minimum mean-square error (WMMSE) algorithm \cite{yifei21gnn}.

\subsubsection{\textit{Scalability and Generalizability}} 
With enhanced learning capacity, MOAs can be used to solve large-scale and complex optimization problems. Incorporating the properties of target task into the NN architectures further improves the scalability and generalizability of MOAs. For example, message passing-based GNNs can be safely generalized to solve large-scale problems even when trained on small-scale samples \cite{yifei21gnn}, thereby leading to reduced training cost. The decentralized nature of some specialized MOAs enables the efficient training in large-scale networks, such as the multi-agent reinforcement learning (MARL) \cite{khan2018scalable}. Advanced ML techniques, such as  transfer learning \cite{yifei20lorm} and meta-learning \cite{zhang2020meta} can tackle the task mismatch problems with fewer training samples, thereby improving the generalizability of MOAs.

\subsubsection{\textit{Computational Efficiency}} 
ML inference only requires a small number of simple operations and can be realized in real time. 
By shifting the computations from online to offline, ML is highly attractive for computational intensive optimization tasks in 6G \cite{Yu2022RoleOD}. The online deployment of well-trained MOAs can effectively reduce the system delay and improve the overall performance.

\subsubsection{\textit{Robustness}} 
ML-based approaches shall be robust to the imperfect model assumptions and dynamic wireless environment due to the data-driven nature. Through learning from experiences, MOAs work well even under unknown environment when the mathematical model is unavailable. 
For example, continual learning \cite{sun2022learning} is robust to the dynamic environments by sequentially handling different wireless parameters (e.g., different CSI distributions).


Despite of the recent successful tries of MOAs, the important questions in this context are what kind of optimization problems can be effectively solved by ML techniques and which ML techniques would provide reliable, timely and effective solutions for these optimization problems.  
In this paper, we will try to answer these questions by focusing mainly on some exemplary ML design frameworks applied to solve various large-scale optimization problems in 6G networks.
Within each framework, we highlight the motivations, the NN design principles, type of optimization problems, toy examples of its applications in 6G wireless networks, the theoretical analysis, the related research challenges as well as the summary of advantages and disadvantages.  
Specifically, the algorithm unrolling \cite{shi2022twc4ma}, learning to branch-and-bound (LBB) \cite{yifei20lorm}, GNN for structured optimization \cite{yifei21gnn}, deep reinforcement learning (DRL) for stochastic optimization \cite{9650909}, end-to-end learning for semantic optimization \cite{guangming21semcom} as well as federated learning (FL) for distributed optimization \cite{elbir2021federated} will be covered in the following sections to shed light on the excellent performance of ML compared with the conventional optimization algorithm in a variety of practical domains and provide guidance on the usage of ML techniques in 6G networks, followed by the summary of network design philosophies, theoretical tools, implementation issues and discussion of future directions to drive forward the research in this area. 
Before the detailed elaborations of specific MOA designs, we firstly summarize the existing design paradigms of MOAs from different perspectives in the following subsections.   
\begin{figure*}[h]
	\centering
	\includegraphics[width=1\textwidth]{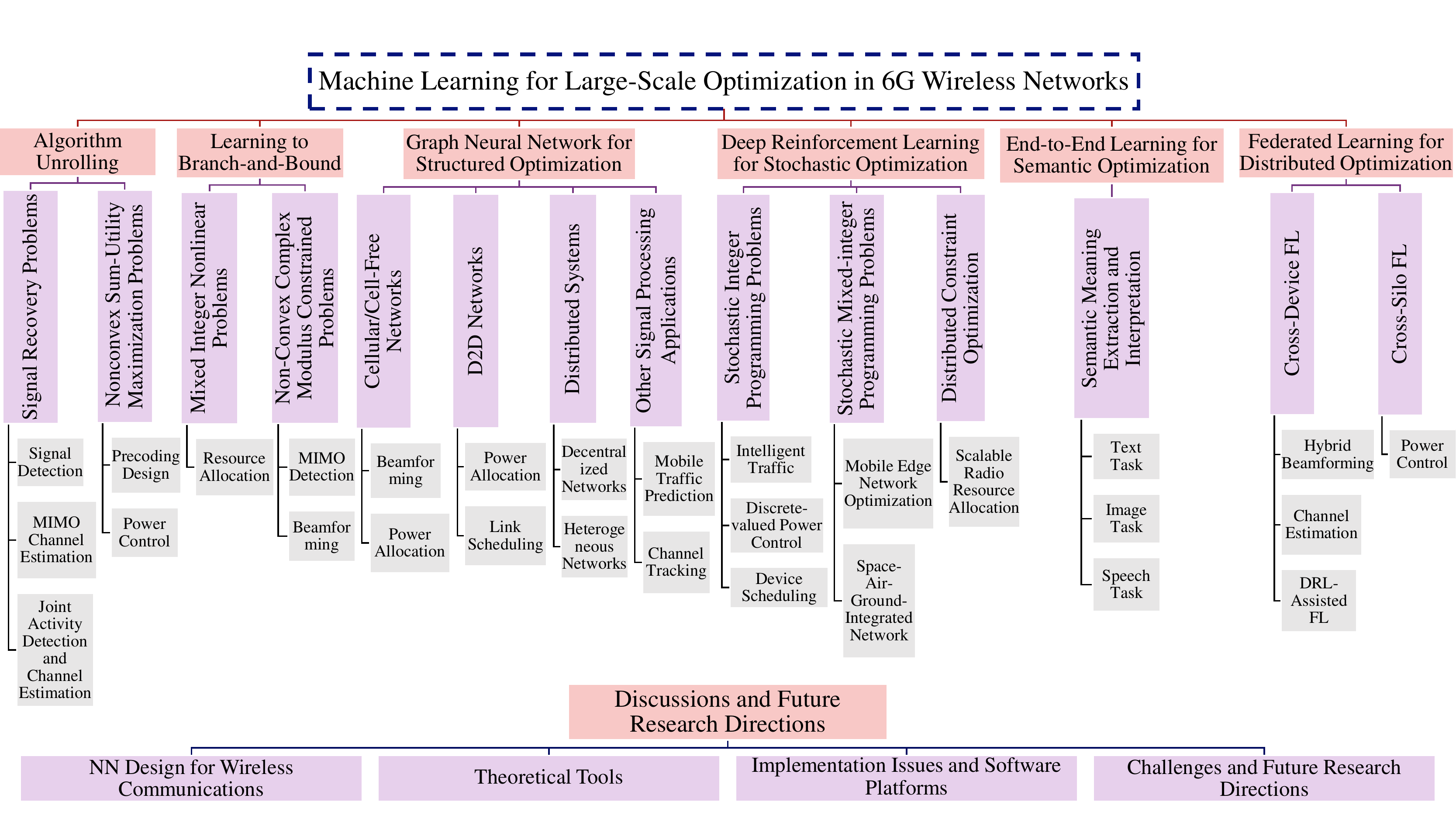}
	\caption{An overview of the main topics of ML-based optimization algorithms, related problems and the wireless applications.}
	 \label{fig:intro}
\end{figure*}

\subsection {Category for ML-Based Optimization Algorithms} 
\subsubsection{Learning Principle}
From the perspective of learning principle, \textit{learning to optimize} can be divided into supervised and unsupervised learning.
With the labeled training data, supervised learning directly learns the nonlinear mapping between problem parameters and optimal solutions of optimization problems through generic NNs (e.g., multi-layer perceptron (MLP) for channel estimation \cite{solt19DL4ce}) or specialized NNs (e.g., algorithm unrolling for joint active device and channel estimation (JADCE) \cite{shi2022twc4ma}).
The universal approximation theorem states that feed-forward NN with one single hidden layer can approximate continuous functions to arbitrary precision \cite{haykin2004comprehensive}, which provides theoretical support for supervised MOA designs by treating NN as a universal function approximator.
With the unlabeled training data, unsupervised approach learns the optimal solutions by adopting the objective function of original optimization problem as the training loss function. Then the NN is evaluated and updated by means of any gradient-based optimizer (e.g., GNN takes minus weighted sum rate as loss function for scalable radio resource management \cite{yifei21gnn} and DRL aims to maximize expected cumulative discounted rewards for resource allocation in vehicle-to-vehicle communications \cite{8633948}). 

Compared with the COAs, supervised MOAs enjoy similar or superior optimization performance by leveraging the strong representation power of NN. 
By constructing connections with COAs, the supervised MOAs show better interpretability and enables performance analysis for certain network architectures. 
However, the original optimization problem needs to be solved repeatedly with varying problem parameters to collect the training labels, which can induce huge computational costs in data acquisition. 
Besides, in supervised MOA, the NN is trained as a universal approximator of an existing optimization algorithm, and thus the performance greatly depends on that of existing algorithms.
Instead, the unsupervised MOAs can greatly simplify the process of data acquisition and can be effectively adopted to solve non-convex or NP-hard problem, where no tractable COAs can be found, thereby greatly eliminating the dependence on the COAs and facilitating great flexibility to search for optimal solutions. 
However, its performance is primarily restricted by the type of optimization problems. 
For highly non-convex problem suffering from the ``curse" of local minima, unsupervised MOA may be trapped into a spurious solution with poor performance. 
In addition, the unsupervised MOA is usually computational complex and requires a larger training set to produce expected outcomes.
\subsubsection{NN Architecture}
From the perspective of NN architecture, there are mainly two design principles as summarized below. 
\begin{itemize}
\item \textit{Generic NN:}
Most of the MOAs adopt generic NNs, such as MLP, convolutional neural network (CNN), autoencoder (AE), etc, which are not tailored for specific tasks. 
The generic NNs are not specific to a particular task or dataset, but can be applied for general tasks and datasets to achieve an acceptable performance.
For example, with tremendous training data and large NN architecture, MLP is usually adopted as a benchmark algorithm for performance
comparison in multi-input multi-output (MIMO) detection \cite{neev19l2detect}, power control \cite{Shen2022GraphNN}, semantic communication \cite{xie2021deep}, etc. 
However, the flexibility comes with the cost of poor data efficiency (high training overhead), poor robustness and poor generalization ability. 
\item \textit{Task-Specific NN:}
Another design principle of NN architecture is customized implementation by incorporating the structure of target task, datasets and domain-specific knowledge into the NN architecture. 
The specialized NN demonstrates some unique advantages empirically and theoretically, such as robustness to model uncertainties, scalability and generalizability in large-scale problem, and high training efficiency.  
Some of specialized NNs can handle tricky constraints in the optimization problems (e.g., integer or constant envelope constraints) and enable performance guarantees under certain conditions.
Nonetheless, the task-specific NN is highly customized and different problems require separate NN designs. 
Besides, for complex problems, it can be hard to construct a specific NN. 
The design of specific NN highly depends on the structure of problem itself, or the property of existing algorithms that can be used to solve the problem. 
For example, the NN architecture of algorithm unrolling comes from parameterizing the original iterative algorithm \cite{monga2021algorithm}.   
Therefore, the key to designing a specialized NN is to identify the special characteristics of problem/datasets/classic algorithms and sophisticatedly integrate them into the design of the NN architecture and training algorithm.

\end{itemize}

\subsubsection{Theoretical Analysis}
From the perspective of theoretical analysis, most of the MOAs treat the NN as a black box without interpretations (e.g., MLP, CNN and AE), whose performance can only be demonstrated numerically. 
However, in wireless communication systems, transparentness and reliability are of pivotal importance for a practical algorithm design \cite{wen2022challenges}. 
Therefore, it is paramount to understand the learning mechanism of NN and its applicable conditions. 
Inspired by the pertinent COAs, some of the MOAs in the ``supervised learning" category enable theoretical analysis by constructing a relationship of performance between these two types of methods. 
If equivalence can be proved, the performance analysis of MOAs can be developed based on that of COAs  (e.g., the performance analysis of algorithm unrolling approaches \cite{shi2022twc4ma}, LBB approach \cite{yifei20lorm}, GNN approach \cite{Shen2022GraphNN}, etc.). 

\subsection{Related Works and Our Contributions}
There exist several survey papers on ML for wireless communications \cite{chen2017machine, zappone2019wireless, zhang2019deep, qIAN_CST18, sun2019application}.
All these studies provide visions of artificial intelligence-based wireless network designs, enumerating on the applicable cases and scenarios in wireless networks where the ML can make a viable impact. Particularly, the vast majority of surveys \cite{chen2017machine, zappone2019wireless, zhang2019deep, qIAN_CST18, sun2019application} started with the introduction of the fundamentals of ML, including the ML theory, the framework of generic deep networks as well as its update rules and training methods, followed by envisioning the wireless applications of different ML techniques in future wireless networks from different aspects. 
Among them, Zappone \textit{et al.} \cite{zappone2019wireless} provided an in-depth quantitative analysis for each use-case of DL-based wireless network design. 
Mao \textit{et al.} \cite{qIAN_CST18} focused on the discussion of DL applications in different wireless communication layers (i.e., physical layer, data link layer, network layer and upper layer), while Zhang \textit{et al.}. \cite{zhang2019deep} considered the mobile, sensor networks and their related applications. 
However, none of existing works provides a thorough survey of ML/DL techniques for large-scale wireless networks from the perspective of optimization. 
In light of the underlying different optimization problems involved in a variety of practical fields, various customized ML paradigms are extensively reviewed in this paper. 
By identifying the task-specific structures of large-scale optimization problems and classifying the existing ML frameworks accordingly, this paper bridges the elusive ML algorithms and well-grounded optimization theory to improve the interpretability and transparency of deep neural networks (DNNs), and inspires a game-changing new perspective for solving large-scale optimization problems in wireless networks.
The major contributions are summarized as follows:
\begin{itemize}
	\item The challenges of wireless network optimizations in 6G systems, as well as the restrictions of traditional optimization algorithms are discussed to motivate the ML-based wireless network optimizations. 
	The existing design paradigms of MOAs are systematically elaborated from different perspectives, such as learning principles, NN architectures and theoretical foundations, which are presented in Section \ref{sec:1}.
	\item The connections between large-scale optimization problems and specialized DL algorithms are thoroughly discussed to reveal the potential of MOA for improving the system performance and provide insightful guidance on the use of ML in 6G networks. 
	Specifically, some promising learning to optimize frameworks (i.e., algorithm unrolling, LBB, GNN and DRL) are thoroughly discussed from Section \ref{sec:2} to Section \ref{sec:5}, detailing properties and classifications of large-scale wireless optimization problems, NN design patterns catering to the features of optimization problems and the use-cases in various wireless applications.
	\item Semantic communication is elaborated for end-to-end optimization of communication systems in Section \ref{sec:6}, which provides a classic semantic communication architecture to endow the NNs with the ability of semantic information extraction and recovery to optimize the transmission efficiency and reliability trade-offs.
	\item Federated learning, as a prominent distributed ML scheme to effectively solve the model optimization problem in ML over hyper-scale wireless networks, is described in Section \ref{sec:7}, where the issues of network/data heterogeneity and instability, as well as the deployment in different wireless communication systems are illustrated.
	\item The guidelines for MOA designs in terms of the choice of loss functions, network architectures, training methods, and the ways to handle optimization constraints are given in Section \ref{sec:8}. 
	The issues pertaining to the theoretical progress of various MOAs, implementations as well as challenges and future research directions are also presented in Section \ref{sec:8}. 
\end{itemize}

We summarize the main topics of ML-based optimization algorithms, related problems and the wireless applications in Fig. \ref{fig:intro}.

\section{Algorithm Unrolling}\label{sec:2}
In this section, we introduce one widely adopted ``learn to optimize" algorithm design framework, termed algorithm unrolling, which constructs a layered network to mimic each iteration of a classic iterative algorithm. 
We start from the motivation of algorithm unrolling and its design framework, which provides a guidance on how to unroll an iterative algorithm, followed by the case studies in wireless communication networks. 
The advantages and disadvantages of algorithm unrolling are summarized at the end of this section. 

\subsection{Motivations and Design Frameworks}
\begin{figure*}[h] 
  \centering
  \includegraphics[width=1\textwidth]{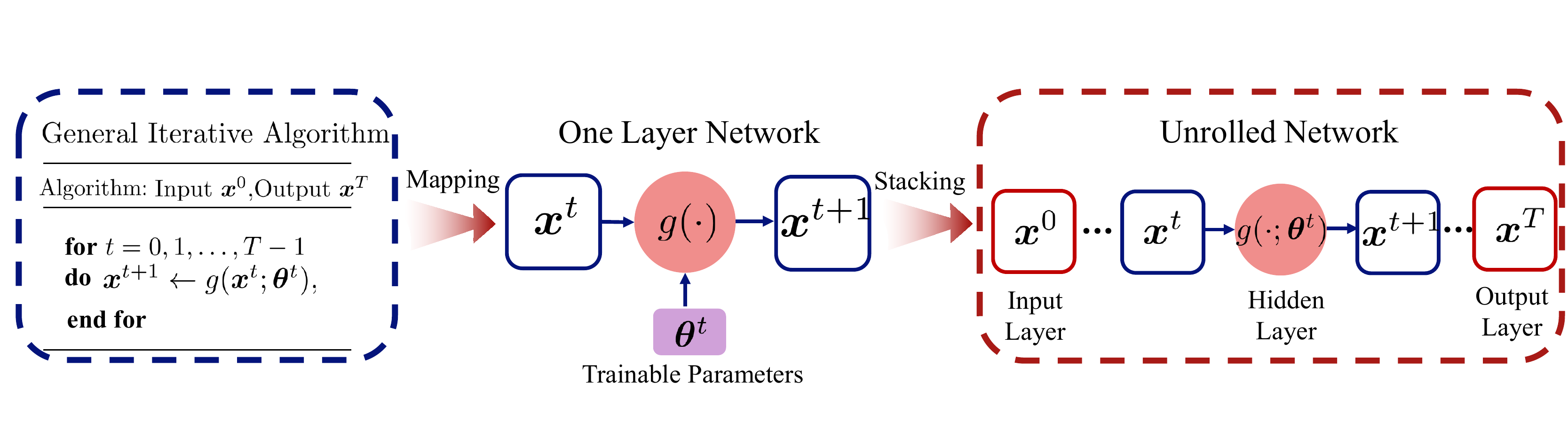}
  \caption{The general framework of algorithm unrolling method.}
  \label{fig:au}
\end{figure*}
The success of ``end-to-end learning" framework requires huge training datasets and significant computational cost due to a large number of training parameters of generic NNs while serving as a universal function approximator. 
The low training efficiency of generic NN hinders its application for dynamic and large-scale wireless networks. 
Moreover, in future 6G wireless networks with scenarios where abundant high-quality training samples such as CSI are unavailable, the performance of general DNN may significantly degrade and even underperform traditional algorithms. 
Furthermore, the generic NN is treated as a “black-box”, where the functionality of each layer and the performance guarantees of NN are hard to obtain. 
The lack of interpretability of black-box DNNs can be a serious limitation in contrast with optimization-based approaches with theoretical guarantees in wireless networks, where reliability and predictability are of vital importance. 
To address these limitations, algorithm unrolling is an emerging method that provides a concrete and systematic connection between classic iterative algorithms and deep neural networks. 
By unfolding an iterative algorithm with algorithm parameters transferred to training parameters of NN, the unrolled NN enables interpretability of each layer and even theoretical guarantees are possible. 
Due to the potential in developing efficient, high-performance and theoretical guaranteed NN using reasonably sized training sets, it is a pleasant surprise that algorithm unrolling rapidly grows in both theoretic investigations and practical applications \cite{monga2021algorithm}.

Algorithm unrolling was first introduced by Gregor and LeCun \cite{gregor2010learning} to accelerate the ISTA for improving the computational efficiency of sparse coding. 
The basic idea is to map each ISTA iteration to a neural network layer and then stack the layers together, which can be viewed as executing an ISTA iteration multiple times by a layer-wise neural network. 
The same techniques can be further applied to general iterative algorithms, in which the update form is given by
\begin{align}
\bm{x}^{t+1} = g(\bm{x}^t; \bm{\theta}^t), \quad t = 0, 1, 2, \cdots, T.
\end{align}
Here, $\bm{x}^t \in \R^{n}, t=1,\cdots, T$ are the iterative variable vector (e.g., the signal to be recovered or the variable to be optimized), $g(\cdot;\cdot): \R^n \rightarrow \R^n$ is the iterative function of a specific iteration algorithm, and $\bm{\theta}^t \in \R^{m}, t=1,\cdots, T$ are trainable parameters (including model parameters and regularization coefficients) of the algorithm.
The overall principle of algorithm unrolling is to unroll a specific iterative algorithm into a deep network by mapping each iteration function $g(\cdot;\cdot)$ into a single network layer and stacking a finite number of layers together.
The forward procedure of NN is equivalent to the execution of the iterative algorithm. 
The unrolled network architecture thus depends on the original iterative algorithm (e.g., the unrolled ISTA turns to be an recurrent neural network (RNN) \cite{gregor2010learning}), as the single network layer shares the same structure of iteration function $g(\cdot;\cdot)$.
The details for the algorithm unrolling are illustrated in Figure. \ref{fig:au}. 
The trainable parameters $\bm{\theta}^t \in \R^{m}, t=1,\cdots, T$ can be learned through end-to-end approach:
\begin{align}\label{fnc:loss}
\mathop{\rm{minimize}}\limits_{\bm{\Theta}_T}~ \mathcal{L}\left(\bm{x}^{T+1}(\bm{\Theta}_T)\right),
\end{align}
where $\mathcal{L}(\cdot)$ is the loss function for training, $\bm{\Theta}_T=\{\bm{\theta}^t\}_{t=0}^{T}$ is the entire trainable parameters of total $T$-layer network and $\bm{x}^{T+1}(\cdot)$ is the output function of the unrolled network.
Due to the customized structure of NN, the end-to-end training may suffer from spurious local minima and gradient explosion or vanishment during the training. 
Instead of directly solving \eqref{fnc:loss}, the common adopted training strategy for unrolled networks is layer-wise method \cite{mark2017amp}, which can achieve more efficient training due to the better parameter initialization.
That is, the whole training process can be divided into $T$ sequential sub-training processes. 
For the $t$-th sub-training process, we aim to refine the trainable parameters $\bm{\Theta}_t$, where a two-stage method is used. 
The first stage is dedicated to optimize parameter $\bm{\theta}_t$ individually, while the latter learns the whole $\bm{\Theta}_t$ jointly by fixing the learned $\bm{\theta}_t$ as initialization.
In the testing stage, feeding the data forward through the unrolled network with learned parameters is equivalent to executing the parameter-optimized iterative algorithm for a finite number of iterations.

In recent years, algorithm unrolling approaches have been extensively applied to solve various signal processing problems, such as sparse and low rank regression, probabilistic graphical model, differential equations and quadratic optimization \cite{chen2021learning}, and have been applied to a wide range of applications including but not limited to compressive sensing (CS) \cite{yang2018admm}, deconvolution \cite{li2019algorithm}, and image processing \cite{chen2016trainable}. 
The superior performance and high computational efficiency of algorithm unrolling have been proved in various domains. 
At a high level, algorithm unrolling methods take advantages of both optimization-based priors and data-driven learning ability of NN.  
Compared with generic black-box NNs, unrolled NNs have much fewer parameters due to the inheritance
of the structure and domain knowledge from a specific iterative algorithm, allowing it to benefit in terms of the computational efficiency, interpretability, generalizability and reliability. 
Theoretical tools from the traditional optimization theory can be explored to describe the convergence behavior and performance guarantees of unrolled NN. 
Compared with iterative counterparts, algorithm unrolling enables improved performance due to its expanded representation capability by tuning trainable parameters of model-based algorithm through data-drive learning.

Due to the stringent requirements of large-scale 6G wireless networks on efficiency and reliability, algorithm unrolling has recently attracted much attention in wireless communications for solving sparse optimization problems and some non-convex optimization problems.
In the following subsections, we review several representative cases of algorithm unrolling in wireless communication applications. 
Table \ref{ta:au} summarizes the types of problems we will cover, the corresponding application topics, the underlying iterative algorithms and the type of trainable parameters of the unrolled methods. 


\begin{table*}[htp]
\centering
\caption{Algorithm Unrolling Methods for Wireless applications}
\begin{tabular}{ccccl}
\hline
Target Problems                                  & Applications                                                     & Unrolled Networks                                              & \tabincell{c}{Original \\ Algorithms} & Type of Trained Parameters                                        \\ \hline
\multirow{10}{*}{\tabincell{c}{Signal Recovery \\ Problems}}   & \multirow{4}{*}{MIMO Detection}                                  & DetNet \cite{neev19l2detect}                  & PGD                 &\tabincell{c}{DNN weight and gradient step-size}                             \\ \cline{3-5}                    &      & OAMPNet \cite{hengtao2020MIMOdetection}       & OAMP                & \tabincell{c}{Step-size and nonlinear estimator factor}                        \\ \cline{3-5} 
  &  & MMNet \cite{mehrdad20MIMOdetect}              & ISTA                & Scale factor                                                     \\ \cline{3-5} 
   &        & CMDNet \cite{edagr21CMDNet}                   & CMD                 & Step-size and gradient scale                                      \\ \cline{2-5} 
  & \multirow{3}{*}{Channel Estimation}& GM-LAMP \cite{xiuhong21GM-LAMP}  & AMP     & \tabincell{l}{Linear transform coefficient and shrinkage \\ parameter}       \\ \cline{3-5} 
 &          & mpNet \cite{taha22mpNet}      & MP                  & Linear transform coefficient      \\ \cline{3-5}    &          & ADMM-OGChannelNet   \cite{liangtian21ADMMnet} & ADMM                & Linear transform coefficient and step-size                      \\ \cline{2-5}            & \multirow{3}{*}{JCADE} & DADMM \cite{zhendong2022JCEAUiot}             & ADMM                & \tabincell{l}{Step-size, shrinkage parameter and auxiliary \\ variable}            \\ \cline{3-5}         &                                                                  & FAT-DL \cite{xiaodan21ampnet}                 & AMP                 & Denoiser factor and scale factor                                \\ \cline{3-5}       &           & LISTA-GS \cite{shi2022twc4ma}                 & ISTA-GS             & \tabincell{l}{Linear transform coefficient, step-size and \\ shrinkage parameter} \\ \hline
\multirow{4}{*}{\tabincell{c}{Non-convex Sum-Utility \\ Maximization Problems}} & \multirow{4}{*}{Precoding Design}                             & IAIDNN \cite{Hu2020unrolling}                 & WMMSE               &\tabincell{l}{Linear transform coefficient and trainable \\ offset}               \\ \cline{3-5}     &                                       & PDD-TJAPB \cite{mingmin21twcwmmse}            & WMMSE               & Long-term variable                                               \\ \cline{3-5}     &   &  UPGDNet  \cite{he2022unsupervised}          & PGD              & DNN weight and scale factor                   \\ \cline{2-5}
& Power Control

& PDD-SSCA \cite{an21tspdeepunfolding}          & WMMSE               & Short-term variable                      \\ \hline
\end{tabular}
	\label{ta:au}
\end{table*}

\subsection{Application 1: Signal Recovery Problems} 
With the emergence of large-scale signal processing techniques in 6G systems, such as big data, massive IoT, massive access network, large-scale antennas, etc., signal recovery, especially sparse signal recovery, has been a widely encountered optimization problem in diverse wireless applications \cite{qin2018sparse, ShiZCL18}. 
Consider the following widely accepted model for signal recovery in wireless networks
\begin{align}\label{eq:srp}
\bm{y} = \bm{A} \bm{x} + \bm{w}, 
\end{align}
where $\bm{y}$ is the received vector or matrix at the BS, $\bm{A}$ is the measurement matrix (e.g., the channel matrix in signal detection, beam selection matrix in beamspace channel estimation and pilot matrix in JADCE), $\bm{w}$ is Gaussian noise and $\bm{x}$ is the unknown signal to be recovered (e.g., symbols in signal detection, channels in channel estimation and channels of active devices in JADCE). 

When the $\bm{x}$ in \eqref{eq:srp} is a sparse signal, the well known ISTA has the following simple iterative expression: $ \bm{x}^{t+1} = \eta_{\lambda} (\bm{x}^t + \frac{1}{C}\bm{A}^T(\bm{y} - \bm{A}\bm{x}^t) )$, where $\eta$, $\lambda$ and $C$ are threshold operator, regulation parameter and step size, respectively.
However, traditional ISTA suffers from high computation complexity for recovering sparse signals.
By a simple variable substitution to separate the inputs and output of network (i.e., $\bm{W}^t_1 = \frac{1}{C}\bm{A}^T$, $\bm{W}^t_2 = \bm{I} -  \frac{1}{C}\bm{A}^T \bm{A}$, and $\theta^t = \lambda$) and parameterizing them as trainable factors, the algorithm unrolling approach maps the original ISTA into an unrolled RNN as follows with $t = 0, 1, \ldots, T-1$,
\begin{align}\label{eq:LISTA}
	\bm{x}^{t+1} = \eta_{\theta^t}\left(\bm{W}_1^t \bm{y} + \bm{W}_2^t \bm{x}^{t}\right).
\end{align}
This approach inherits the structure and domain knowledge of the ISTA-based algorithm, also improves the convergence rate and computational efficiency of original algorithm through end-to-end training, which can be extended for recovering signals with different sparse patterns (e.g. group sparse pattern \cite{shi2022twc4ma}).

In summary, the imperfectness of practical data links and inherent various sparsity structures of practical signals render many conventional (sparse) signal recovery algorithms, such as ISTA, AMP, OMP, etc., suffering from the unsatisfactory performance, slow convergence and high complexity when employed in practical large-scale wireless networks.
Due to the superior performance, algorithm unrolling methods have been applied to solve signal recovery problems in the applications of  signal detection
 \cite{neev19l2detect,hengtao2020MIMOdetection,mehrdad20MIMOdetect,edagr21CMDNet}, channel estimation \cite{xiuhong21GM-LAMP,taha22mpNet,liangtian21ADMMnet} and JADCE \cite{zhendong2022JCEAUiot,xiaodan21ampnet,shi2022twc4ma}.

\subsubsection{Data Detection}
Data detection at the receiver has been a challenging task due to the wireless channel fading. 
Conventional data detection techniques, such as linear detectors \cite{bjornson2017massive}, SDR \cite{jalden08srd}, sphere decoding \cite{viter1999sphere}, AMP \cite{jeon2015optimality}, etc., depend on the mathematical models of wireless channels and usually assume perfect CSI at the receiver which is replaced by its estimate in the practical system. 
The channel dependence of data detection undermines the detector performance in practical wireless systems where the channels can be highly complex, poorly understood, hard to be modeled or with estimation errors. Moreover, the traditional detection techniques suffer from significant complexity-reliability trade-off, which cannot be efficiently implemented at the scale required by 6G massive MIMO systems.
To overcome these limitations, 
ML-based receivers have been extensively studied, which learn the mapping from the channel outputs to the transmitted symbols in a data-driven manner, and several algorithm unrolling approaches for MIMO detection including DetNet \cite{neev19l2detect}, OAMPNet \cite{hengtao2020MIMOdetection}, MMNet \cite{mehrdad20MIMOdetect} and CMDNet \cite{edagr21CMDNet} were proposed.

When assuming perfect CSI at receiver, Samuel \textit{et al.} \cite{neev19l2detect} unfolded the projected gradient descent algorithm called DetNet by treating the gradient step sizes as learned parameters, followed by common MLPs for improving expressive power.
However, the architecture of DetNet does not contain the properties of iterative methods, leading to high complexity and poor interpretability. 
To further improve detection performance with imperfect CSI, a model-driven unrolled orthogonal AMP (OAMP) network called OAMPNet \cite{hengtao2020MIMOdetection} was proposed to jointly estimate channel and detect signal by taking channel statistics and channel estimation errors into consideration, which has only a few trainable parameters and can be trained with a much smaller data set compared with DetNet.
DetNet and OAMPNet are both trained offline with simple model assumptions (e.g., i.i.d. Gaussian channels, low-order modulation schemes) and can suffer a large performance gap for realistic channels \cite{mehrdad20MIMOdetect}.
To overcome these limitations, MMNet \cite{mehrdad20MIMOdetect} was proposed by unrolling ISTA, which enables online training by exploiting the locality of realistic channels in both frequency and time domains.
In particular, MMNet parameterizes the linear transformation in ISTA, followed by the estimation of noise variance for different transmitted symbols in the nonlinear denoiser.
To further support fast online training, an unrolled concrete maximum a-posteriori detection network (CMDNet) was proposed by \cite{edagr21CMDNet}, which is theoretically revealed to be able to learn the approximate probabilities of the individual optimal detector.

\subsubsection{Massive MIMO Channel Estimation}
In order to optimize the data rate and energy consumption trade-off, channel estimation is crucial in communication systems.
As there are only a few dominant propagation paths despite the high dimension of the channel, massive MIMO channel estimation problems turn out to be sparse signal recovery problems. 
Classical CS-based algorithms suffer from high computational complexity and poor estimation accuracy in low signal to noise ratio (SNR) regions, especially for large-scale antenna arrays in wide-band system with complex sparse structures.
Recent years witness an emergence of a number of unrolled channel estimation networks benefiting from both domain knowledge and DL, such as GM-LAMP \cite{xiuhong21GM-LAMP}, mpNet \cite{taha22mpNet} and ADMM-OGChannelNet \cite{liangtian21ADMMnet} for MIMO channel estimation problems.

In millimeter wave (mmWave) MIMO systems, GM-LAMP \cite{xiuhong21GM-LAMP} unrolls AMP by deriving a new shrinkage function based on the Gaussian mixture prior information of beamspace channels to improve the estimation accuracy of AMP.
To address the basis mismatch issue in off-grid mmWave channel estimation problem, deep unrolled network architecture ADMMOGChannelNet \cite{liangtian21ADMMnet} was proposed by mapping the data flow to the iterative procedures of ADMMOG algorithm, which is computationally more efficient with performance guarantees.
However, due to the intrinsic supervised nature, these methods all require collecting a database of clean channels for offline training, which may hinder their practical applicability.
When ground-truth channel data are unavailable, an unrolled matching pursuit mpNet \cite{taha22mpNet} was designed for MIMO channel estimation in an unsupervised way, which can automatically correct its channel estimation algorithm based on incoming data with the advantage of training online due to its simple network structure.

\subsubsection{Joint Activity Detection and Channel Estimation}
The JADCE problem in grant-free massive access scenario \cite{shi2022twc4ma} is a typical sparse optimization problem in wireless networks, as the sporadic transmission leads the joint activity and channel matrix to exhibit group-sparse pattern.
On the other hand, solving JADCE problem also becomes more challenging for large-scale networks with large-scale antenna arrays and massive number of IoT devices.
Considering a multi-antenna BS and a large number of devices, the signal model in JADCE can be written as $\bm{Y} = \bm{A}\bm{X} + \bm{W}$, where $\bm{A}$ denotes the pilot matrix and $\bm{X} = \bm{\Lambda} \bm{H} $ is the device state matrix with diagonal
activity matrix $\bm{\Lambda} $ and channel matrix $\bm{H}$.
To recover the group-sparse device state matrix $\bm{X}$ with improved estimation accuracy and low computational complexity, extended unrolled versions of ADMM-based \cite{zhendong2022JCEAUiot}, AMP-based \cite{xiaodan21ampnet} and ISTA-based \cite{shi2022twc4ma} frameworks were proposed, respectively, for JADCE problems in massive access scenarios.

Assuming the device state matrix enjoys Bernoulli-Gaussian mixture distribution, an AMP-based unrolled network with dimension reduction module was proposed by \cite{xiaodan21ampnet} to reduce the length of pilot sequences and computational complexity for JADCE problems.
To directly exploit group sparsity, other studies \cite{zhendong2022JCEAUiot,shi2022twc4ma} focused on solving the $\ell_{2,1}$ regularized group least absolute shrinkage and selection operator (LASSO) problem in a model-driven DL approach, which does not depend on any prior distribution.
Specifically, Mao \textit{et al.} \cite{zhendong2022JCEAUiot} unrolled ADMM to solve the group LASSO, where the network parameters are optimally learned using the stochastic gradient descent algorithm.
With the advantages of fast convergence rate, high robustness and theoretical guarantees, ISTA-based \cite{shi2022twc4ma} algorithm unrolling framework was proposed by extending LASSO-based decoder to group LASSO to circumvent the high computational cost of classic ISTA and poor algorithm robustness of AMP simultaneously.

\subsection{Application 2: Non-Convex Sum-Utility Maximization Problems}
A wide variety of resource management problems are directly or indirectly reliant on the sum-utility maximization problems, which aim at maximizing the system sum-utility (e.g. sum-rate) subjecting to the transmit power constraint. 
The general formulation with $K$ devices and one BS can be expressed as
\begin{subequations}
\begin{align}
	\mathop {\rm{maximize}}_{ \bm{V}}&~~\mathcal{U}\left(R_1(\bm{V}), \cdots, R_K(\bm{V}) \right)  \\ 
	{\rm{subject~to}}&~~ Q(\bm{V}) \leq P ,
\end{align}
\end{subequations}
where $\bm{V}$ denotes the resource (e.g., precoding matrix at the BS) to be optimized, $\mathcal{U}(\cdot)$ is the network utility function (e.g., weighted-sum function), $R_k(\cdot) $ represents the achievable rate for device $k$ and $Q(\cdot)\leq P $ is the power constraint.
Unfortunately, most of the sum-utility maximization problems are non-convex and very difficult to solve.

In addition to the direct parameterization of the iterative algorithm into a neural network as in Section II-B, the algorithm unrolling methods can also use trainable parameters to approximate the complex operators (e.g., matrix inversion) in the iterative algorithm to achieve the purpose of reducing the computational complexity for wireless applications such as precoding design \cite{Hu2020unrolling,mingmin21twcwmmse,he2022unsupervised} and power control \cite{an21tspdeepunfolding}.
The iterative WMMSE algorithm is one of the most representative algorithms for sum-rate maximization problems \cite{qingjiang11wmmse}, which is guaranteed to converge to a stationary point. 
The general form of WMMSE is given by,
\begin{align}
	\bm{U}^{t} = F_t(\bm{V}^{t-1}), \bm{W}^{t} = G_t(\bm{U}^{t},\bm{V}^{t-1}), \bm{V}^{t} = J_t(\bm{U}^{t},\bm{W}^{t}).
\end{align}
$\bm{U}$, $\bm{W}$ are introduced auxiliary variables and $F_t(\cdot), G_t(\cdot)$, $J_t(\cdot)$ are the iterative mapping functions at the $t$-th WMMSE iteration, where $F_t(\cdot)$ and $J_t(\cdot)$ contain computationally intensive matrix inversion operations. 
By using trainable parameters to approximate the matrix inversion and matrix multiplication functions, unrolled WMMSE enjoys low computational complexity, whose $t$-th layer network can be written as follows,
\begin{subequations}
\begin{align}
	&\bm{U}^{t} = \tilde{F}_t(\bm{V}^{t-1},\bm{X}^{u,t},\bm{Y}^{u,t},\bm{Z}^{u,t}) + \bm{O}^{u,t} \\
	&\bm{W}^{t} = \tilde{G}_t(\bm{U}^{t},\bm{V}^{t-1}, \bm{X}^{w,t},\bm{Y}^{w,t},\bm{Z}^{w,t}) \\
	&\bm{V}^{t} = \tilde{J}_t(\bm{U}^{t},\bm{W}^{t},\bm{X}^{v,t},\bm{Y}^{v,t},\bm{Z}^{v,t}) + \bm{O}^{v,t},
\end{align}
\end{subequations}
where trainable parameters $\{ \bm{X}^{u,t},\bm{Y}^{u,t},\bm{Z}^{u,t} \}$, $\{  \bm{X}^{w,t},\bm{Y}^{w,t},\bm{Z}^{w,t} \}$ and $ \{  \bm{X}^{v,t},\bm{Y}^{v,t},\bm{Z}^{v,t} \}$ are used to approximate the corresponding inversed matrix in original $F_t(\cdot), G_t(\cdot)$ and $J_t(\cdot)$ with $\bm{O}^{u,t}$ and $\bm{O}^{v,t}$ as the trainable offsets.

A deep-unfolding framework IAIDNN was proposed in \cite{Hu2020unrolling}, where a number of trainable parameters are introduced to replace the high-complexity matrix inversion operations in classic WMMSE algorithm.
Beneficial from both optimal performance of WMMSE algorithm and extensive representation power of DNNs, IAIDNN achieves the performance of the iterative WMMSE algorithm with much lower computational complexity and a smaller number of training samples.
In \cite{mingmin21twcwmmse}, an unrolled WMMSE approach was also proposed to solve a short-term sub-problem decomposed by original non-convex stochastic problem with low complexity for precoding design in reconfigurable intelligent surfaces (RIS)-aided communication systems.
Such unrolled WMMSE network was also adopted in part of \cite{an21tspdeepunfolding} to approximate the iterative WMMSE algorithm with low training complexity and reduced memory overhead, which is adopted as a short-term sub-algorithm in a two-stage stochastic optimization problem (e.g., power minimization for two-timescale hybrid beamforming).
The above unrolled networks are all trained under a supervised way due to the complex structure of WMMSE.
An unsupervised deep unrolling framework based on projection gradient descent called UPGDNet was proposed in \cite{he2022unsupervised} to solve the sum-rate maximization problems in the scenario of multiuser ultra-reliable low-latency communications (URLLC) with finite block length transmission, which demonstrates a satisfactory generalization ability and low complexity.

\subsection{Advantages and Disadvantages of Algorithm Unrolling}
The advantages of algorithm unrolling methods are summarized as follows.
\subsubsection{Higher Computational Efficiency}
Compared with the end-to-end learning based on generic NN, reduced number of training parameters in unrolled NN can significantly boost the computational efficiency of NN. 
Besides, due to the incorporation of domain-specific knowledge through unrolling, the training of unrolled NN is faster and requires fewer training data.  
Algorithm unrolling methods can significantly improve the convergence rate of original iterative algorithm (e.g., LISTA-GS \cite{shi2022twc4ma} converges less than $10$ iterations while ISTA needs more than hundreds iterations), and also can reduce the complexity of one-step iterative process (e.g., IAIDNN \cite{Hu2020unrolling} reduces the computational complexity of WMMSE from $\mathcal{O}(n^3)$ to $\mathcal{O}(n^{2.73})$ with $n$ denoting number of system parameters).

\subsubsection{Better Learning Performance}
By extending the iterative counterparts and training using datasets, the algorithm unrolling can achieve superior performance compared with the conventional iterative algorithm. 
For example, mpNet \cite{taha22mpNet} and LISTA-GS \cite{shi2022twc4ma} achieve more accurate estimation performance (more than $5$dB normalized mean-squared error enhancement) compared with ISTA-based algorithms.

\subsubsection{Interpretability and Theoretical Analysis}
Inherited from the traditional iterative algorithm, the behavious of each unrolled network layer is interpretable. 
Algorithm unrolling methods, to some extent, build a bridge between deep learning and iterative formulations, where the optimization tools can be used to define the optimal learned parameters that leading to fastest convergence rate (e.g., the optimal parameters defined in LISTA-GS guarantee the linear convergence rate for recovering group-sparse matrices \cite{shi2022twc4ma}).

However, there are some disadvantages of these methods.
First, for complex iterative algorithms when highly nonlinear or non-smooth operations are involved, it is hard to develop efficient NN to unfold the complicated iterative operations. 
Second, for iterative algorithms with slow convergence, the depth of unrolled NN will be large, which can easily suffer from gradient explosion or vanishment during the training stage.  
Third, even if the extended representation ability of algorithm unrolling, its convergence is hard to be guaranteed for complex iterative algorithms (e.g. unrolled WMMSE and unrolled ADMM). 
Moreover, its performance is restricted by the iterative algorithms. Analyzing the impact of trainable parameters on the convergence and learning accuracy is also challenging.

\section{Learning to Branch-and-Bound}\label{sec:3}
Many important applications in wireless networks involve complicated combinatorial problems, whose optimal solutions are hard to obtain efficiently. 
A learning-based BB algorithm, namely LBB, is introduced in this section to tackle the combinatorial problem with low computational complexity. 
Instead of end-to-end learning, the LBB replaces the complex pruning step of traditional BB with NN to accelerate searching for optimal solutions in feasible region. 
We first introduce the traditional globally-optimal BB algorithm to solve a general combinatorial problem. 
Then we introduce a learned BB to learn the optimal pruning policy of BB in a data-driven manner, which will be followed by the case studies in wireless communication systems. 
The advantages and disadvantages of LBB are summarized at the end of this section. 

\subsection{Global Optimal Branch-and-Bound}
BB algorithm can find global optimal solutions for non-convex combinatorial problems, e.g., discrete and mixed combinatorial optimization problems. It implicitly enumerates all possible solutions by dividing the original problem into a series of sub-problems (\textit{branch step}) and systematically discards the non-promising sub-problems based on lower bounds or upper bounds (\textit{bound step}).
Specifically, the \textit{branch step} is to partition the feasible region into smaller subregions in a tree structure, where the root node is the original problem and the leaf node represents the subproblem over the corresponding subregions.
The \textit{bound step} uses the features obtained at each node to prune off the subregions that do not contain the optimal solution, until BB eventually converges on an exact result \cite{land2010automatic}.
To guide and accelerate the searching process, the \textit{branch step} involves node and variable selection determining which node to explore and the fractional variables to branch on in next iteration, whereas the \textit{bound step} consists of two main policies: evaluating the bounds of selected nodes by solving the subproblems and pruning policy which determines whether to explore the subregion corresponding to the node.  
The pruning policy at bound step depends on the optimal solution and optimal objective value of relaxed subproblem at each node, where the constraints for the undetermined discrete variables are relaxed into convex continuous constraints and then various convex solvers can be applied to solve the relaxed convex subproblem, e.g., linear programs (LPs), second order cone programs (SOCPs) or semidefinite programs (SDPs).
For example, binary constraints can be relaxed into box constraints and the non-convex complex modulus constraints can be relaxed to their convex envelopes \cite{yifei20lorm}.

By learning the time-consuming components of BB algorithm using NN, the learning-based BB can significantly reduce computational complexity with near optimal performance. Authors in \cite{huang2021branch} provided a survey of learning-based BB techniques regarding to the key decisions in BB, such as learning-based branching and learning-based pruning, and summarized the merits and flaws of different learning methods. 
In original branch step of BB algorithm, various branching rules (e.g., strong branching, hybrid branching) are used in branch variable selection by calculating the score of candidate variables to indicate their qualities and then picking the variable with the highest score.
In this way, enormous branch decisions are required while a single bad one could sharply increase the size of search tree without improvement in the learning performance.
To accelerate the branching rules, imitation learning was adopted in \cite{DBLP:journals/informs/AlvarezLW17,DBLP:journals/disopt/MorrisonJSS16} to learn a fast auxiliary branching policy by approximating the traditional branching rules using expert experience, which can outperform the initial expert with carefully designed features. 
In addition, such approach leads to a simpler learning task with smaller Vapnik-Chervonenkis dimension, which only needs a few training samples, and thereby speeding up the training process consequently. 
To overcome the complex feature calculation at each selected node, authors in \cite{DBLP:conf/nips/GasseCFC019} encoded branching policies into a graph convolutional network (GCN), where features on the graph can be efficiently extracted by various message passing approaches. 
To solve large-scale mixed integer programming (MIP), the authors in \cite{nair2020solving} constructed two corresponding GCN components (i.e. neural diving and neural branching) to learn a branching policy for enhancing the performance.
 
Even with the learned branching policy, the computational complexity of BB algorithm is generally exponential w.r.t. the number of optimize variables due to the inefficient pruning policy.
Based on this observation, other studies \cite{he2014learning, ai2020010} focused on the supervised learning of optimal pruning policy to solve large-scale MIP efficiently. In the next subsection, we shall introduce the motivations and design frameworks of pruning policy learning-based LBB in large-scale 6G wireless networks.

\subsection {Motivations and Design Frameworks of LBB}
In 6G wireless networks, typical resource allocation problems, such as subcarrier allocation in orthogonal frequency division multiple access (OFDMA) \cite{wong1999ofdma}, user association \cite{ye2013user} and access point selection in could-radio access network (C-RAN) \cite{shi2014group}, can be formulated into mixed combinatorial optimization problems. 
With the rapid growth of wireless network scales (e.g., the massive number of IoT devices and the massive number of antennas), the dimension of optimization variable becomes very large, which makes learning-based BB a promising approach to tackle the exponential complexity of traditional BB while maintaining the global optimality. 
In this subsection, we highlight a promising learning strategy for node pruning in BB algorithm proposed in \cite{he2014learning, ai2020010}, termed LBB, which has been shown to achieve low computational complexity with near-optimal performance. 

The main idea of LBB is to model the tree search process as a sequential decision problem because whether or not exploring the subregion of the tree node corresponds to the \textit{preserve} decision or \textit{prune} decision.
Such problem can be efficiently solved by a binary classifier with problem features as the input and decision states as the output.
The procedure of LBB includes training data generation, feature design, binary classifier learning and searching space controlling, as described below.
\subsubsection{Training Data Generation and Feature Design}
The original BB algorithm is directly applied to generate the training dataset.
Specifically, the problem parameters of each randomly generated wireless
network are collected.
For each set of problem parameters, the original BB algorithm is adopted to obtain the optimal solution and  the features of all explored nodes are recorded.
Then the nodes whose feasible sets contain the optimal solution are labeled as \textit{preserve} while the others are labeled as \textit{prune}.
The input features of the NN are also of vital importance for training a good classifier in LBB.
Features are divided into tow categories, i.e., problem-independent features and problem-dependent features\cite{yifei20lorm,lee2019learning}.
Problem-independent features correspond to the structure of the
binary tree generated by the BB algorithm, which includes node features computed from the current node (e.g., the depth of current node), branching features computed from the branching variable of the current node (e.g., the branching variable's value at the current node) and tree features computed from the BB search tree (e.g., global upper and lower bounds) \cite{he2014learning}. 
While problem-dependent features correspond to the domain knowledge of different specific problems. 
Typically in wireless communications, the CSI feature and some descriptions of the radio resources (e.g. power feature) are usually considered as problem-dependent features to exploit the domain knowledge for efficient policy learning. 
By feeding the problem dependent features into the classifier learning, the LBB can avoid solving relaxed problems at each branching node, which can further reduce the computational burden. 

\subsubsection{Binary Classifier Learning}
The binary classifier determines whether to preserve a node or not and a good classifier can prune as many non-optimal nodes as possible to minimize the search time.
Standard classifiers, such as logistic regression \cite{DeMaris1995ATI} and support vector machine (SVM) \cite{DBLP:journals/jmiv/Bayro-CorrochanoA07}, are inefficient for high-dimensional data classification with tangled mapping between input and output \cite{davide10svm}. 
To effectively capture the complicated relationship between the input features and output classification labels, MLP can be employed for binary classifier learning in LBB due to its powerful expression ability \cite{yifei20lorm}.
In each layer of MLP, the input is multiplied with a learned weight matrix, followed by a rectified linear unit function (Relu) as activation function.
The output layer employs the soft-max function to calculate the probability of each class, while the weighted cross entropy is adopted as loss function.
\begin{figure}[h]
  \centering
  \includegraphics[width=0.5\textwidth]{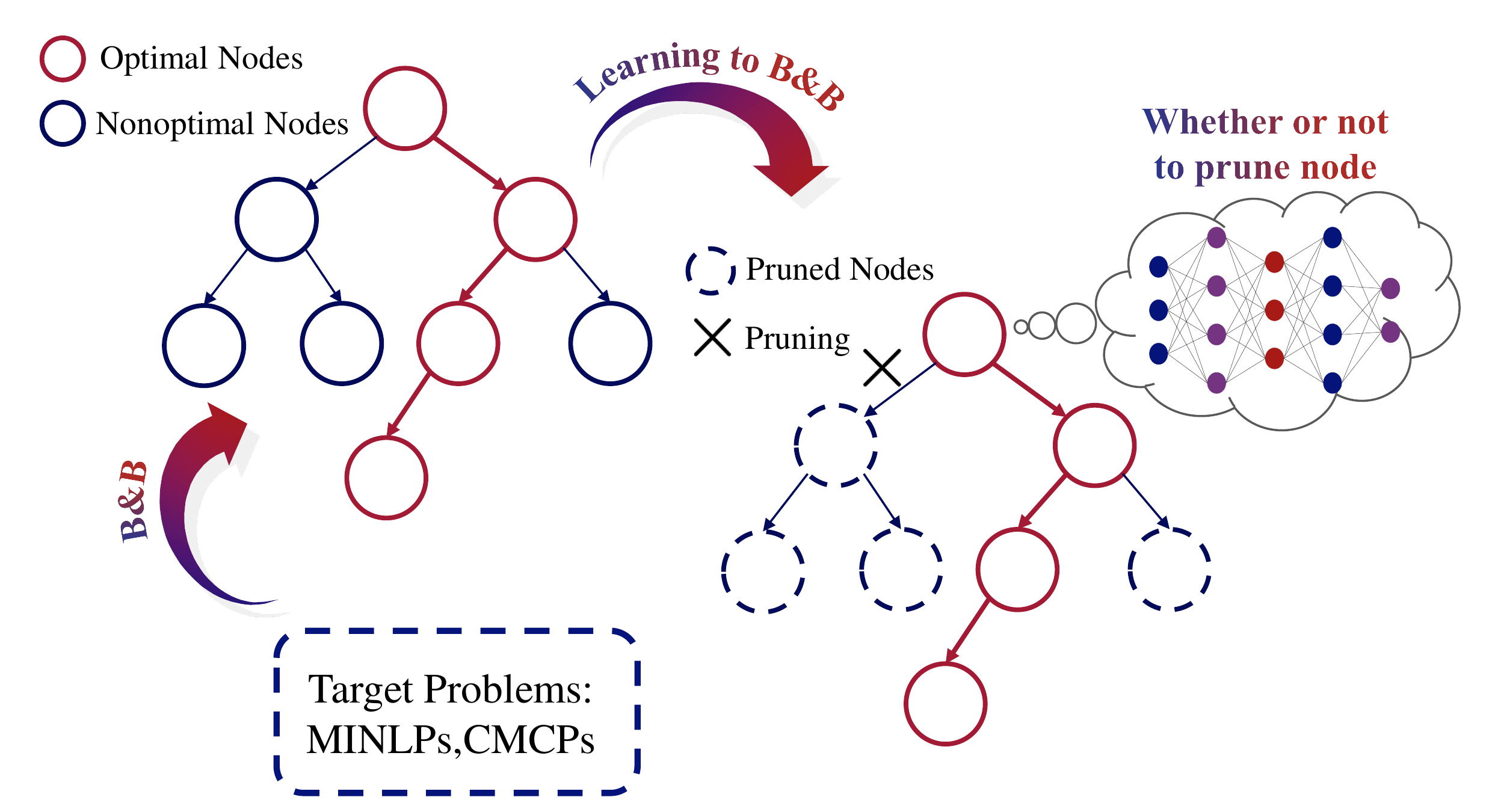}
  \caption{Learning to branch-and-bound method.}
  \label{fig:lbb}
\end{figure}

\subsubsection{Training Sample Unbalance and Searching Space Controlling}
It is observed that the number of non-optimal (pruned) nodes is much larger than the number of optimal (preserved) nodes and the mistakes at early stages can have greater impact on the performance during BB searching process.
Thus, to enable efficient network training and achieve a good classification performance, weighted cross entropy was adopted in \cite{lee2019learning}, where the higher weights are assigned to the nodes labeled as \textit{preserve} and the nodes with small depth in the training dataset to raise the priority of these training samples in the learning of NN \cite{yifei20lorm}.
Since too aggressive pruning policy may lead to no feasible solutions and too moderate pruning policy may lead to abundant preserved nodes, a threshold was adopted to control the searching space dynamically.
For instance, a higher threshold for the class pruning will preserve more nodes than that with a lower threshold, leading to a larger searching space and better performance.  
The searching space controlling can guarantee the feasibility of LBB with reduced computational complexity.

The overall frameworks of BB and LBB are illustrated in Figure \ref{fig:lbb}.
With the advantages of near-optimality and low computational complexity for challenging non-convex
and combinatorial problems, LBB methods have gain much traction in the context of large-scale wireless networks to solve various resource allocation problems. 
In the following subsections, we review several representatives of non-convex combinatorial optimization problem with extensive applications in wireless networks to fully reveal the power of LBB for efficient and high-quality resource management.

\subsection{Application 1: Mixed Integer Nonlinear Problems}
In wireless networks, typical resource management problems, such as user selection, user association, spectrum allocation, computational offloading, interference and power management, can be formulated into a mixed integer nonlinear problem (MINLP), which is general NP-hard. 
The generic formulation is given by \cite{yifei20lorm}
\begin{subequations}
\begin{align}
\text{MINLP}: \mathop {\rm{minimize}}_{\bm{a}, \bm{w}}&~~f(\bm{a}, \bm{w}) \\ 
{\rm{subject~to}}&~~ Q(\bm{a}, \bm{w}) \leq 0, \\
&~~a_i \in \N, w_i \in \C, \forall i,
\end{align}
\end{subequations}
where $f(\cdot,\cdot)$ is the objective function (usually non-linear), such as power consumption, sum rate, communication delay, etc, discrete-valued variable $a_i$ and continuous-valued variable $w_i$ are the elements of $\bm{a}$ and $\bm{w}$, such as sub-carrier index, user index, device index for discrete variable and beamforming, power for continuous variable, and $Q(\cdot,\cdot)$ denotes certain constraints, e.g., the quality of service (QoS) and power constraints.

A typical example of MINLP is network power minimization problem in C-RAN, which consists of binary variables (i.e., the selection of remote radio heads and front-haul links), continuous variables (i.e., downlink beamforming coefficients), QoS constraints and power constraints.
To get the near-optimal performance with affordable complexity, LBB via imitation learning was proposed in \cite{yifei20lorm}, where the depth-first-search was adopted as the branch variable selection rule which always choses the first undetermined node during variable selection process.
In feature design, besides problem-independent features such as the depth of node, the local upper bound of node, current global lower bound and so on, the CSI feature and power feature are designed as problem-dependent features.
LBB algorithm was shown to achieve success using only tens to hundreds of training samples for solving MINLPs in the applications of resource allocation problem in device-to-device (D2D) communications \cite{lee2019learning} and mobile edge computing \cite{yurong22L2ORA}.
Specifically, in \cite{yurong22L2ORA}, a learning strategy for node pruning in BB algorithm was proposed for the offloading resource assignment in mobile edge computing, where DNN was applied to approximate the unknown mapping between the attributes of BB tree nodes and the pruning decisions in the BB tree search.
To further reduce the sample complexity, imitation learning was adopted in \cite{lee2019learning} to solve MINLPs of resource allocation in D2D systems, where DAgger algorithm was employed to correct the mistakes the learned policy makes by iteratively collecting data. Specifically,
instead of using the training data only once, the pruning policy $\pi^t$ learned in imitation learning shall explore all the tree nodes and generate their corresponding features at iteration $t$.
Then based on these datasets, a new prune policy $\pi^{t+1} $ can be learned at next iteration to correct the mistakes made by $ \pi^t$.
To adapt to time-varying network settings, a transfer learning via self-imitation method was adopted in \cite{yifei20lorm} to quickly adapt the learned pruning policy in LBB to the new task with reduced training samples.

\subsection{Application 2: Non-Convex Complex Modulus Constrained Problems}
Many resource management problems in wireless networks can be formulated as non-convex complex modulus constrained problems (CMCPs), which is general NP-hard. 
The generic formulation of CMCP is given by \cite{zhe22LBB},
\begin{subequations}\label{CMCP}
\begin{align}
\text{CMCP}: \mathop {\rm{minimize}}_{\bm{a}, \bm{w}}&~~g(\bm{w}) \\ 
{\rm{subject~to}}&~~| \bm{b}^{H}_i \bm{w} + c_i| \geq 1, \forall i, \label{CMCP:constr1} \\
&~~  \arg{ (\bm{b}^{H}_i \bm{w} + c_i)} \in \mathcal{A}_i, \forall i , \label{CMCP:constr2}
\end{align}
\end{subequations}
where $g(\cdot): \C^N \rightarrow \R$ is a convex objective function (e.g., power consumption), constraint \eqref{CMCP:constr1} denotes the non-convex minimum complex modulus constraints on $N$ linear transformations of the optimization variable $\bm{w} \in \C^{N}$ (e.g., the transmitted symbol vector and the beamformimg vector) with coefficients $\bm{b}_i \in \C^{N}$ and $c_i \in \C$, and constraint \eqref{CMCP:constr2} represents the corresponding argument constraints where each argument set $\mathcal{A}_i$ can be continuous or discrete (e.g., box constraints or discrete phase sets).

One toy example of CMCPs in wireless networks is the QoS-constrained multi-cast beamforming optimization problem in multiple-input single-output (MISO) downlink transmission, which minimizes the total transmit power at the BS (i.e., $\min_{\bm{w}} \Vert \bm{w} \Vert_2^2 $) subject to individual SNR constraints for all devices (i.e., $ | \bm{h}^{H}_k \bm{w}| \geq 1, \forall k$).
The formulated problem corresponds to the standard CMCP in \eqref{CMCP} by letting $\bm{b}_k = \bm{h}_{k}$, $c_k = 0$ and $\mathcal{A}_{k}=[0,2 \pi] ,\forall k$.
Other applications of CMCPs include MIMO detection and passive beamforming in RIS-assisted systems \cite{zhe22LBB}.
Unlike the BB algorithms for solving MINLPs with discrete branching variables, the branching of BB algorithm for solving CMCP is based on argument sets to deal with the continuous variables \cite{cheng17bb}. 
As a result, the continuous argument set can lead to unbounded extension of tree search in BB algorithm and tremendous number of binary classification tasks in the searching process.
Therefore, it is difficult to find the optimal solution of the CMCP using only one classifier \cite{lee2019learning,yifei20lorm}. 
To address this challenge, ensemble learning was applied in \cite{zhe22LBB} to train multiple classifiers in LBB, which are further combined to achieve better performance. 
To facilitate the multi-classifier learning and address the data unbalance in LBB, \cite{zhe22LBB} devised the under-sampling training and ensemble testing. In under-sampling training, individual classifier is trained on each of the data subsets sampled both from the majority set and minority set. In ensemble testing, LBB is executed multiple times using learned multiple classifiers in parallel to choose the best solution from all the results. 
Regarding to the feature design, besides problem-independent features such as local node features (e.g., the argument set $\mathcal{A}$) and global tree features (e.g., the global lower bound), the CSI, the SNR, the received signal and the complex modulus constraint can be designed as problem-dependent features.

\subsection{Advantages and Disadvantages}
LBB algorithm was shown to achieve near-optimal performance and meanwhile substantially reduce computational complexity using only tens to hundreds of training samples, due to the inheritance of the structure of traditional BB algorithm and the exploitation of DL techniques.
If the parameters and features are properly chosen, the number of relaxed problems to be solved can be reduced from $\mathcal{O}(2^L)$ in BB to $\mathcal{O}(L)$ in LBB for MINLPs \cite{yifei20lorm}, where $L$ denotes the number of integer variables.
However, the training of LBB depends on the labels generated from traditional BB, which can induce great computational burden to generate the training set. 
Meanwhile, feature design in LBB is not supported by theory and hence different designs can lead to different results of the algorithm.

\section{Graph Neural Network for Structured Optimization}\label{sec:4}
The graph-structured topology of wireless network enables the successful usage of GNN to solve a broad range of design problems over the wireless networks \cite{suarez2022graph}. 
As a specialized NN for graph-structured data, GNN can exploit the domain knowledge of various applications to achieve near-optimal learning performance with good scalability and generalizability. 
In this section, we commence with the framework of GNN for structured optimization. 
Then we illustrate the graph modeling of optimization problems in wireless networks, after which, several applications of GNN are discussed. 
The advantages and disadvantages of GNN-based solution in wireless networks are summarized at the end of this section.

\subsection{Principles of Graph Neural Network}
Recently, a growing number of applications have emerged possessing graph-structured data, e.g., social networks and transportation networks, with high dimensional features, active interactions between graph nodes and potentially time-varying structures. 
The emerging GNN can effectively incorporate the graph structure into the architecture of NN to model the node properties and the relationships between nodes to explore the hidden features in graph-structured data. Hence, GNN not only features for good scalability to large-scale graphs and good generalizability to dynamic graph structures, but also can achieve near-optimal performance with more efficient training.
Traditionally, a graph-structured data can be mathematically represented as a pair $G = (V,E)$, where $V$ is the set of nodes and $E$ is the set of edges.
For a node, its (1-hop) neighborhood is defined as the set of all nodes with edges connected to it.
These edges can be represented by an adjacency matrix $\bm{A}$, where $\bm{A}_{i,j} = 1$ if and only if edges $(i,j) \in E$ for all nodes $i,j \in V$.
The graph is undirected if $\bm{A}$ is symmetric, otherwise it is directed. 
The properties associated to the nodes and edges are important for learning over graph. 
Denote $\bm{z}_i $ as the node features associated with node $ i \in V$, $\bm{e}_{i,j}$ as the edge features associated with edge $(i,j) \in E$. 
Since the graph structure may be changed by the permutation of nodes, a key desideratum for designing GNNs is that the devised GNN should satisfy \textit{permutation invariance} or \textit{permutation equivariance} \cite{zonghan21GNN}, which can precisely capture the internal structure of graph data.
As illustrated in Figure \ref{fig:pipe}, permutation invariance means that the output of GNN does not depend on the node order used to encode adjacency matrix. 
Permutation equivariance means that the output of GNN is permuted according to the same permutation as the input node order.  
Such properties enable faster training, less training samples, better scalability and generalizability of GNN. 
For example, compared with MLP, the parameters of GNN can be shared for graphs with varying sizes and permuted input, thereby achieving good generalizability, while an MLP has to be retrained when the topology is changed. 
More precisely, authors in \cite{Shen2022GraphNN} theoretically showed that the GNN's generalization error and required number of training samples are $\mathcal{O}(n)$ and $\mathcal{O}(n^2)$ lower than those of MLPs, where $n$ is the number of nodes in the graph. 

GNN implements the learning over graph by exacting the neighbor information to enrich the feature of each node and spreading these features over the graph according to the graph structure. 
The framework of GNN is illustrated in Figure \ref{fig:pipe}.
Generally in a vanilla GNN with multiple hidden layers, the output of last GNN layer contains the processed information of all nodes and edges, which can be used for classification or prediction.
In each GNN layer, each node aggregates the information of its neighbors to update its hidden state.
Specifically, let $\bm{d}_{k}^{(\ell)}$ be the hidden state of the $k$-th node at the $\ell$-th GNN layer. Each GNN layer contains the aggregation and combination step as follows.
\subsubsection{Aggregation Step}
The aggregation step aims to update the node's hidden state with its neighbors' information.
Typically, the $k$-th node uses an NN to aggregate its neighbors’ outputs of the previous layer (the $(\ell - 1)$-th layer), followed by a pooling function (e.g., element-wise max-pooling or element-wise mean-pooling) that is invariant to the permutation of the inputs.
The update is given by
\begin{align}
    \bm{a}_{k}^{(\ell)} = \text{PL}_{j \in \mathcal{N}(k)} \left(  f_1^{(\ell)} (\bm{d}_{k}^{(\ell-1)}, \bm{d}_{j}^{(\ell-1)}, \bm{e}_{j,k} | \bm{\Theta}_1^{(\ell)} ) \right),
\end{align}
where $\text{PL}_{j \in \mathcal{N}(k)}(\cdot)$ denotes the pooling function used for aggregating the outputs of the nodes in $\mathcal{N}(k)$, $\mathcal{N}(k)$ denotes the set of neighboring nodes of node $k$, $f_1^{(\ell)}(\cdot|\bm{\Theta}_1^{(\ell)} )$ is the aggregation function implemented using MLP with model parameters $\bm{\Theta}_1^{(\ell)}$ at the $\ell$-th layer and $\bm{e}_{j,k}$ is the feature of edge $(j,k)$.


\subsubsection{Combination Step}
After obtaining the aggregated information, another combination function is applied to process information and update the hidden state at each node. Specifically, the aggregated information is combined with the node's own information as follows:
\begin{align}
    \bm{d}_{k}^{(\ell)} =   f_2^{(\ell)} \left( \bm{d}_{k}^{(\ell-1)}, \bm{a}_{k}^{(\ell)}, \bm{z}_{k} | \bm{\Theta}_2^{(\ell)}  \right),
\end{align}
where $f_2^{(\ell)}(\cdot|\bm{\Theta}_2^{(\ell)} )$ represents parameterized combination function
with model parameters $\bm{\Theta}_2^{(\ell)}$ and $\bm{z}_{k}$ is the feature of node $k$.

To improve the learning ability of GNN over different graphs, the aggregation function $f_1$, the combination function $f_2$ and the pooling function $\text{PL}$ should be carefully designed. 
Authors in \cite{Shen2022GraphNN} gave general guidelines for designing these functions.  
First, the message aggregation and combination should be simplified to reduce the complexity of calculations. 
Second, different pooling functions are suitable for different graphs and optimization problems. 
For example, the max-pooling function is suitable when the neighbors’ influence is sparse or the problem parameters are noisy, because it focuses on the neighbors that are most influential. 
While the sum-pooling function gives summary statistics of the neighbors, which works well when we aim to obtain a summary of the neighbors. 
Third, the aggregation and combination function should be properly designed to satisfy the permutation invariance or permutation equivariance property.
Last, the input embedding is of vital importance for information representation. 
Therefore, it is often preferable to employ an input embedding neural network to lift or compress the features into a proper dimension to search for a balance between training complexity and learning performance. 

\begin{figure}[t]
        \centering
        \begin{minipage}{.46\textwidth}
                \centering
                \includegraphics[width=1.0\columnwidth]{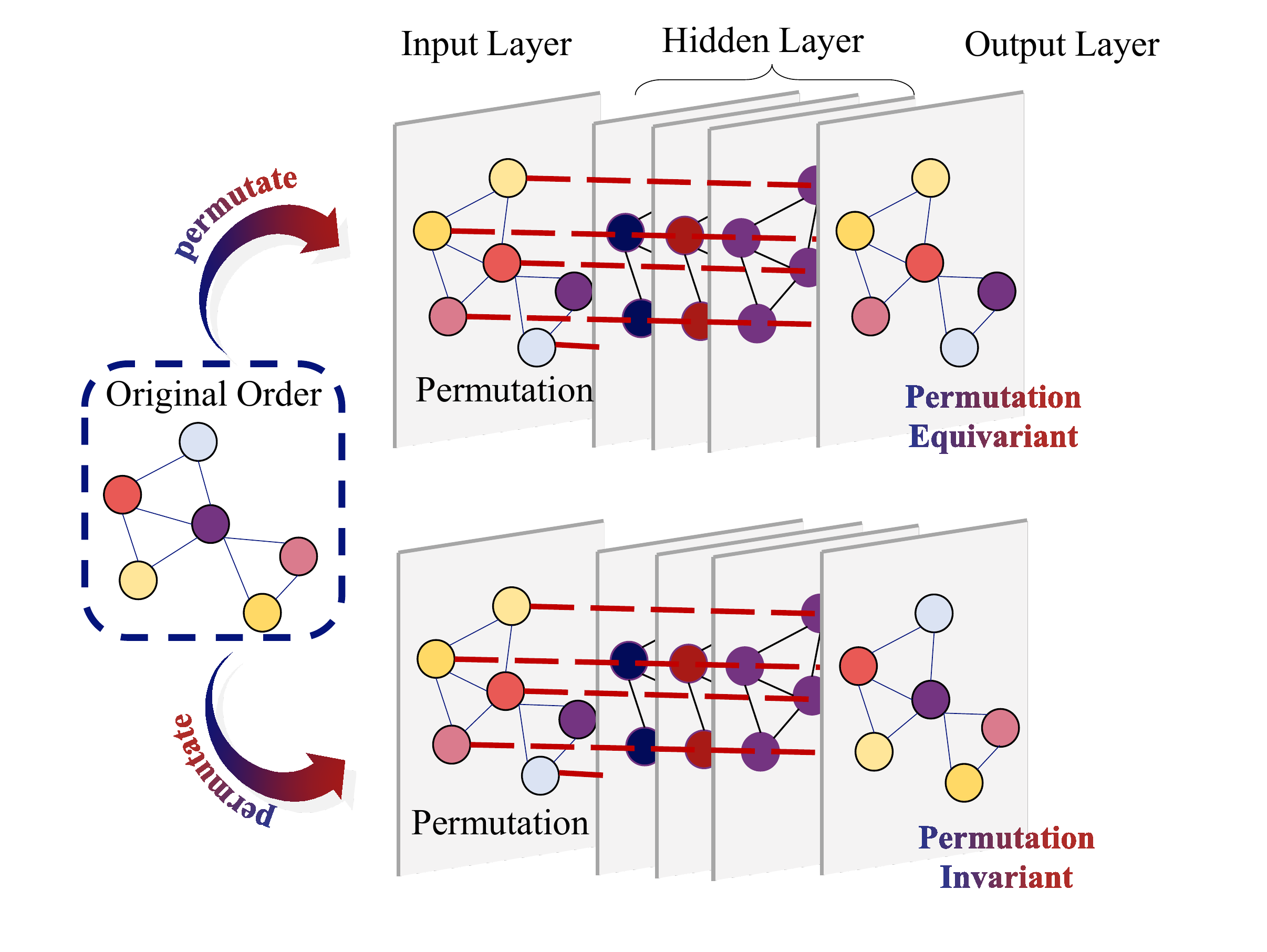}
                \caption{Illustration of permutation equivariance and permutation invariance. Different colors of nodes represent different orders of nodes.}\label{fig:pipe}
        \end{minipage}
        \hspace{4mm}
        \begin{minipage}{.46\textwidth}
                \centering
                \includegraphics[width=1.0\columnwidth]{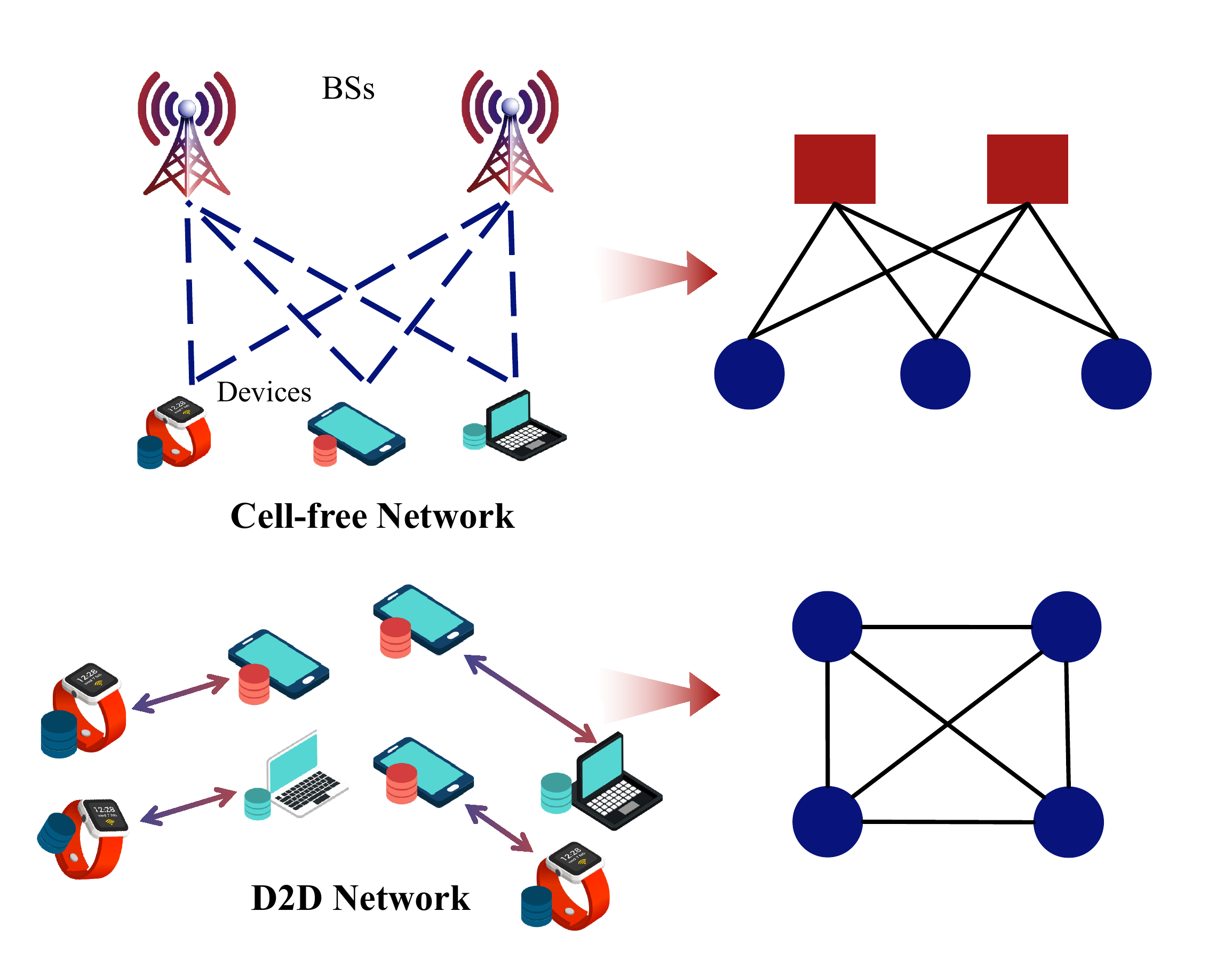}
                \caption{Examples of the wireless communication graph modeling (i.e. cell-free networks and D2D networks). }\label{fig:gnn}
        \end{minipage}
\end{figure}



\subsection{Graph Optimization Problems in Wireless Networks}
In wireless networks, the relationship between users, BSs and even antennas spontaneously yields to a wireless communication graph.
Specifically, a wireless network can be modeled as a directed/undirected graph with node and edge features, where communication devices (users and BSs) can be treated as nodes, channels between devices can be treated as directed edges. 
An edge $(i,j)$ exists if there are interdependencies, e.g., communication link, interference link or other connectivity patterns, from node $i$ to node $j$. 
The features associated with the node represent the properties of devices, e.g., user's weight in weighted sum rate maximization, while the features associated with the edge stand for the link properties, e.g., channel gains. 
Mathematically, the general form of optimization problem on wireless communication graph can be expressed as \cite{yifei21gnn}
\begin{subequations}\label{eq:wgp}
\begin{align}
\underset{\boldsymbol{X}}{\operatorname{minimize}} ~&f(\bm{X}, \bm{Z}, A) \\
\text { subject to } &Q(\bm{X}, \bm{Z}, A) \leq 0,
\end{align}
\end{subequations}
where $f(\cdot)$ is the objective function (usually non-convex), $\bm{X}$ denotes the collection of optimization variables assigned to all nodes, $\bm{Z}$ denotes the node feature matrix which can model the heterogeneity of node, and $A$ is the edge feature tensor, which can incorporate more comprehensive link properties, $Q(\cdot)$ is the constraint.
The graph optimization problem is general enough to cover most of the resource management problems in wireless networks, e.g., power control, beamforming design, link scheduling, etc. 

The graph-structured optimization problem \eqref{eq:wgp} defined on wireless communication graph satisfies the permutation equivariance and permutation invariance properties \cite{shen2019graph}. 
Namely, the objective function $f$ and constraint function $Q$ are invariant to the permutation of the device indices, while the output of GNN, i.e., the optimal variable $\bm{X}$ would be permuted in the same way as the permutation of device orders. 
To solve the graph optimization problem effectively, the GNNs which satisfy the permutation equivariance/invariance properties are favorable. 
To fully reveal the advantage of GNN-based solution for \eqref{eq:wgp}, authors in \cite{Shen2022GraphNN} bridged the GNN with a class of COAs, i.e., distributed message passing, to justify that for any graph optimization problem, there exists a GNN that can solve it from the algorithmic perspective. 
Compared with the COAs, the data driven GNN can achieve near-optimal performance with reduced computational complexity. 
Compared with MLP, the superior performance of GNN in solving graph optimization problem in wireless networks is theoretically proven in \cite{Shen2022GraphNN} in terms of the optimization performance and sample efficiency.  

Table \ref{ta:gnn} summarizes the wireless applications of GNN. 
Based on the topology of graph, the applications of GNN may cover resource management problems in D2D/cellular/cell-free/distributed network and other signal processing fields. 
In the following subsections, we introduce several application examples of GNN, detailing the graph modeling of different problems and the GNN architecture developed to solve the respective optimization problems.
\begin{table*}[]
	\centering
	\caption{Wireless Applications of Graph Neural Network}
	\resizebox{\linewidth}{!}{
	\begin{tabular}{ccccccc}
		\hline
		Applications                                                                                           & Wireless Issues                           & \multicolumn{1}{c}{Ref} & Node                      & Edge                            & Node Feature                                                                                                         & Edge Feature                                                                          \\ \hline
		\multirow{3}{*}{\tabincell{c}{Cellular/Cell-free \\ Networks} }                                                         & \multirow{2}{*}{\tabincell{c}{Beamforming \\ Optimizations}  } &          \cite{kim2022bipartite}               & Antenna / User            & Communication link              & Beam feature                                                                                                         & Channel coefficient                                                                   \\ \cline{3-7} 
		&                                           &              \cite{Jiang2021Learning}           & User / RIS                & Link between nodes              & Received pilots / Mean of all user features                                                                          & \textbackslash{}                                                                      \\ \cline{2-7} 
		& Power Allocation                          &       \cite{eisen2020large}                  & BS / User                 & Communication link              & State information                                                                                                    & Channel coefficient                                                                   \\ \hline
		\multirow{3}{*}{D2D Networks}                                                                          & Power Allocation                          &        \cite{shen2019graph,yifei21gnn}                 & Transceiver pair          & Directed interference link      & \begin{tabular}[c]{@{}c@{}}The state of the direct channel   and \\      the weight of transceiver pair\end{tabular} & \begin{tabular}[c]{@{}c@{}}The states of the\\      interference channel\end{tabular} \\ \cline{2-7} 
		& \multirow{2}{*}{Link Scheduling}          &     \cite{mengyuan21LSGNN}                    & Transceiver pair          & Directed interference link      & Channel gains or distance                                                                                            & Channel gains or distance                                                             \\ \cline{3-7} 
		&                                           &          \cite{zhao2021distributed, zhao2022delay}               & Communication link        & Interference link               & Queue length and link rate                                                                                           & \textbackslash{}                                                                      \\ \hline
		\multirow{5}{*}{Distributed Systems}                                                                   & \multirow{3}{*}{\tabincell{c}{Decentralized \\ Networks} }   &       \cite{wang22decentraGNN}                  & Transmitter               & Communication link              & Application state                                                                                                    & Channel coefficient                                                                   \\ \cline{3-7} 
		&                                           &           \cite{naderializadeh2022learning}              & Receiver                  & Direct link / Interference link & Proportional-fairness ratio                                                                                          & \textbackslash{}                                                                      \\ \cline{3-7} 
		&                                           &           \cite{lee2021decentralized}              & Device                    & Communication link              & Binary internal   characteristics                                                                                    & \textbackslash{}                                                                      \\ \cline{2-7} 
		& \multirow{2}{*}{\tabincell{c}{Heterogeneous \\ Networks}   } &           \cite{li2022heterogeneous}              & Different device          & Meta-path                       & Subchannel gain and power budget                                                                                     & Distance between nodes                                                                \\ \cline{3-7} 
		&                                           &          \cite{jia22GNN}               & Antenna or user           & Communication link              & State information                                                                                                    & Channel coefficient                                                                   \\ \hline
		\multirow{2}{*}{\begin{tabular}[c]{@{}c@{}}Other   Signal Processing\\      Applications\end{tabular}} &   \tabincell{c}{ Mobile Traffic \\  Prediction}              &       \cite{kaiwen22GNN4TP,feiyang22dtGNN,guo2020optimized}                  & Region                    & Road connecting the regions     & Historical traffic                                                                                                   & \textbackslash{}                                                                      \\ \cline{2-7} 
		& Channel Tracking                          &           \cite{yang2020graph}              & Element of channel vector & Link between channel element    & Channel coefficient                                                                                                  & Channel spatial correlation                                                           \\ \hline
	\end{tabular}}
    \label{ta:gnn}
\end{table*}

\subsection{Application 1: Graph Neural Networks in Cellular/Cell-Free Networks}  
In cellular or cell free networks, we can treat the different communication devices, including antennas, user equipments (UEs), RISs, and BSs, as graph nodes and their interdependencies as graph edges, as illustrated in Fig. \ref{fig:gnn}. 
In the following, we will provide some examples of GNN-based resource allocation in cellular or cell-free networks.
\subsubsection{Beamforming Optimization} 
The multi-antenna beamforming optimization problems were considered in \cite{kim2022bipartite}, where a bipartite GNN framework consisting of antenna/user vertices and channel edges was proposed to improve the scalability and the generalization ability of DNN approaches.
When considering the RIS-assisted cellular networks, GNN is proposed in \cite{Jiang2021Learning} to directly map the received pilots to the beamformers at the BS and reflective patterns at the RIS by solving sum-rate maximization problem.
To model it as a graph optimization problem, the RIS and users are taken as graph nodes and the communication links between users and RIS are taken as the graph edges.
The input features of user nodes are the received pilots and the input feature of RIS node is the mean of all user features.
After updating the input features through multiple GNN layers, the output features of user nodes will be mapped to the beamforming matrix at BS and the output feature of RIS node will be mapped to the reflective pattern of RIS. 
To capitalize on the special properties of GNN, the GNN layers are designed such that the beamforming vectors corresponding to different users are permutation equivariant and  the reflective pattern at RIS is permutation invariant for the change of the order of user channels, thereby enjoying the beneficial performance of GNN.

\subsubsection{Power Allocation}
Consider a cellular/cell-free network consisting of one/multiple BSs and multiple users. 
The cellular/cell-free system can be modeled as a fully connected bipartite graph, where the BSs and users are viewed as nodes, the communication links including direct links and interference links between BSs and users are regarded as edges. 
In the power allocation problem, the edge features are typically the channel coefficients and node features are the state information of user demand (e.g., the data arrival rate for traffic demand \cite{eisen2020large}).
After updating through GNN layers, the hidden states at the user nodes are mapped to the power values through learnable MLPs.
Many existing works \cite{eisen2020large} further developed advanced GNN algorithms to handle various adverse factors in wireless systems. 
For example, a random edge GNN (REGNN) was proposed in \cite{eisen2020large} to handle the fast fading channels in power allocation problem, where the underlying graph structure is a random variable drawn from a specific distribution. 
The authors in \cite{eisen2020large} further proposed an unsupervised model-free primal-dual learning method to address the general utility-constrained (e.g., binary power constraint) wireless resource allocation problems (e.g., the sum-rate maximization problems).

\subsection{Application 2: Graph Neural Networks in D2D Networks}
In D2D networks, supposing there are multiple transceiver pairs, each transceiver pair usually can be treated as one node, where the node features include the direct link CSI, the weight of each transmission pair and other related information. 
The interference link between transmission pairs can be treated as the edge, where the edge features include interference CSIs. 
The underlying graph, as shown in Fig. \ref{fig:gnn}, can be directed or undirected according to the definition of edge features. 
The GNNs can be effectively adopted to solve various resource allocation problems in D2D networks, such as power allocation \cite{shen2019graph,yifei21gnn,arinadm21wmmsegnn}, link scheduling \cite{mengyuan21LSGNN,zhao2021distributed, zhao2022delay}  and so on.

\subsubsection{Power Allocation}
Based on a standard D2D graph, IGCNet \cite{shen2019graph} and MPGNN \cite{yifei21gnn} were proposed to apply GNN over the D2D graph for the power control problem, where the wireless graph modeling are designed to match the permutation equivariance property of interference channels, and the aggregation and combination functions are realized using MLPs. 
To inherit the advantages of classic algorithm, an unrolled WMMSE algorithm was parameterized in \cite{arinadm21wmmsegnn} to design the GNN architecture for solving power allocation problem in a single-hop ad hoc interference network. 

\subsubsection{Link Scheduling} 
By exploiting and incorporating the underlying topology of wireless networks to learning algorithms, GNNs are well-suited for efficient scheduling of transmission links in wireless communications \cite{mengyuan21LSGNN,zhao2021distributed, zhao2022delay}.
To eliminate the expensive channel estimation stage, Lee \textit{et al.} \cite{mengyuan21LSGNN} firstly constructed a graph embedding process for link scheduling in D2D networks, where each D2D pair is designed as a node while interference links among D2D pairs are the edges for efficient graph design.
Based on Structure2Vec architecture, each node in the graph is represented by a low-dimensional feature vector, where the link classifier can be trained with high efficiency to decide whether a D2D pair should be activated.
Other studies \cite{zhao2021distributed, zhao2022delay} focused on link scheduling in wireless multi-hop networks, where the stream of packets from a source user (one node) to a destination user (the other node) may pass through multiple edges on the designed graph.
In \cite{zhao2021distributed}, a trainable graph convolutional network (GCN) module was proposed to improve the optimality gap with the traditional greedy approaches for the NP-hard maximum weight independent set (MWIS) problem in link scheduling.
To reduce the delay of scheduling, a delay-oriented distributed
scheduler based on GCNs was proposed in \cite{zhao2022delay}, in which multi-step lookahead backlogs and the network topology were fully captured by the node embedding layers.

\subsection{Application 3: Graph Neural Networks in Distributed Systems} 
The typical characteristics of distributed systems are decentralized setting (i.e., the transceivers only have knowledge of their local radio environment and make local decisions based on this information) and heterogeneous nature (i.e., there are different types of nodes with different node features in a communication graph), where GNNs also have shown their superior performance and scalability for these systems.

\subsubsection{Decentralized Networks}
To address the intrinsic information delay and asynchrony between devices in decentralized cooperative wireless systems, a primal-dual learning based aggregation GNNs (Agg-GNNs) was proposed in \cite{wang22decentraGNN} to design localized resource policies with delay and asynchronous constraints.
This can be achieved by adding multiple network layers to process delayed information after signal aggregations in Agg-GNNs, which gather spatial and temporal-correlated information of the global wireless networks.
Similar primal-dual learning method was adopted in \cite{naderializadeh2022learning} to learn resilient radio resource management (RRM) policies with adaptive per-user minimum-capacity constraints, which can adapt to the current network conditions via optimized slack variables. 
By parameterizing the RRM policies using scalable GNNs based on the graph topology of wireless networks, negligible duality gap can be proved and superior trade-off between average rate and user fairness can be achieved. 
The parallel implementation of GNNs was discussed in \cite{kim2022bipartite} to facilitate the deployment of decentralized GNN in distributed MIMO configurations, e.g., cell-free MIMO and fog radio access networks.
Furthermore, authors in \cite{lee2021decentralized} analyzed the robustness of a decentralized GNN-based binary classifier for inference considering the imperfect fading channels and wireless noises in the exchange of local information between neighboring nodes, where a novel retransmission mechanism to enhance the prediction robustness was proposed under different communication systems.

\subsubsection{Heterogeneous Networks}
In heterogeneous networks, different types of communication subjectives (e.g., users, access points, mobile stations, etc.) can be modeled as different types of nodes connected by different types of edges (e.g., transmission edges and inference edges) to indicate a more complex wireless communication network.
A heterogeneous graph neural network (HGNN) was firstly proposed in \cite{li2022heterogeneous} to characterize two different types of nodes (access points (APs) and mobile stations (MSs)) and various types of edges between nodes (e.g., uplink/downlink transmission path and inter-AP/inter-MS interference path) in distributed cell-free massive systems. 
To address the impact of heterogeneous nodes, an adaptive node embedding layer was proposed, where the node features of AP and MS are transformed by two embedding matrices to handle the varying input feature dimensions before the process of GNN layers.
Similar HGNN was considered in \cite{zhang2021scalable} for D2D resource allocation, which sets individual aggregation/update functions according to different relations between nodes to address network heterogeneity.
Different from the traditional GNN-based power allocation, whose outputs are permutation equivariant to arbitrary permutations of users, author in \cite{jia22GNN} constructed a permutation equivariant heterogeneous GNN (PGNN) to learn the optimal power allocation policy in cellular networks whose outputs are only equivariant to some permutations of user nodes to precisely match the identified properties in the power allocation policy. 
It shows that the PGNN achieves better learning performance in terms of sample efficiency, computational complexity and performance optimality due to the exploitation of the well-matched policy properties and the heterogeneous design of GNN.

\subsection{Other Signal Processing Applications}
\subsubsection{Mobile Traffic Prediction}
Graph-based methods can also be applied to address large-scale mobile traffic prediction \cite{kaiwen22GNN4TP,feiyang22dtGNN}, where the challenge is to exploit the time-evolving nature of mobile movements and the spatial relations of mobile traffic demand. 
Typically, a graph is constructed to characterize the spatial structure of the traffic data in different geometric regions by dividing the area into discrete grids, where node represents the region and edge represents the road connecting the regions. 
To capture the temporal correlations of traffic data, RNN-based \cite{guo2020optimized,kaiwen22GNN4TP} and CNN-based \cite{feiyang22dtGNN} methods can be employed. 
For example, Guo \textit{et al.} in \cite{guo2020optimized} proposed graph convolutional RNN for traffic prediction, where GCN and gated recurrent unit were used to exploit the spatial and temporal structure of the traffic data.
He \textit{et al.} in \cite{kaiwen22GNN4TP} constructed a spatial relation graph of traffic data to capture the near-far spatial correlation, and utilized an attention-based structural RNN to capture the temporal dependency and spatial relationship simultaneously. 
Besides the spatial-temporal structures of traffic sequences, the user's mobility patterns have also been exploited in \cite{feiyang22dtGNN} by a graph-based temporal convolutional network for accurate traffic prediction, where each node denotes a wireless AP, the directed edges indicate the movements of mobile users during the time steps of interest, and the temporal convolutional network layers further model the temporal trend of mobile data traffic on each AP.

\subsubsection{Channel Tracking}
The massive MIMO channels can be modeled as a graph, where GNN can extract the spatial correlations within the large-scale channels for efficient channel estimation.
Specially, different antennas are considered as nodes with their channel coefficients being node features, while the spatial relationships are modeled as edges with the channel spatial correlations being edge features for graph modeling.
In \cite{yang2020graph}, a GNN-based channel tracking framework was designed, which contains an encoder fed with historical channel samples, a core network performing GNN updates and a decoder to decode the node and edge attributes. 
It shows that the graph-structure captured GNN significantly outperforms feed-forward NNs for channel tracking.

\subsection{Advantages and Disadvantages}
The advantages of GNNs are summarized as follows.
\subsubsection{High Scalability and Generalizability}
GNNs enjoy promising scalability and generalization ability for two reasons.
First, the permutation invariance and permutation equivariance properties of GNNs enable the learned NN to adapt to large-scale and dynamic scenarios by exploiting the analogies or equivalent patterns between the training network topology and dynamic testing conditions automatically. 
Second, GNNs leverage the distributed message passing architecture to learn local relationships among graph nodes and combinatorial generalization over graphs \cite{suarez2022graph}.
As a result, GNNs can generalize to large-scale communication networks with varying sizes and permuted structures (e.g., more users, antennas, BSs, etc.).

\subsubsection{Good Learning Performance and High Computational Efficiency}
Compared with generic NNs, GNNs are more suitable for graph-structured data and distributed systems, which can exploit the task-specific knowledge for better learning performance with less training samples \cite{yifei21gnn}.

However, GNNs also suffer from explanation, generalization and representation limitations.
For example, GNNs cannot compute some important graph properties such as the longest or shortest cycle, diameter, or certain motifs \cite{garg2020generalization}, which are crucial for the theoretical performance analysis over graph. 
GNNs also show a poor learning performance when aggregating messages across a long path, and this situation cannot be improved by increasing the number of aggregation network layers in practice \cite{alon2021on}.
The theoretic understandings of GNNs are still in an early stage.

\section{Deep Reinforcement Learning for Stochastic Optimization}\label{sec:5}
The long-term performance optimization in dynamic and uncertain wireless networks can be cast as a stochastic optimization problem, where the network entities need to learn the optimal policies over time under system uncertainties. 
DRL has emerged as an efficient and powerful ML tool in addressing the sequential decision-making problems in dynamic and large-scale networks without explicit models of transmission environment, which allows the agents to update the decision policies through interactions with the unknown environment. 
In this section, we start with the motivations of DRL. 
Then we present the basic concepts of reinforcement learning (RL) 
and the categories of DRL techniques, followed by the case studies of DRL in wireless communication networks to fully reveal the power of DRL in solving stochastic optimization problems in dynamic large-scale wireless networks. 
The pros and cons are summarized at the end of this section.
\subsection{Motivations of DRL}
With the emergence of diversified application scenarios and the ever-growing density of wireless devices in modern networks, such as IoT, unmanned aerial vehicle (UAV), and massive machine-type communication systems, the performance optimization in these networks becomes extremely complicated due to the dynamic and uncertain environment. 
The general trends towards intelligent, autonomous, self-adaptive, and decentralized networks have been indispensable, where the agents learn the optimal decision policies automatically based on local or minimal exchanged information to maximize the network performance over time in dynamic and large-scale modern emerging networks.  
This kind of problem can be modeled as a Markov decision process (MDP) and various techniques, such as dynamic programming \cite{bertsekas2011dynamic} and RL can be employed to solve the MDP problem. 
However, in stochastic and dynamic environments, it is infeasible to build an explicit mathematical model to fully capture the characteristics of the time-varying environments, which renders many model-based methods impractical. 
RL is a machine learning technique that aims at maximizing the accumulated discounted reward of an MDP with collected experiences in a data-driven manner. 
Specifically, through trial-and-error interactions between the agent and the environment, the RL enables the agent to establish a general long-term optimal control policy while keeping track of the real-time environmental dynamics for sequential optimization problems.
\begin{figure}
	\centering
	\includegraphics[width=0.45\textwidth]{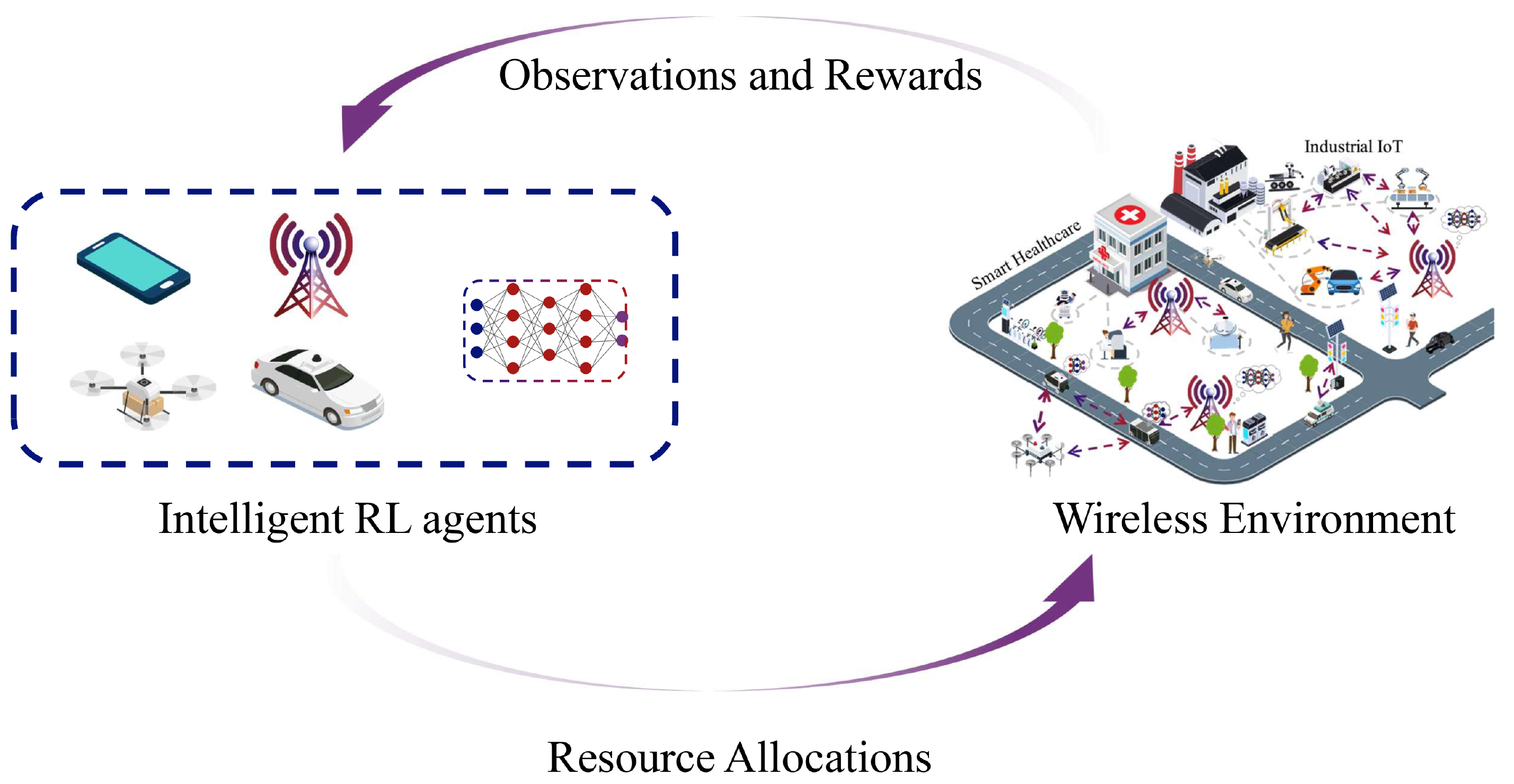}
	\caption{Reinforcement learning method in wireless communications.}
	\label{fig:rl}
\end{figure} 

Even though RL can effectively solve the MDP without explicit state transformation models, the exploration of an unknown environment consumes lots of time, especially for a highly dynamic and large-scale network. 
Motivated by the strong learning ability of DNN as a universal function approximator, the combination of RL and DNN, namely DRL, demonstrates great success in complex and dynamic wireless networks. 
The DRL enjoys the fast learning ability of DNN to speedup the learning process of RL and the transfer ability of DNN to be scalable to large-scale networks. 
On the other hand, DRL benefits from the RL techniques to be adaptive to the dynamic condition and amenable to distributed implementation.    
The DRL technique is illustrated in Figure \ref{fig:rl}. Before discussing the technical details of DRL, we will first introduce the basic knowledge of RL in the following subsection. 

\subsection {RL and Categories of DRL}
MDP can be described with $\langle\mathcal{S, A, P, R},\gamma\rangle$, where $\mathcal{S}$ and $\mathcal{A}$ denote the state space and action space, respectively, $\mathcal{P}$ denotes the state transition probability, $\mathcal{R}$ and $\gamma$ denote the reward function and the discount factor over the future reward.
The Markov property of state transition is expressed as
\begin{align}
	&\mathcal{P}\left[s' = s(t+1)\;|\;s(1),\ldots,s(t), a(1), \ldots, a(t)\right]\nonumber \\
	= &\mathcal{P}\left[s' = s(t+1)\;|\;s(t), a(t)\right].
\end{align}
The intelligent agent determines its next move $a(t)\in\mathcal{A}$ according to its current state $s(t)\in\mathcal{S}$ as well as its learned policy $\pi$, which involves in the state transition of the environment, i.e., $s(t+1) \sim P(s'\;|\;s(t),a(t))$.
At the end of each step, the agent receives a reward $r(t) = \mathcal{R}(s(t), a(t))$ from the environment, stores the tuple $\langle s(t), a(t), r(t), s(t+1)\rangle$ and updates its policy according to the corresponding state-action pair.
Note that $\mathcal{P}$ and $\mathcal{R}$ are generally unknown due to the randomness and the complexity of the environment, therefore we shall learn the optimal control policy with model-free RL.
The objective of RL is to find an optimal policy to maximize the expected accumulated discounted reward, i.e., 
$
	\max_{\pi}\mathbb{E}_{\pi}\left[\sum_{j=0}^{\infty}\gamma^j r(t+j)\right],
$
where $\pi$ is the policy denoting the mapping from a state to an action, $r(t) = R(s(t),\pi(s(t))$ is the instantaneous reward obtained after the action $a(t)=\pi(s(t))$ is performed under state $s(t)$. 
In particular, a good policy shall balance the instant reward in the current round and the potential dividend in the future rounds to achieve long-term optimality in sequential decision-making.

In each round of the decision process, the intelligent agent evaluates the value of the state under a policy $\pi$, which is defined as 
$
	V_{\pi}(s) = \mathbb{E}_{\pi}\left[\sum_{j=0}^{\infty}\gamma^{j}r(t+j)\;|\; s\in\mathcal{S}\right].
$
The state value function $V_{\pi}(s)$ represents the discounted reward that could be achieved at initial state $s$ following the policy $\pi$. Similarly, the state-action function can be defined as 
$
	Q_{\pi}(s,a) = \mathbb{E}_{\pi}\left[\sum_{j=0}^{\infty}\gamma^{j}r(t+j)\;|\; s\in\mathcal{S}, a\in\mathcal{A}\right], 
$
which characterizes the expected discounted reward when starting from initial state $s$ and action $a$, then following the policy $\pi$ thereafter. 
In particular, there is a conversion relationship between state value function $V_{\pi}(s) = \sum_{a}\pi(a|s)Q_{\pi}(s,a)$ and state-action function $Q_{\pi}(s,a) = \sum_{s'}\mathcal{P}(s'|s,a)\left(R(s,a) + \gamma V_{\pi}(s')\right)$, where $\pi(a|s)$ denotes the conditional probability of each action $a$ under given state $s$ following policy $\pi$.
By substituting $Q_{\pi}(s,a)$ into $V_{\pi}(s)$, we have the Bellman expectation equation
$
	V_{\pi}(s) =\sum_{a}\pi(a|s)\left(R(s,a) + \gamma\sum_{s'}\mathcal{P}(s'|s,a)V_{\pi}(s')\right).
$
Given the optimal policy $\pi^*$, we have the following Bellman optimality equation  for $V_*$ and $Q_*$, respectively, i.e., 
\begin{equation}\label{bellman_opt}
	V_{\pi^*}(s) = \max_{a}\left\{R(s,a) + \gamma\sum_{s'}\mathcal{P}(s'|s,a)V(s')\right\},
\end{equation}
\begin{equation}\label{bellman_opt2}
	Q_{\pi^*}(s,a) =\max_{a}\left\{R(s,a) + \gamma\sum_{s'}\mathcal{P}(s'|s,a)V_{\pi^*}(s')\right\},
\end{equation}
where the optimal action in each round shall be found to achieve the Bellman optimality equation.

When the dimension of state and action space is small, Q-learning\cite{mnih2015human} is a widely used algorithm in RL to find the optimal action sequences by visiting each action-state pair. However, in complex and large-scale networks, the state and action space can be very large. To accelerate the learning process, DNN can be embedded into the RL framework to approximate computational-expensive quantities with parameterized functions. Based on the types of quantities fitted by the DNN, the DRL techniques can be categorized into value-based and policy-based algorithms. 
\subsubsection{Value-Based DRL}
The value-based DRL can be applied to solve MDP with discrete state/action space, where the value functions are approximated using DNN. The widely used method is deep Q-learning (DQL)\cite{mnih2015human}, which implements a deep Q-network to fit the value of $Q_\pi^\star (s,a)$ in \eqref{bellman_opt2}.  
In particular, DQL accepts the features of states as input and outputs the fitted action-state values for all actions, where the action with the largest state-action value is adopted, i.e.
$
	a(t) = \underset{a\in\mathcal{A}}{\arg\max}\;Q(s, a).
$
To train the NN, a loss function defined as the mean square error (MSE) between a target output, calculated using target network parameters $\theta^\prime$, and the actual output, characterized by parameters $\theta$, is utilized. 
To reduce the oscillations, the target network $\theta^\prime$ is updated much slower than the actual network $\theta$. 
With the learned Q values, the actions chosen at each time slot follow a widely used $\epsilon$-greedy strategy, where the agent chooses the action randomly with probability $\epsilon$ and the action that maximizes the current action-state value with probability $1-\epsilon$ to balance the exploration and exploitation in the training process. 
\subsubsection{Policy-Based DRL}
The agents in the policy-based algorithms learn the policy mapping from the current state to the optimal action or the probability of each action directly through the DNN technique instead of evaluating the state-action values, 
which can be applied to both continuous and discrete action spaces.
In particular, the neural networks are optimized by minimizing the following loss function,
\begin{equation}
	\rho(\theta) = \lim_{T\to\infty}\frac{1}{T}\sum_{t =1}^{T} r_t= \sum_{s\in\mathcal{S}}d_{\theta}(s)\sum_{a\in\mathcal{A}}\pi_{\theta}(a|s)R(s,a),
\end{equation}
where $d_{\theta}(s) = \lim_{T\to\infty} \mathcal{P}(s_t = s|s_0)$ denotes the steady-state probability distribution under policy $\pi$, which is parameterized by $\theta$. 
To facilitate the model parameter update, actor-critic framework, as in deep deterministic policy gradient (DDPG) \cite{DDPG2015}, can be employed.
In particular, the critic learns a parameterized state-action function $Q^\omega_{\pi}(s, a)$ by using the Bellman equation as in DQL, where a copy of the actual critic network can be adopted to calculate the target values with slowly updated weights to improve the learning stability. On the other hand, the actor learns a parameterized function $\pi_{\theta}(a\;|\;s)$ specifying the current policy through mapping the states to a specified action, which can be obtained by maximizing the learned value function of the critic. 
\begin{table}[htp!]
	\caption{DRL-enabled stochastic optimization in wireless networks}
	\resizebox{\linewidth}{!}{
		\begin{tabular}{cccc}
			\toprule
			Prolem Types &DRL Methods &\tabincell{c}{ Application Scenarios}& Refs\\ 
			\toprule
			\multirow{3}{*}{IP}&\multirow{3}{*}{Value-Based}&Intelligent Traffic&\cite{8633520, 8686046, wei2018intellilight, 9417262, 8569568,8317839, YE2019155, DU2022122523, 8605023}\\ 
			\cline{3-4}
			&&\tabincell{c}{Discrete-Valued Power Control}&\cite{7997286, 8633948, 9079819, 8770243, 9326357}\\ 
			\cline{3-4}
			&&Device Scheduling&\cite{10.1145/3132847.3133045,9251950,10.1145/3288599.3288634}\\ 
			\hline
			\multirow{2}{*}{\tabincell{c}{Stochastic \\ MINLP}}&\multirow{2}{*}{Policy-Based}&\tabincell{c}{Mobile Edge Network Optimization}&\cite{8932481, 9295428, 9492254, 9112328, 9333595, 9093962, 9635652, 9385791}\\ \cline{3-4}
			&&Space-Air-Ground-Integrated Network&\cite{9351533, 9641753, 9749852, 9749175, 9652086}\\ 
			\hline
			DCOP&\tabincell{c}{MADRL}&\tabincell{c}{Scalable Radio Resource Allocation}&\cite{8792117,9120241,9650909,8792382,9346039}\\  
			\bottomrule
	\end{tabular}}
	\label{ta:DRL_opt}
\end{table}

\begin{table*}[t]
	\caption{DRL-assisted Intelligent Transportation Systems}
	\resizebox{\linewidth}{!}{
		\begin{tabular}{cccccc}
			\toprule
			\tabincell{c}{Application Scenarios}&DRL Methods & Refs&State Space&Action Space &Reward Function\\ \toprule
			\multirow{4}{*}{ATSC}&\tabincell{c}{DQN}&\cite{8633520}&\tabincell{c}{Handcraft features of local traffic}&\tabincell{c}{Traffic signal}&\tabincell{c}{Accumulated waiting time}\\ \cline{2-6}
			&\tabincell{c}{LSTM+DQN}&\cite{8686046}&\tabincell{c}{Number of vehicles traffic flow}&\tabincell{c}{Duration of traffic signal}&\tabincell{c}{Waiting rate of vehicles}\\  \cline{2-6}
			& \tabincell{c}{CNN+DQN}&\cite{wei2018intellilight}&\tabincell{c}{Local transportation state}&\tabincell{c}{Traffic signal}&\tabincell{c}{Mixed function of delay,\\ queue length and waiting time}\\  \cline{2-6}
			&\tabincell{c}{CNN+DQN}&\cite{9417262}&\tabincell{c}{Local transportation state,\\ number of intelligent vehicles}&\tabincell{c}{Traffic signal,\\vehicle detouring}&\tabincell{c}{Mixed function of waiting \\ time and system influence}\\  
			\hline
			\multirow{3}{*}{Autonomous Driving}&\tabincell{c}{DQN}&\cite{8569568}&\tabincell{c}{Current speed, land index \\ and road information}&\tabincell{c}{Lane change, speed change}&\tabincell{c}{Self-defined indicator}\\ \cline{2-6}
			&\tabincell{c}{DQN}&\cite{8317839}&\tabincell{c}{Current pedestrain status}&\tabincell{c}{Pedestrain detection}&\tabincell{c}{Self-defined indicator}\\ \cline{2-6}
			&\tabincell{c}{DQN}&\cite{YE2019155}&\tabincell{c}{Real-time traffic flow scenes}&\tabincell{c}{Car following, lane change}&\tabincell{c}{Self-defined indicator}\\
			\hline
			\multirow{2}{*}{On-Board Management}&\tabincell{c}{DQN}&\cite{DU2022122523}&\tabincell{c}{State of charge, power demand \\ and generator speed}&\tabincell{c}{Throttle of the engine}&\tabincell{c}{Mixed function of state of charge\\ and instantaneous fuel consumption rate}\\  \cline{2-6}
			&\tabincell{c}{DQN}&\cite{8605023}&\tabincell{c}{State of charge}&\tabincell{c}{Operation rate parameter \\of energy source}&\tabincell{c}{Deviation
				between \\ SOC of all units}\\  
			\bottomrule
	\end{tabular}}
	\label{ta:DRL_IP}
\end{table*}

Recently, DRL-based approaches have attracted attentions of the wireless community to solve different types of wireless resource allocation problems, such as integer programming, mixed-integer linear programming and sequential optimization problem, in a wide range of applications including but not limited to multi-access scheduling, power control, beamforming designs, and bandwidth allocation.
In general, DRL-based approaches inherit the advantages of conventional control theorem with theoretical guarantees for convergence and optimality \cite{beggs2005convergence, littman1996generalized, DDPG2015, mnih2015human, silver2014deterministic} in the long-term performance optimization, and the advantages of data-driven DL with high computational efficiency and excellent learning performance in complex, dynamic, and large-scale wireless networks. 
In the following subsections, we shall review several representative cases of the application of DRL in wireless resource allocation according to the types of optimization problems. The covered cases are summarized in Table \ref{ta:DRL_opt} based on the optimization problem types, application scenarios, DRL methods, MDP elements, etc.

\subsection{Application 1: Stochastic Integer Programming Problems}
The stochastic integer programming (IP) problems involving discrete variables in dynamic scenarios have been quite common in wireless resource allocation, which have the following general forms: 
\begin{subequations}\label{integer_pro}
\begin{align}
		\underset{x_1,\ldots, x_T}{\rm{maximize}}~&\;\sum_{t=1}^{T}c_tf_t(\bm{x}_t; \bm{y}_t),\\
		\rm{subject~to}~&\;g_t(\bm{x}_t)\leq \bm{b}_t,\\
		\and~&\; \bm{y}_t \sim P(\bm{y}\;|\;\bm{y}_{t-1}, \bm{x}_{t-1}),\\
		\and~&\;\bm{x}_t\in\mathbb{Z}^{n},
\end{align}
\end{subequations}
where the objective function $f_t(\cdot)$ can be the utility function of wireless networks,  $g_t (\cdot)$ can be the performance constraint functions, and the discrete-valued variables $\bm{x}_t$ can be resource allocation variables, such as device scheduling, $y_t $ stands for the dynamic state of the environment with Markov evolutions. 
As a non-convex problem with NP-hardness, the COAs for solving IPs suffer from unintended performance degradation, high computational complexity and high requirements for datasets in the actual implementation.
DL-based algorithms, such as LBB in Section \ref{sec:3}, can be employed to solve the IP problem. 
However, LBB and DRL focus on different application scenarios.
For example, LBB aims at solving one-shot IP problems, which cannot handle long-term constraints. 
Instead, DRL is specifically used to cope with long-term constrained stochastic optimization problems by adaptively interacting with the environment to learn the long-term optimal policies. 
Besides, LBB requires explicit models to achieve a good learning performance. Instead, DRL approaches are model-free, which can successfully perform for unknown dynamics through autonomous exploration. 
In conjunction with the good robustness, DRL provides great potentials of scalability in handling high-dimensional problems and flexibility in embedding diverse task-specific features for decision making compared to the LBB-based algorithms.

Therefore, the value-based DRL algorithms can be effectively adopted to solve the challenging stochastic IP problem by transforming the original IP problem into an MDP with discrete action spaces. Specifically, the objective function or the metric strongly correlated with the optimization objective can be regarded as a reward function in MDP. The state should include the features relevant to the decision policies while the action can be the discrete optimization variables. Then the value-based DRL algorithms can be employed to search for the optimal actions in discrete action space.
We introduce several applications of value-based DRL in solving the IP problem.

\subsubsection{Intelligent Traffic}
With a growing increase of autonomous vehicles and intelligent roadside units, establishing an intelligent transportation system (ITS) is becoming important to improve transportation efficiency in the pursuit of smart cities, which has been attracting considerable attentions from researchers, and plenty of DRL-based methods have been proposed thereafter.
Some of the hottest applications for DRL-assisted ITS are adaptive traffic signal control (ATSC), autonomous driving, and on-board energy management, the details of which, including DRL algorithms and MDP modeling, are summarized in the table \ref{ta:DRL_IP}.

For example, Choe \textit{et al.} \cite{8633520} developed a deep Q network (DQN)-based approach to maximize transportation efficiency by minimizing local accumulated waiting time according to handcraft features of local traffic.
Further, Jin \textit{et al.} \cite{8686046} took traffic flow into account to minimize the waiting rate of vehicles for higher transportation efficiency.
Wei \textit{et al.}\cite{wei2018intellilight} integrated multi-source factors into the design of the reward function to improve the overall performance in reducing the network latency. 
Beyond optimizations over traffic signals, the need for autonomous driving such as lane changing and auto-breaking rises with the rapid development of intelligent vehicles and the Internet of Vehicles (IoV).
For instance, Hoel \textit{et al.}\cite{8569568} jointly optimized lane change policy and speed change policy of autonomous driving vehicles according to their locations and speeds to minimize their travel time.
Ye \textit{et al.}\cite{YE2019155} further extended the state from abstract features to real-time traffic flow scenes and enabled autonomous vehicles to optimize car following and land-changing policies for higher travel efficiency.
The joint control of roadside units and vehicles was firstly investigated to improve traffic efficiency in \cite{9417262} by jointly controlling the traffic signal of an intelligent traffic light and the detouring behavior of intelligent vehicles connected to the IoV according to the real-time traffic flow information.
Through the unique design of the MDP modeling, the agent learned to maximize the accumulated reduced waiting time while preventing the side roads from high congestion as well, thus maximizing the traffic efficiency of the whole traffic network. 
The numerical simulations in \cite{9417262} show that the DQN-based algorithm proposed therein can achieve better performance than that of conventional strategies as well as the DRL methods that only control the traffic signals with affordable computational consumption for highly time-sensitive real-time traffic control scenarios.

\subsubsection{Discrete-Valued Power Control}
To maximize network utility, the discrete-valued power control as a classic IP problem can be effectively solved by DQN as proposed in \cite{7997286, 8633948, 9079819, 8770243, 9326357}.
For example, Xu \textit{et al.} \cite{7997286} firstly adopted DQN to dynamically allocate discrete-valued transmit powers of BSs in C-RANs, which manifests superior performance in terms of energy efficiency and the adaptability to highly dynamic environments.
Ye \textit{et al.} \cite{8633948} further extended the DQN to address the joint allocation of sub-band and transmit power in V2V networks, where the proposed DQN-based scheduling strategy can be executed in distributed systems to meet the stringent latency requirements.
The authors in \cite{9079819} applied the DQN in the mobile edge computing scenarios to jointly optimize the transmit power and device scheduling, which shows significant enhancement in aspects of network delay and resource consumption.
Chu \textit{et al.} \cite{8770243} considered an energy-harvesting assisted communication system without any prior knowledge assumed of the energy dynamics and developed a two-stage strategy to solve the joint optimization problem of battery prediction and sum-rate maximization, where a long short-term memory (LSTM)-based network is used to improve the prediction accuracy of battery status in the first stage and the DRL is employed to optimize real-time transmit power and access policy for sum-rate maximization.
\subsubsection{Device Scheduling}
In wireless networks, as each device can only be in the state of scheduled or non-scheduled, the device scheduling optimization turns out to be an IP problem and DQN can be effectively used to address it. For example, Lee \textit{et al.} \cite{8809381} proposed a circumstance-independent DQN-based scheduler to maximize the network utility under various conditions and QoS constraints. The unconstrained Lagrangian function was adopted as a reward function to cope with various constraints. The DQN-based scheduler was also successfully used in FL systems and mobile edge computing systems \cite{10.1145/3132847.3133045,9251950, 10.1145/3288599.3288634} to schedule participating devices.
For example, Zhou \textit{et al.} \cite{10.1145/3132847.3133045} adopted DQN to schedule training batches to optimize the quality of query response for a cloud-enabled DNN inference system.  
Young \textit{et al.} \cite{9251950} further considered the cloud-edge hybrid inference system and proposed AutoScale, an automatic DQN-based scheduler, to dynamically schedule the inference execution target for improving the prediction accuracy. 
To tackle the problem of non-i.i.d distribution of heterogeneous data in the inference of FL system, Young \textit{et al.} \cite{kim2021autofl} proposed AutoFL, a heterogeneity-aware DQN-based scheduler, to schedule the FL participants for overall energy efficiency enhancement. 

\subsection{Application 2: Stochastic Mixed Integer Nonlinear Problems}
Stochastic MINLPs involving both continuous-valued and discrete-valued variables in dynamic environments have been widely encountered in the wireless communication systems, 
which have the following general forms:
\begin{subequations}\label{mixed_integer_pro}
\begin{align}
\underset{\{\bm{x}_i\}_{1}^{T},\{\bm{z}_i\}_{1}^{T}}{\rm{maximize}}~&\;\sum_{t=1}^{T}c_tf_t(\bm{x}_{t}, \bm{z}_{t}; \bm{y}_{t}),\\
\rm{subject~to}~&\;g(\bm{x}_{t})\leq \bm{b}_t,\label{constraint_rl_minlp1}\\
~&\;h(\bm{z}_t)\leq \bm{d}_t,\label{constraint_rl_minlp2}\\
\and~&\;\bm{x}_t\in\mathbb{Z}^{n},\\ 
\and~&\;\bm{z}_t\in\mathbb{R}^{n},\;\;\forall t,\\
\and~&\; \bm{y}_t \sim P(\bm{y}\;|\;\bm{y}_{t-1}, \bm{x}_{t-1},bm{z}_{t-1}),
\end{align}
\end{subequations}
where $f_t(\cdot)$ denotes the utility function of wireless networks, $g_t(\cdot)$ and $h_t(\cdot)$ can be the performance/resource constraint functions, such as the QoS constraints and the maximum/average power constraints,  
$\bm{x}_t$ and $\bm{z}_t$ denote discrete-valued and continuous-valued network resources, respectively, $\bm{y}_t$ denotes the dynamic states of the environment with Markovian properties. 
Obviously, it is computationally expensive and analytically 
intractable to solve such an NP-hard stochastic non-convex problem with COAs. 
On the other hand, the value-based DRL algorithms can only deal with the discrete action space to evaluate the state-action values for each possible action. 
Hence, bypassing the evaluation of the state-action values and directly choosing the action under the currently learned policy, the policy-based methods can be effectively employed to solve the challenging stochastic MINLPs.
To model the stochastic MINLP as an MDP, the state can be defined as the features related to the decision-making, 
the discrete-valued and continuous-valued variables yet to be optimized can be considered as actions of MDP, which are sampled from the learned policy mapping function.
The reward can be defined as the objective function or the metric strongly correlated with the optimization objective.
Then, the policy-based methods can be employed to learn the optimal policy function directly in the continuous action space.
Note that the discrete-valued variables in the stochastic MINLPs can be obtained by rounding the optimized continuous-valued variables in the testing.
We introduce several applications of the policy-based DRL in solving the stochastic MINLPs.

\subsubsection{Mobile Edge Network Optimization}
Applications migrated from cloud to edge have been a prevalent trend in the era of 5G and beyond. 
By enabling a large number of access points and integrating the widely distributed computing, caching, and communication resources, the mobile edge network (MEN) can significantly reduce the communication latency and improve the network performance by jointly optimizing the heterogeneous resources at the edge.
DRL-based methods\cite{8932481, 9295428, 9492254, 9112328, 9333595, 9093962, 9635652, 9385791} have shown promising performance in the multi-resource joint optimization in MEN attributed to their self-adaptation to dynamic environment, the generalization power for high-dimensional problem and the real-time inference in practice. 
For instance, Ke \textit{et al.} \cite{8932481} first formulated a joint optimization problem for task offloading, bandwidth allocation, and energy sensing in IoT networks, and then proposed a DDPG-based joint design scheme to minimize the transmission delay and energy consumption, which can well handle the time-varying channel and dynamic bandwidth.
Further, the authors in \cite{9295428} developed the Wolpertinger DDPG to eliminate the possible performance degradation of DDPG induced by rounding discrete variables in a similar context.
Li \textit{et al.} \cite{9112328} proposed an LSTM-assisted DRL algorithm to allocate resources across slices under varying service demands, where the LSTM mechanism was adopted for higher tracking accuracy of user mobility and the system utility.
Xu \textit{et al.} \cite{9333595} further considered the joint optimization of channel allocation and continuous energy harvesting time while taking energy consumption and queue length into account, where the proposed DRL algorithm can achieve higher throughput with stringent performance constraints.
Recently, the block-chain application, as a computational intensive scenario, has been widely combined with the MEN to achieve higher mining efficiency.
For example, Du \textit{et al.} \cite{9385791} developed an asynchronous DRL-based algorithm to adaptively allocate the channel resources and establish the pricing policy by maximizing the rational profit among all miners while taking the wireless fading channel into account.
Specifically, the advantageous actor-critic algorithm (A3C) was adopted to avoid the overestimation or underestimation of the chosen action, which can achieve better performance than that of the DDPG as numerically verified.
\subsubsection{Space-Air-Ground-Integrated Network}
Space-air-ground integrated (SAGI) network provides ubiquitous communication and computing services from cloud to edge. 
Due to the physical distance among layers (e.g., space, air and ground), it is essential to develop a joint optimization framework to coordinate the resources across different layers and timescales to satisfy stringent QoS constraints, where various DRL-based approaches have been proposed\cite{9351533, 9641753, 9749852, 9749175, 9652086}.
For instance, Liao \textit{et al.} \cite{9351533} developed an actor-critic-based DRL algorithm to jointly optimize the task offloading and computational resource allocation by minimizing the cross-layer queuing delay, where a queuing-aware agent was developed to balance the instantaneous queuing boosting and long-term latency constraints.
Further, they proposed a block-chain and semi-distributed learning-based DRL algorithm by minimizing latency while guaranteeing long-term security requirements.
Wang \textit{et al.} \cite{9749175} took the multi-dimensional resource heterogeneity and network dynamics into account and proposed a soft actor-critic-based DRL algorithm by minimizing the energy consumption and the queuing latency of the offloading tasks.

\subsection{Application 3: Distributed Constraint Optimization Problems}
Distributed constraint optimization problems in multi-agent systems (MASs) involving multiple nodes are challenging but very common in wireless systems, which have the following general forms:
\begin{subequations}\label{MADRL}
\begin{align}
\underset{\{\bm{x}_{i,t}\}_{1}^{N}}{\rm{maximize}}&~\;\sum_{t=1}^{T}c_tf_t(\{\bm{x}_{i,t}\}_{1}^{N};\{\bm{y}_{i,t}\}_{1}^{N}),\\
\rm{subject~to}&~\;g_{i,t}(\bm{x}_{i,t})\leq \bm{b}_{i,t},\;\forall i,\\
\and&~\{\bm{y}_{i,t+1}\}_{1}^{N} \sim P(\{\bm{y}_{i,t+1}\}_{1}^{N}|\{\bm{x}_{i,t}\}_{1}^{N}, \{\bm{y}_{i,t}\}_{1}^{N}),
\end{align}
\end{subequations}
where $\{\bm{x}_{i,t}\}_{1}^{N}$ and $\{\bm{y}_{i,t}\}_{1}^{N}$ denote the set of variables and system parameters of the wireless systems, respectively, $f_t(\cdot)$ denotes the objective function of the MAS, $g_{i,t}(\cdot)$ denotes the performance constraint function of each variable. Note that the evolution of system states $\{\bm{y}_{i,t+1}\}_{1}^{N}$ can be modeled as a Markov process.
Specifically, the agent can be the BS or edge device in wireless networks, the system parameters can be wireless fading channels or edge resource status, while the set of variables can be the corresponding resource allocation policies.
It is worth noting that the system parameters of MAS at the current time slot depend on the system parameters and action variables of MAS at the last time slot, therefore the interactions among agents determine the system performance.
Due to the coupling among the agents and the non-convexity of the objective function, it is challenging to achieve the desired performance within acceptable computational delay using COAs.
Fortunately, built on the MAS, the multi-agent deep RL (MADRL) shows unmatched performance in dealing with the multi-agent co-learning problems, which allows multiple agents to learn multiple individual policies and one global policy collaboratively based on the interactions among agents and environment.

MADRL enables each agent to develop its own decision-making policy which can be executed decentralized, thereby improving the scalability of network significantly.
In particular, the agents in MAS can build either competitive relationships or cooperative relationships.
As a result, the objective of MADRL algorithms can be divided into three categories:
maximizing the global reward by coordinating all agents, maximizing the reward of each agent by constructing an equilibrium among agents, or constructing a competitive equilibrium among different groups of cooperative agents. In wireless networks, it is more general that the MAS operates in cooperative mode, which is the focus of the following discussion. Note that the learned policy in MADRL can be unstable due to the fact that the state of each agent depends not only on its own states and actions but also on the actions and states of other agents. Such mutual coupling depends on the communication range of the network. In some scenarios, the agent can only communicate with the surrounding agents, thereby leading to a partially observable MDP. Therefore, besides the consistent trial-and-error explorations as in a single-agent system, an efficient communication mechanism among agents is also important to achieve the long-term optimal policy in dynamic MAS. In the following, we illustrate the application examples of MADRL for cooperative MAS design. 
 

To overcome the issue of huge CSI overhead for centralized RRM design in large-scale wireless networks, Yasir \textit{et al.} \cite{8792117} firstly adopted the MADRL technique to dynamically allocate transmit power at each BS through mutual coordination for sum-rate maximization of wireless networks.
By modeling each BS as an intelligent agent, each agent determines its transmit power individually according to its local and neighboring channel information, achieving the same performance as that of COA with full CSI.
Further, the authors in \cite{9120241} extended it to the continuous-valued power control, where three different RL algorithms were proposed to feature a promising performance of MADRL.
However, the above methods require additional CSI exchange between the neighboring BSs, which can downgrade the spectrum efficiency.
To address this, DEC-MAPC proposed in \cite{9650909} achieved fully decentralized power control only using local CSI while maximizing the sum-rate of the network.
Specifically, to achieve fully distributed implementation, DEC-MAPC was proposed to decompose the global state-action value into a monotonic increasing non-linear function of all local state-action values. 
As a result, each BS only needs to determine its transmit power by maximizing the local state-action value, then the network utility can be maximized. 
To address the continuous power control while cooperation, an actor-critic framework with a double critic network was adopted in DEC-MAPC, leading to more accurate estimation of state-action values and higher network utility.

\subsection{Advantages and Disadvantages}
The advantages of DRL methods are summarized as follows.

\subsubsection{High Adaptability}
The training data for policy updates are collected from historical interactions with the environment.
As a result, the training of DRL policy can keep track of real-time wireless dynamics.
Compared with generic NNs, the training data in DRL methods are label-free, making it unrestricted to the traditional algorithms and allowing more degrees of freedom to improve the learning performance. 
Additionally, the DRL is model-free, thus enabling the exploration of unknown and complicated environments.

\subsubsection{Suitable for Long-Term Optimization}
The DRL-based methods can learn a long-term optimal policy that takes the potential reward in the future and long-term system constraints into account rather than just considering instantaneous system reward and one-shot constraint.

However, the DRL methods also have some limitations.
First, it is hard to train a global optimal policy because the action space is generally too large to be exhaustively traversed.
Besides, to obtain a good policy, numerous training experiences shall be stored, which however is challenging due to the limited storage capacity of local devices.
Second, while the DRL methods can be deployed with a relatively simple network structure, there are lots of hyperparameters involved in the training and execution, whose values are chosen manually by costly trials. The fine-tuning of hyperparameters is labor-intensive and time-consuming, and improperly chosen values can significantly degrade the performance of DRL. 
Third, DRL is developed based on the MDP modeling. 
For some complicated applications, the definition of MDP can be challenging. 
For example, it can be hard to acquire some state information efficiently in practice. 
It also can be quite difficult to define a highly featured state space and choose a good reward function in some scenarios.

\section{End-to-End Learning for Semantic Optimization}\label{sec:6}
The joint optimization of physical layer transmitter and receiver in wireless communication systems is extremely challenging due to infinitely large searching space for modular functions, the complex interactions among modules and the highly non-convexity of optimized global performance in terms of communication rate, transmission reliability, resource consumption, transmission delays, etc., in conventional block-based communication systems. 
DL-enabled end-to-end learning has been studied to merge the transmission blocks and jointly design the transmitter and receiver in a data-driven manner. 
To boost the system capacity and improve transmission efficiency and reliability, semantic communication has been envisioned as a new transmission paradigm by delivering the semantic meaning rather than bit stream of transmitted messages. 
In this section, we identify the motivation of semantic communication and review the classic framework of a semantic system, followed by the DL-enabled semantic system design and the overviews of its applications for different types of transmission tasks. 

\subsection{Motivation and Challenges}
In conventional wireless communication systems, message compression and message error correction are achieved by source coding and channel coding, respectively, aiming for high transmission efficiency and reliability. 
In view of this, conventional communication networks have been designed to optimize data-oriented performance metrics such as communication data rate, spectrum/energy efficiency, symbol or bit level accuracy and latency, while ignoring the semantic meaning behind the transmitted messages \cite{shannon1949mathematical}.
For instance, the bit-error rate (BER) or symbol-error rate (SER) is usually taken as performance metric in communication systems to measure the bit or symbol level accuracy and effectiveness of transmit symbols \cite{shannon1948mathematical}.
With the communication system capacity approaching Shannon limit and the booming development of ML, it is an increasing belief in the community that classical Shannon's information theory needs to be upgraded for the next evolution of wireless communication networks.
A variety of services emerged in 6G wireless systems are service/content-centric, which means they are more concerned about the semantic-related information instead of physical data symbols, which sparks a paradigm shift from the symbol transmission to the semantic meaning transmission in communication systems. 
By delivering the semantic meaning of the message relevant to the transmission task directly rather than its exact copy, semantic communication is expected to break through the classic design paradigms of Shannon which is targeting at the accurate transference of source signals to the destination receiver.  
Specifically, the semantic communication can increase the system capacity by identifying and extracting the semantic meaning and eliminating the irrelevant information from transmitted messages to realize compression.  
The semantic communication can guarantee the reliability of transmission through exact semantic meaning recovery/interpretation at the receiver. However, to reap the benefit of semantic communication, semantic-aware optimization for wireless  techniques and network structures should be activated to accommodate to the new requirements in semantic communication systems. 

There are several challenges for semantic-aware optimization, which are summarized as follows. 
\subsubsection{New Metric Design}
To enable semantic-aware optimization, it is indispensable to design metrics for both exact semantic meaning extraction and accurate semantic meaning transmission. 
The first metric measures the meaning behind the transmitted symbols mathematically. 
The second metric characterizes the semantic errors between recovered and transmitted semantic meaning to guarantee successful semantic meaning inference at receiver. 
\subsubsection{Joint Design of Transmitter and Receiver}
Instead of the separate design, the coordinated design of transmitter (source) and receiver (destination) can achieve high system capacity and reliable transmission simultaneously by exploiting semantic side information. 
Such joint design is expected to compress the transmitted signals maximally while reserving the semantic meaning at transmitter and recover the semantic meaning at receiver to combat the channel fading and semantic noise. 
\subsubsection{Mathematical Theories}
It is still an ongoing research direction to develop efficient and elegant mathematical theories to evaluate the overall performance of a semantic communication system.   

Motivated by recent ML tools, DL-enabled semantic communication system has received considerable attention, where the transmitter and receiver implemented by DNNs can be jointly learned targeting at good overall performance \cite{guangming21semcom}. 
In the following, we will introduce the architecture of a classic semantic communication system, after which we will review the existing techniques to address the challenges for semantic-aware optimization.
\subsection{Architectures of Semantic Communication Systems}
Semantic communication system usually contains three components including semantic transmitter and receiver, knowledge base, as well as semantic noise and error \cite{luo2022semantic, guangming21semcom}.

\subsubsection{Semantic Transmitter and Receiver}
The desired semantic transmitter and receiver are expected to be agents with intelligence (e.g. humans and smart devices), aiming to perform the functions of semantic communication terminals, e.g., executing highly intelligent compression/extraction/interpretation algorithms, sensing the environment to obtain high-level data and updating the knowledge bases, etc. 
Semantic encoders are typically deployed in the semantic transmitter, which are able to extract the meaning of the source (e.g. text, speech and image messages) and encode these features into symbols (bits) for transmission.
The receiver with semantic decoder should be able to recover the compressed features sent by semantic transmitter as well as perform various intelligent tasks based on the inferred semantic information (e.g., automatic speech recognition when transmitting speech signals).

\subsubsection{Knowledge Base}
The semantic transmitter and receiver contain certain knowledge bases (KBs) to capture the meaning of the knowledge entities and their complex relationships. 
The KBs at transmitter and receiver are expected to be constantly updated by self-learning and both contain the knowledge elements involved in the current communication, which constitute the core of a semantic communication system \cite{luo2022semantic}.
The KBs are the knowledge models that the transmitter and receiver observed previously and can be shared through communications.
With the transmitter KB, the semantic transmitter extracts the features of the transmitting messages and then the semantic receiver can interpret and infer the meanings of them based on the receiver KB.
Based on different types of source messages such as text, image or audio, the KBs could be different for various applications.
In addition, the KBs at the semantic transmitter and receiver may also be different due to the different abilities for understanding (e.g., the transmitter is Chinese language system while the receiver only uses English).

\subsubsection{Semantic Noise and Error}
Semantic noise that interferes with the interpretation of the semantic information during encoding, data transportation, and decoding processes is introduced as one of the semantic communication components \cite{guangming21semcom}.
Semantic communication system contains two kinds of noises, namely channel noise and semantic noise.
In addition to physical channel noise such as additive white Gaussian noise (AWGN), fading channels, and multi-path effect, which are introduced by channel impairments and can cause the signal attenuation and distortion, the semantic noise is defined as a type of disturbance in message interpretation processes due to the ambiguity in words, sentences or symbols used in the message transmission \cite{xie2021deep}.
Semantic noise can lead to semantic errors in the receiver and misunderstanding of the received message.
The semantic noise may occur when KBs between the semantic transmitter and receiver are mismatched.
These two kinds of noises will eventually lead to semantic understanding errors at the receiver and it is hard to distinguish which factor causes the errors.
\begin{figure}[t]
  \centering
  \includegraphics[width=0.5\textwidth]{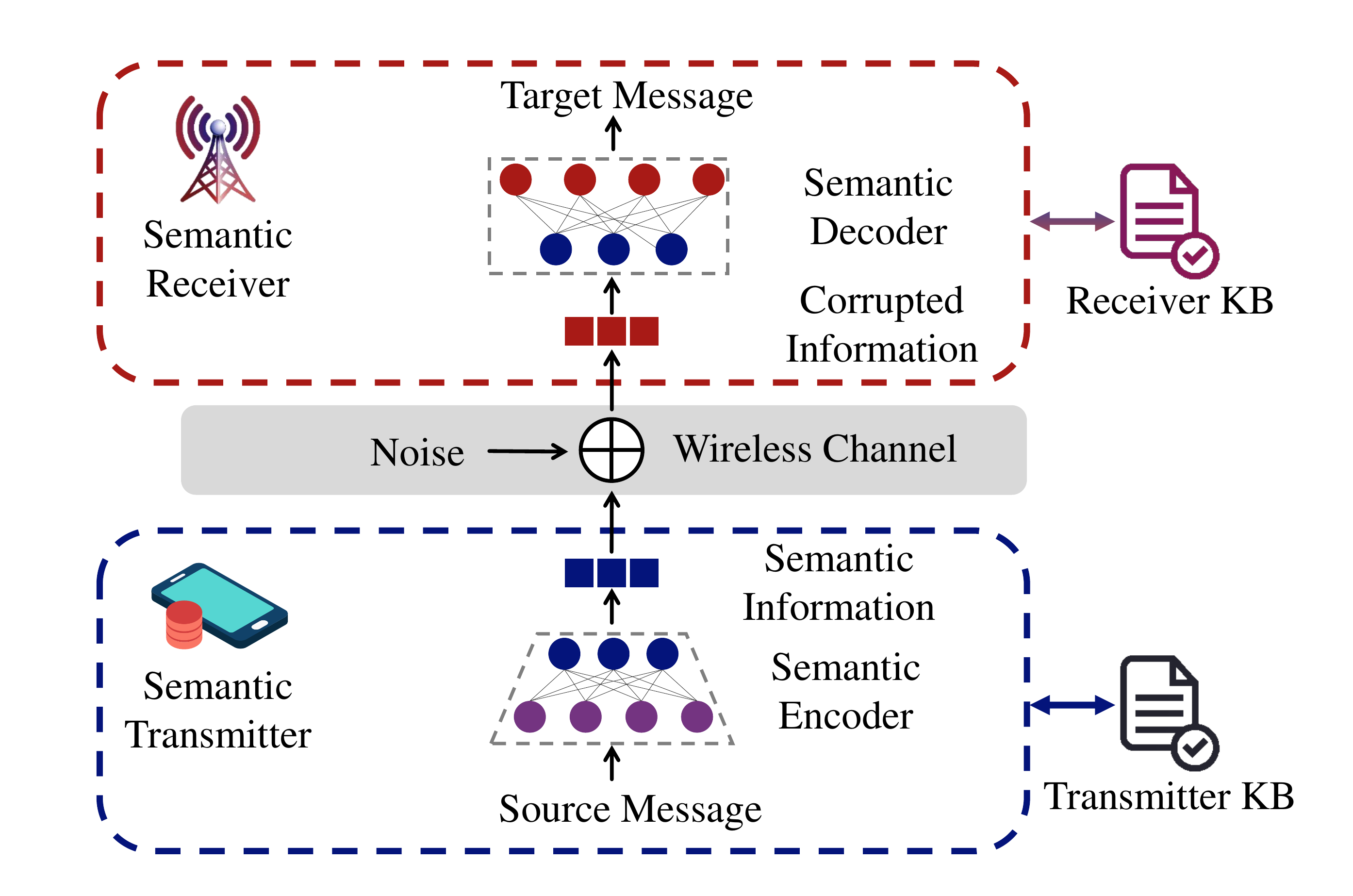}
  \caption{Semantic communication system.}
  \label{fig:5}
\end{figure}

\subsection{Semantic Meaning Extraction and Interpretation}
In wireless networks, the core issue of semantic communication is how to extract the semantic information and then perfectly recover it after data transportation, which can be expressed mathematically under information-theoretic view as follows.
For the source signal $\bm{x}$ and its corresponding received signal $\bm{y}$, which belong to a pair of random variables $(\bm{X},\bm{Y})$, the probability model involved in semantic communication is represented as a Markov chain $\bm{Y} \leftrightarrow \bm{X}  \leftrightarrow \bm{Z} \leftrightarrow \hat{\bm{Z}}$.
The random variable $\bm{Z}$ and $\hat{\bm{Z}}$ represent the semantic information after semantic encoder and the received semantic information at semantic decoder, respectively.
Assuming a wireless fading channel from the semantic encoder to the semantic decoder, then we have $\hat{\bm{Z}} = \bm{HZ} + \bm{W}$, where the random variables $\bm{H}$ and $\bm{W}$ represent the channel model and the channel noise model, respectively.
The semantic encoder $C_{\bm{\theta}}(\cdot)$  can be represented by parameter $\bm{\theta}$, while the semantic decoder $C_{\bm{\phi}}(\cdot)$ is parameterized by $\bm{\phi}$ at the receiver side.
Then the communication probabilistic model satisfies $p(\bm{y}|\bm{x}) = p_{\bm{\phi}}(\bm{y}|\hat{\bm{z}}) \cdot  p_{\bm{\theta}}(\hat{\bm{z}}|\bm{x}) $, where $ p_{\bm{\theta}}(\hat{\bm{z}}|\bm{x}) = p_{c}(\hat{\bm{z}}|\bm{z}) \cdot p_{\bm{\theta}}(\bm{z}|\bm{x})$ denotes the transmitter and channel probabilistic encoder and $p_{\bm{\theta}}, p_{c}, p_{\bm{\phi}} $ denote the transition probabilities of the semantic encoder, the wireless channel, and the semantic decoder, respectively.
The Markov chain of semantic communication thus reduces to $\bm{Y} \leftrightarrow \bm{X} \leftrightarrow \hat{\bm{Z}}$.
Since the goal of semantic communication is to maximize expected faithfulness in representing observed messages (that is to say to minimize the semantic errors) and minimize the amount of data to be transmitted \cite{beck2022semantic}.
Then the general form of optimization problem of semantic communication can be expressed as
\begin{subequations}
\begin{align}\label{eq:semantic}
    \underset{P_{\hat{\bm{Z}}| \bm{X}}}{\operatorname{minimize}} ~&\big[f(\bm{Y}, \bm{X},\hat{\bm{Z} }), g(\bm{X},\hat{\bm{Z}} )\big] \\
\text { subject to } & (\bm{X},\bm{Y}) \in \mathcal{K},
\end{align}
\end{subequations}
where $P_{\hat{\bm{Z}}| \bm{X}}$ denotes a statistical mapping of source information to received semantic information, function $f(\cdot, \cdot,\cdot )$ measures the semantic error, function $g(\cdot,\cdot )$ characterizes the number of symbols to be transmitted in semantic communication, and $\mathcal{K}$ denotes the background knowledge.

The key of semantic communication is to define the semantic-aware optimization metrics $g$ to quantify the semantic information and $f$ to measure the semantic error in \eqref{eq:semantic} and jointly design the semantic encoder and decoder to obtain the optimal mapping $P_{\hat{\bm{Z}}| \bm{X}}$ in a task-oriented sense. 
Inspired by the powerful representation ability of DNN and its successful employment in natural language processing (NLP), DL-based end-to-end semantic system has gain much traction, in which the semantic encoder and decoder are implemented by DNNs to represent and interpret the semantic meaning, and are jointly trained in an end-to-end manner to achieve the global optimality. 
In the following, we will detail an information-theoretic framework for semantic communication system design by theoretically characterizing the trade-off between compression of semantic feature extraction and distortion of semantic meaning transfer. 

\emph{An Information Bottleneck Optimal Semantic System}: 
The IB framework was proposed in \cite{chechik2003information,goldfeld2020information} as a principle approach to characterize the trade-off between information compression and target signal reconstruction. 
Therefore, IB can be used to provide theoretical guidance for the semantic system design. 

From the perspective of reliable transmission, as long as the encoding information entropy remains unchanged, that is, when $I(\hat{\bm{Z}};\bm{Y}) = I(\bm{X};\bm{Y})$, the semantic communication can recover the target information $\bm{Y}$ completely and losslessly through the semantic decoder theoretically, where the mutual information $ I(\bm{X};\bm{Y}) = \sum_{\bm{y} \in \bm{Y}} \sum_{\bm{x} \in \bm{X}} p(\bm{x}, \bm{y}) \log (\frac{p(\bm{x}, \bm{y})}{p(\bm{x}) p(\bm{y})})$ obtained from the joint probability distribution $p(\bm{x}, \bm{y})$ and the marginal probability distribution $p(\bm{x}),  p(\bm{y}) $ is a measure of the mutual dependence between two random variables.
However, the loss of information is inevitable in the practice due to the signal compression and system noise, therefore it is natural to maximize $I(\hat{\bm{Z}};\bm{Y}) $ while restricting the information flow from source signal $\bm{X}$ to compressed feature $\hat{\bm{Z}}$ in semantic communication.
As a result, the IB-based optimization problem can be formulated as:
\begin{subequations}\label{eq:IB}
\begin{align}
	\underset{P_{\hat{\bm{Z}}| \bm{X}}}{\operatorname{maximize}} ~&I(\hat{\bm{Z}};\bm{Y}) \\
	\text { subject to } &I(\hat{\bm{Z}};\bm{X}) \leq \alpha.
\end{align}
\end{subequations}
Specifically, given samples of $P_{\hat{\bm{Z}}| \bm{X}} $, the objective function of the IB optimization problem is to maximize the mutual information between received semantic information and target signal to minimize the information loss of semantic interpretation for reliable transmission, while the constraint keeps the mutual information between the source signal and the semantic information within a certain range to guarantee a target compression ratio of semantic extraction for improved system efficiency.   
By solving \eqref{eq:IB}, we can learn the parameterized optimal semantic encoder $C_{\bm{\theta}}(\cdot)$. 

To solve the above IB optimization problem, a Lagrangian operator $\beta \geq 0$ can be introduced to maximize its Lagrangian dual equation $\mathcal{L}_{\textit{IB}}(\bm{\theta}) = I(\hat{\bm{Z}};\bm{Y})  - \beta I(\hat{\bm{Z}};\bm{X})$.
The Lagrangian operator $\beta$ controls the trade-off between the compression ratio of the received semantic information $\hat{\bm{Z}}$ with respect to the source information $\bm{X}$ and the amount of semantic information transferred from the transmitter to the receiver.
When $\beta = 0$, the objective $\mathcal{L}_{\textit{IB}}$ aims at minimal distortion (maximal semantic information transfer), whereas for $\beta \rightarrow \infty$, data rate is minimized (compression ratio is maximized).
To overcome the intractability of mutual information in IB optimization, the authors of \cite{DBLP:conf/iclr/AlemiFD017} constructed a lower bound for $\mathcal{L}_{\textit{IB}}(\bm{\theta})$ using some variational distribution $q_{\bm{\psi}}(\hat{\bm{z}})$, which can be easily optimized.
Specifically, exploiting elementary properties of mutual information, entropy and Kullback–Leibler divergence (KLD), the $\mathcal{L}_{\textit{IB}}(\bm{\theta})$ is lower bounded by $\E_{p_{\theta}(\bm{y}, \hat{\bm{z}})}[\log p_{\bm{\phi}}(\bm{y}|\hat{\bm{z}}) ] - \beta \E_{p(\bm{x})}[ D_{\textit{KL}}( p_{\bm{\theta}}(\hat{\bm{z}}|\bm{x}) ||q_{\bm{\psi}}(\hat{\bm{z}}) )] $.
Using the re-parametrization trick with conditions that hold
for variational auto-encoder (VAE) \cite{DBLP:journals/corr/KingmaW13}, this lower bound enables the optimization of the parameters of semantic encoder $\bm{\theta}$, decoder $\bm{\phi}$ and the variational parameters $\bm{\psi}$ via gradient-based methods.
By parameterizing the semantic encoder and decoder as NNs, the IB optimization problem can be effectively solved by ML techniques. 

\subsection{Applications of Semantic Communications}
In Table \ref{ta:semantic}, we summarize the different DL-enabled semantic systems based on the different transmission tasks (text \cite{xie2020lite,xie2021deep,zhang2022context}, image \cite{bourtsoulatze2019deep,kurka2020deepjscc}, speech signals \cite{weng2021semantic,han2022semantic} and general signals \cite{xie2021task,zhang2022deep}), from the perspective of performance metrics, semantic quantity module, semantic error module, loss function, KBs as well as the adaptation to dynamic environment for task-specific applications.
\begin{table*}[htp]
	\caption{DL-enabled semantic systems for different transmission signals}
	\resizebox{\linewidth}{!}{
	\begin{tabular}{cccccccc}
		\toprule
		\multicolumn{1}{l}{\tabincell{c}{ Transmission \\ Signals  }  } & Refs &\tabincell{c}{ Performance Metrics  }  & \tabincell{c}{Semantic Quantity \\ Module (SQM)} &\tabincell{c}{Semantic Error\\ Module (SEM)}   & Loss Function & KBs & \tabincell{c}{Methods for Dynamic \\ Environment} \\ \toprule
		\multirow{3}{*}{Text}                     &  \cite{xie2021deep}    &         \tabincell{c}{ Sentence similarity }             &        Transformer    &     Transformer                        &    Cross-entropy - $\lambda$  LBMI         &  \tabincell{c}{Datasets of words}      &      Transfer learning       \\ \cline{2-8} 
		&   \cite{xie2020lite}   &     BLEU score                &       MLPs                 &                MLPs           &        Cross-entropy       &    \tabincell{c}{Datasets of words} &   $\backslash$           \\ \cline{2-8} 
		&    \cite{zhang2022context}  &    \tabincell{c}{Edge-based similarity, \\ word2vec, hybrid-based similarity, \\ METEOR and BLEU score}                   &       \tabincell{c}{Part-of-speech \\  strategy}               &   \tabincell{c}{Context-based  \\  strategy}                         &      Log-softmax         &  \tabincell{c}{Datasets of \\ words and speech}   &   \tabincell{c}{Context-based \\ dynamic programming \\ algorithm}             \\ \hline
		\multirow{1}{*}{Image}                     &    \cite{bourtsoulatze2019deep,kurka2020deepjscc}  &         PSNR metric            &              Encoder CNN          &         Decoder CNN   &     MSE          &  \tabincell{c}{Datasets of images}   &     $\backslash$            \\ \hline
		\multirow{2}{*}{Speech}                     &   \cite{weng2021semantic} &SDR and PESQ score  &         SE-ResNet module           &   \tabincell{c}{ Multiple SE-ResNet \\ modules}                  &        MSE                   &     \tabincell{c}{Datasets of speeches}         &     $\backslash$            \\ \cline{2-8} 
		&  \cite{han2022semantic}    &      \tabincell{c}{WER and semantic \\ similarity score    }            &  \tabincell{c}{Attention-based \\ NNs    }  &           MLPs                                &      Cross-entropy         &   \tabincell{c}{Datasets of \\ words and speeches} &     $\backslash$          \\ \hline
		\multirow{2}{*}{General}                     &  \cite{xie2021task}    &     \tabincell{c}{Recall@1 for image retrieval, \\ BLEU score for machine translation, \\ answer accuracy for VQA}                 &    Transformer                   &                Transformer           &    Cross-entropy and MSE         &    Different dataset &   $\backslash$            \\ \cline{2-8} 
		&    \cite{zhang2022deep}  &       PSNR and text accuracy              &     $\mathcal{D}_{ob}$                   &             $\mathcal{D}_{pr}$              &         ESD      &  Library data at receiver  &     Data adaptation         \\ \bottomrule
	\end{tabular}}
    \label{ta:semantic}
\end{table*}

\subsubsection{Text Signals}
A DL-based semantic communication system, namely DeepSC, was proposed in \cite{xie2021deep} 
for text transmission, which was the first work clarifying the concept of semantic information and semantic error at the sentence level. 
A new metric, namely sentence similarity, was proposed to reflect the learning performance of semantic system for text transmission. 
To jointly train the deep encoder and decoder, cross-entropy and lower bound of mutual information (LBMI) constitute the loss function, where the first term measures the semantic errors between transmitted message and recovered message and the second term measures the system capacity. 
By minimizing the loss function, the DeepSC can be trained to maximize the system capacity while minimizing the semantic errors. 
A transformer-based DNN in DeepSC further makes it applicable to varying communication conditions. 
Considering the capacity-limited devices in a more practical scenario, authors in \cite{xie2020lite} proposed a distributed semantic communication system for IoT networks, called L-DeepSC, where bilingual evaluation understudy (BLEU) score and cross-entropy are adopted as performance metric and loss function, respectively.
To reduce the model sharing cost on IoT devices, the semantic models are compressed through network sparsification and quantization. 
A refined CSI estimation scheme based on deep denoising network was proposed to eliminate the impact of fading channels on the semantic model training. 

\subsubsection{Image Signals}
When aiming at wireless image transmission, a joint source-channel coding (JSCC) was proposed by \cite{bourtsoulatze2019deep}, which can be regarded as an early semantic communication system.
In JSCC, two CNNs were considered as autoencoder for image feature extraction at transmitter and autodecoder for image reconstruction at receiver, respectively, where a non-trainable layer in the middle represents the noisy communication channel.
Peak signal-to-noise ratio (PSNR) metric was used to measure the learning performance of semantic system for image transmission and the MSE loss function was used to minimize the average distortion between the original input images and its reconstruction. 
To exploit the feedback channel information, Kurka \textit{et al.} \cite{kurka2020deepjscc} further extended deep JSCC to DeepJSCC-f by deploying layered autoencoders, where the transmission of each image signal is divided into multiple layers.

\subsubsection{Speech Signals}
Other series of works focused on learning semantic information directly from the raw speech signals.
Weng \textit{et al.} \cite{weng2021semantic} firstly proposed DL-enabled semantic communication system called DeepSC-S for speech signals, where attention-based SE-ResNets semantic encoder and decoder were proposed to identify the essential speech information and recover signals.
The MSE was used as loss function for training DeepSC-S, and the signal-to-distortion ration (SDR) and the perceptual evaluation of speech distortion (PESQ) score were adopted to measure the performance.
Besides, a novel semantic-oriented speech to text transmission system was considered by \cite{han2022semantic}, where an attention-based network is utilized to identify the semantic representation of the input speech signals and a semantic decoder implemented using MLPs is employed to transform the received speech features to the text form for speech recognition.

\subsubsection{General Signals}
In addition to the semantic communication systems for specific signals, 
general signal transmission was considered in \cite{xie2021task, zhang2022deep}.
A transformer-based semantic communication system was proposed in \cite{xie2021task} for transmitting multimodal data considering three different tasks, 
including image retrieval, machine translation and visual question answering.
However, the training algorithms, such as the loss functions, for different tasks were designed separately.
To address this, Zhang \textit{et al.} \cite{zhang2022deep} firstly designed a semantic-distortion-based universal loss function adapted to general signals, which consists a hyper-parameter-based linear combination of distortion measure functions for observable information $\mathcal{D}_{ob} $ and for pragmatic output $\mathcal{D}_{pr}$. These functions can be designed to be the KLD, cross entropy, MSE, etc., to adapt to the different transmission tasks.

\subsubsection{{Adaptation to Dynamic Environment}}
When the communication environment is dynamic or the transmission task is changed, it will pose great challenges for the design of DL-enabled semantic communication systems. 
First, it requires different KBs to extract and interpret the transmitted messages as the varying of communication environment or transmission tasks. 
For example, if the transmitted message is unseen at the transmitter and receiver, the KBs need to be updated and expanded based on the empirical semantic information, leading to formidable computational costs for the training of semantic encoder and decoder.
The KBs matching the dynamic transmission conditions can be learned from the perceived environments/empirical information and can be shared between transmitter and receiver via communication to minimize the semantic inference errors, which however are complex and long-term processes. 
Existing semantic communication designs \cite{bourtsoulatze2019deep,kurka2020deepjscc,xie2020lite} assumed the shared and fixed KBs, leading to limited scalability and poor generalizability in dynamic environments.
Second, for the newly updated KBs and the dynamic changing environment, the semantic coding strategies should quickly adapt to the new environment and KBs with minimal training cost. 
Transfer learning \cite{xie2021deep,weng2021semantic} was adopted to accelerate the DL model training in dynamic environment by synergizing the past learning experiences to assist the new problem solving. 
However, the re-training of NNs still requires extra communication and computational cost.  
A receiver-leading dynamic semantic communication system with non-shareable KBs at receiver was proposed in \cite{zhang2022deep}, where an individual data adaptation network was configured at the semantic transmitter to tackle the dynamic data transmission environment leaving the semantic encoder adaptive to dynamic environment without retraining.  

\subsection{Advantages and Disadvantages}
Semantic communications are expected to improve the communication efficiency and reliability, enhance the quality of experience for human-oriented services as well as support a more robust and upgrade/evolution-friendly communication systems \cite{guangming21semcom}.
However, both theoretical and practical implementation of semantic communication are in an early stage, which will spark an explosion of research interests in both academia and industry.

\section{Federated Learning for Distributed Optimization}\label{sec:7}
When optimizing a large-scale model with training data scattered across massive number of edge devices, distributed ML has emerged as a key enabling technology to reduce the communication cost and preserve data privacy in resource limited wireless networks. 
FL \cite{Tian_SPM20,yang2020over} has been proposed as a prominent distributed ML scheme to effectively solve the model optimization problem in ML over large-scale wireless networks, where each edge device participates in the training process by exchanging model parameters with data kept locally. 
In this section, we firstly review the FL framework, followed by its applications in wireless communication systems based on different network structures and the summary of its pros and cons.

\subsection{Federated Learning Framework}
Considering more practical scenarios in large-scale wireless networks, where most of training data accounting for a global model learning are generated locally at end devices, in the aforementioned centralized learning framework, edge devices are required to send these data to a central server (e.g. a BS), triggering high communication costs and data privacy issues.
To mitigate these problems, FL, a distributed framework to train a global statistical model, was proposed \cite{kairouz2021advances}.
Under the coordination of a dedicated central server, FL allows multiple devices to participate in the global model training through local model update and model parameters exchange without sharing their raw data \cite{WahabMOT21}, thereby preserving the data privacy and saving the communication cost. 
The unique challenges of FL compared with centralized learning frameworks include system heterogeneity with various end-device features (e.g., transmission environment, communication resources, computational powers, etc.), data heterogeneity with non-identical local data distributions and unbalanced local datasets, and dynamic wireless environment with uncertain wireless channels and access links \cite{khan21fl4iot,yuanming22edgeai,yuhan22irsfl,Wang2022InterferenceMF}, which have stimulated a growing research interest in FL.

\begin{figure}[t]
        \centering
        \begin{minipage}{.46\textwidth}
                \centering
                \includegraphics[width=1.0\columnwidth]{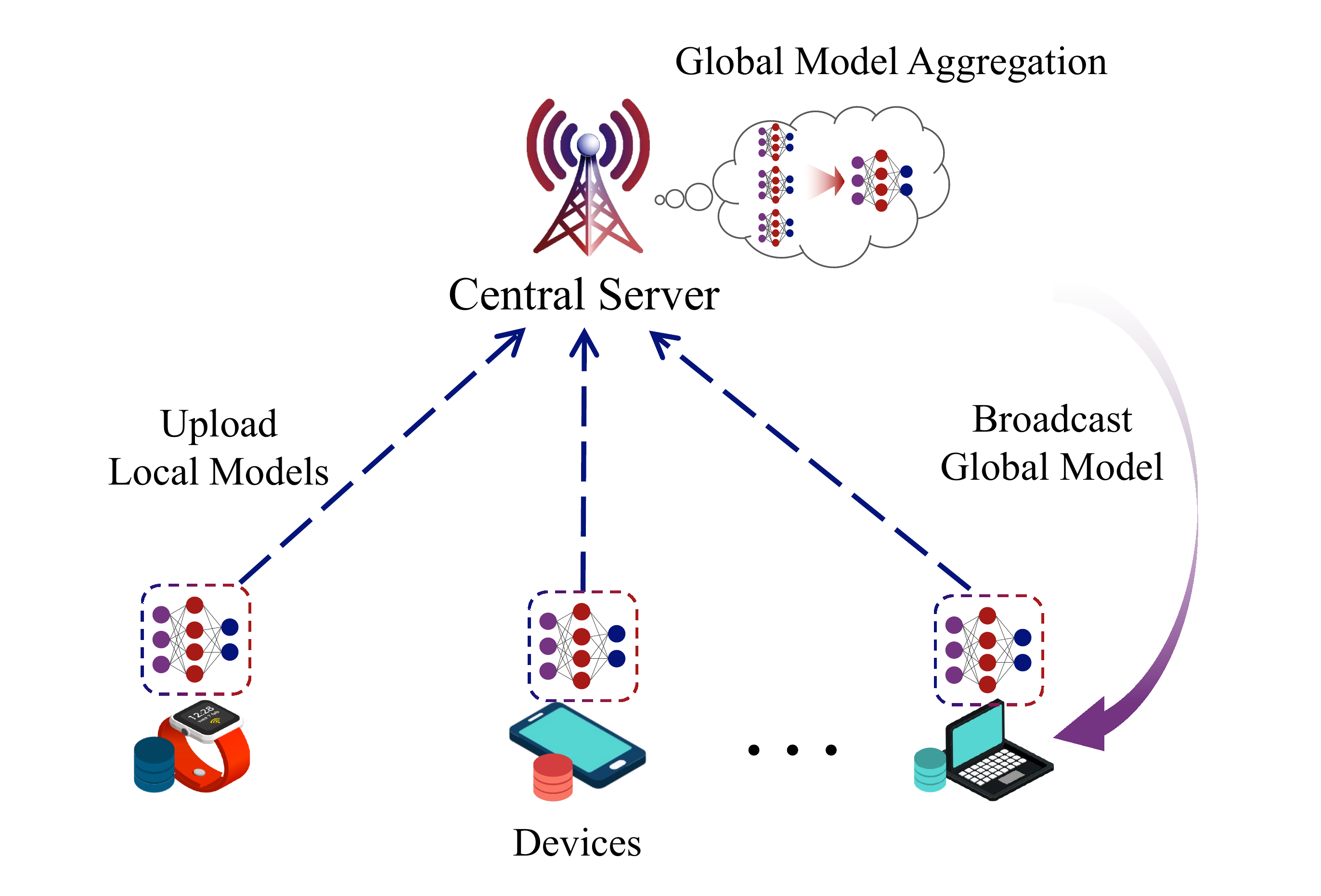}
                \caption{Illustration of cross-device federated learning.}\label{fig:fl1}
        \end{minipage}
        \hspace{4mm}
        \begin{minipage}{.46\textwidth}
                \centering
                \includegraphics[width=1.0\columnwidth]{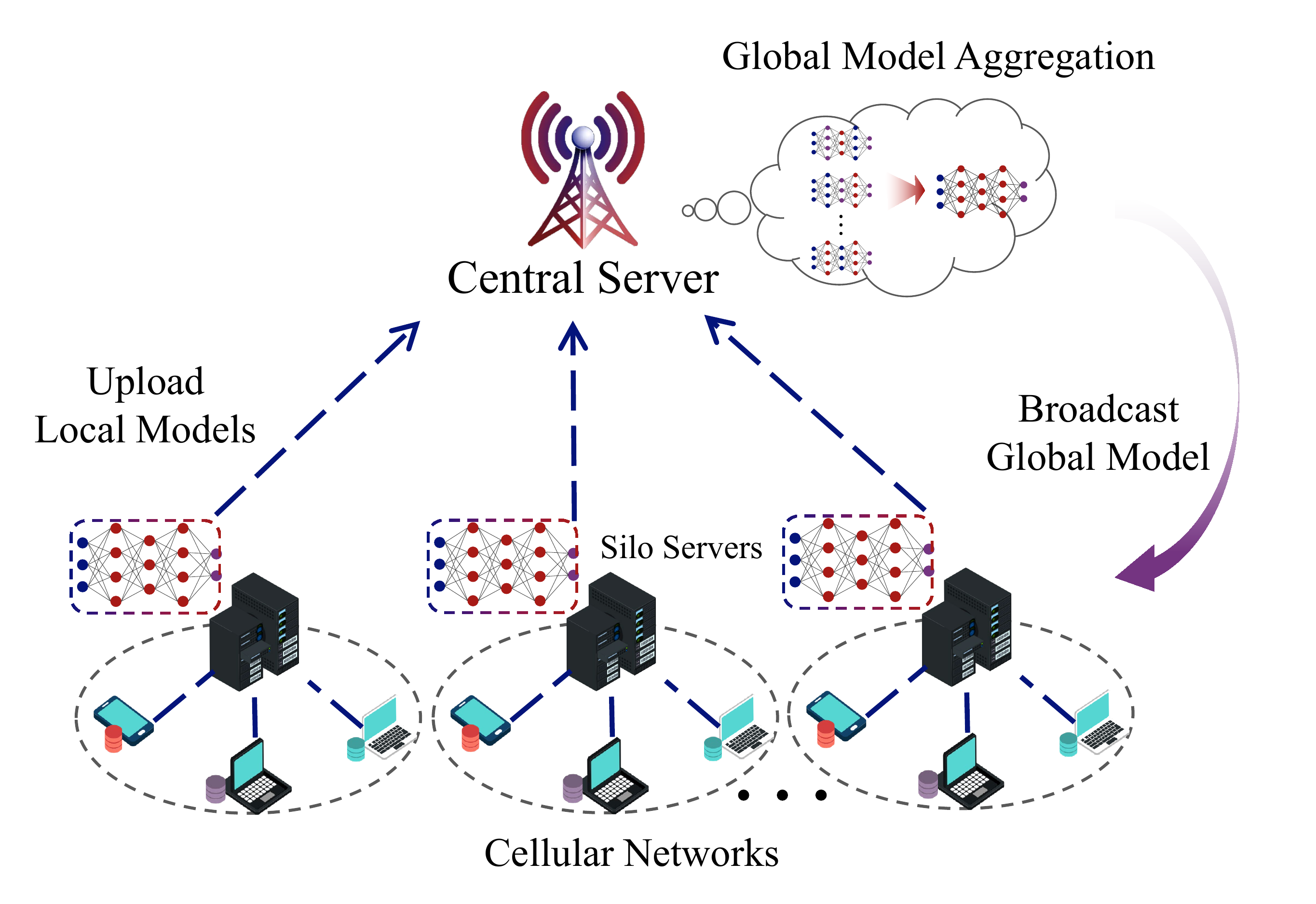}
                \caption{Illustration of cross-silo federated learning. }\label{fig:fl2}
        \end{minipage}
\end{figure}

The goal of FL is to minimize a global loss or empirical risk function $\mathcal{L}_{\textit{FL}}$, i.e., $\text{min}_{\bm{\theta}} \sum_{i \in \mathcal{S}} w_i \mathcal{L}_i(\bm{\theta}; \mathcal{D}_i)$, where $\bm{\theta}$ represents the model weights, $\mathcal{L}_i$ denotes the local loss function of device $i$ over local dataset $\mathcal{D}_i$, $\mathcal{S}$ is the set of participating end-devices and $w_i$ denotes the weight for each local loss function with $w_i \geq 0$ and $\sum w_i = 1$.
In a typical cross-device FL training process, as illustrated in Figure \ref{fig:fl1}, a central server orchestrates and repeats the following steps (referred as federated averaging algorithm) until the convergence of global model. 
\subsubsection{Device Selection}
The central server selects a subset of devices meeting certain eligibility requirements to participate in each training round.
Typical device selection schemes in wireless networks include random scheduling, round robin, proportional fairness as well as incentive mechanism based on auction game \cite{yang2020selection, thi21selection}.

\subsubsection{Global Model Broadcast}
The central server broadcasts the current model set to the selected devices that participate in the training process.

\subsubsection{Local Model Training}
Based on the received global model, each selected device takes a batch of samples from its local dataset and utilizes a local model update algorithm (e.g., stochastic gradient descent algorithm) to obtain the updated local model.

\subsubsection{Model Aggregation}
All the local model updates are aggregated at the central server by
computing the model aggregation function (e.g. weighted average function) to obtain the updated global model.

To implement federated learning in wireless networks with limited resources and unreliable communication links, the efficiency of information transmission becomes the core issue.
In wireless FL, the local model shall be transmitted from end devices to central server, followed by the calculation of the aggregation function at the central server.
Given the structure of model aggregation function, the wireless transmission of local model can be divided into orthogonal transmission and non-orthogonal transmission \cite{Mingzhe_TWC21, li2021delay, yang2019federated, Kaibin_TWC20}.
The practical orthogonal frequency division multiple access (OFDMA) and FDMA techniques are adopted to support interference-free uplink local model transmission and downlink global model transmission in \cite{Mingzhe_TWC21, li2021delay} through orthogonalized frequency allocation, which could increase the communication latency instead.
In order to improve the transmission efficiency, non-orthogonal transmission schemes leveraging the principle of over-the-air computation (AirComp) have been studied in \cite{yang2019federated, Kaibin_TWC20,Wang2022FederatedLV, wang2022over}. 
Unlike the orthogonal transmission which requires the decoding of each local model separately, AirComp allows the central server to receive a desired aggregation function of local models via their concurrent transmissions on same resource block \cite{yang2019federated, Kaibin_TWC20} by exploiting the waveform superposition nature of wireless multiple access channels, therefore the communication latency and the consumed communication resources will not increase with the number of devices, facilitating its usage for large-scale networks. 

Training data is essential for ML model learning.
The centralized ML algorithms assume the training data is independent and identical distribution (i.i.d.) distributed.
In wireless FL, as the training datasets are generated distributed by end devices, the heterogeneous nature of large-scale wireless networks leads to non-independent and non-identical distribution (non-i.i.d.) datasets at different devices.
The dynamic network environments, such as the mobility of devices and the randomness of link connections, further make the sensory data not only heterogeneous but also non-stationary.  
As a result, the non-i.i.d. and non-stationary feature of on-device data becomes one of the key bottlenecks to accomplish efficient and accurate distributed ML tasks in hyper-scale wireless networks.
In the following subsections, we overview the existing wireless FL frameworks to resolve the issue of network/data heterogeneity and instability. 

\subsection{Application 1: Cross-Device Federated Learning}

Cross-device FL involves enormous number of IoT devices or agents, where the data is generated locally and remains typically decentralized \cite{kairouz2021advances}.
In order to deploy DNN models on a larger-scale wireless network, avoid huge communication overhead for training data collection in centralized learning as well as protect data privacy, there are already some studies that leverage FL for wireless communication applications including hybrid beamforming \cite{elbir20FL4bf} and channel estimation \cite{elbir22FL4CE,xuanyu22fldnn}.
Specifically, in these wireless applications, the training data across devices are typically  non-i.i.d. due to the heterogeneity of the transmission environment where devices are located. The non-i.i.d. data can slow down the convergence and deteriorate the learning accuracy of FL.
To improve the overall performance of FL in dealing with various wireless communication problems with non-i.i.d. training data, the deeper and wider CNN models were used in FL-based systems \cite{elbir20FL4bf,elbir22FL4CE} to provide better feature extraction and representation capability. 
In addition to the data heterogeneity, system parameters, such as SNR, antenna numbers, and channel statistics of participated devices are also dynamically changing. As shown in \cite{elbir20FL4bf,elbir22FL4CE}, the decentralized FL-based wireless communication systems are more robust to the channel imperfections and corruptions compared with the centralized learning.
To cope with the varying system conditions, a federated dynamic detection network was proposed in \cite{yang2021federated} to perform the dynamic MIMO detection, where two independent detection networks were built leveraging the algorithm unrolling approach, and a specifically designed route network was built to adaptively select a better detector for every sample under different conditions. However, the multi-network design can induce significant training cost and has the limited adaptability to the changing environment. 


The dynamic and uncertain wireless environment poses great challenges for efficient wireless resource allocation in FL. The impact of resource allocation policies on the learning performance of wireless FL, e.g., the test accuracy and training efficiency, is generally implicit and non-analytic. 
For instance, in the training of the CNN-based wireless FL system, it is hard to obtain an analytical expression of the FL testing accuracy with respect to the resource allocation parameters (e.g., device selection, power allocation, computational resource allocation, etc.). 
As a result, the conventional convex optimization based resource allocation algorithms are infeasible for such scenarios.
Fortunately, the dynamic resource allocation problem in the FL system can be formulated as a stochastic optimization problem by modeling the total training process of FL as an MDP with the resource status at each training round being the state, the resource allocation policies being the action, and FL learning performances (e.g., training latency, learning accuracy, energy consumption, etc.) being the reward.
Therefore, the model-free DRL can be applied to solve the dynamic resource allocation problems in FL while regarding the FL performance changes as a black box \cite{9372789, 9770401, 9205252, 9772619}, given the diverse and dynamic wireless conditions for FL participants.
Specifically, in each training round, the resource allocation policies shall be optimized based on the real-time resource states (i.e., CSI, calculation resources and available bandwidth), leading to the change of the FL performance of DNN model, which is considered as the immediate reward of the current policy. 
Then the state evolves based on the action performed. 
Through trial-and-error method, the long-term optimal resource allocation policy can be learned to maximize the total amount of reward received over time. 

We summarize the representative works of DRL-assisted FL in Table \ref{ta:FL_DRL}, detailing the DRL algorithms, state space, action space and reward function, respectively. 
\begin{table*}[t]
	\caption{DRL-Enabled Federated Learning Frameworks}
	\resizebox{\linewidth}{!}{
		\begin{tabular}{cccccc}
			\toprule
			Algorithm Type &DRL Algorithms & Refs&State Space&Action Space &Reward Function\\ \toprule
			&\multirow{3}{*}{DQN}&\cite{9205252}&\tabincell{c}{Task queue state and the available resource state}&\tabincell{c}{Application partitioning, subcarrier allocation strategy,\\ and service caching placement}&\tabincell{c}{Negative summation of normalized execution time}\\   
			\cline{3-6}
			\multirow{6}{*}{Value-Based}&&\cite{9123956}&\tabincell{c}{Selected action in previous time slot, \\local information, interferers' information \\and interfered neighbors’ information }&\tabincell{c}{Transmit power, beamformer }&\tabincell{c}{Achievable rate minus the sum of\\ the achievable rate losses of the interfered links}\\
			\cline{3-6}
			&&\cite{9629267}&\tabincell{c}{Number of available channels, number of power level, \\length of the binary representation of the feedback signal, \\and indicator of no transmission}&\tabincell{c}{Channel selection policy,\\power allocation policy}&\tabincell{c}{Network utility}\\
			\cline{2-6}
			&\multirow{3}{*}{DDQN}&\cite{9770401}&\tabincell{c}{Channel conditions, resource allocation actions}& Channel selection,  transmit power & Sum-rate of all cells minus QoS based penalty\\
			\cline{3-6}
			&&\cite{9155494}&\tabincell{c}{Compressed model weights}&\tabincell{c}{Device scheduling}&\tabincell{c}{Exponential function of \\achieved test accuracy}\\
			\cline{3-6}
			&&\cite{9478892}&\tabincell{c}{Available CPU, energy unit, and wireless bandwidth}&\tabincell{c}{Device scheduling}&\tabincell{c}{Resource utilization of MEC system.}\\
			\hline
			\multirow{2}{*}{Policy-Based}&A2C&\cite{9772619}&\tabincell{c}{Statistics of data flow }&\tabincell{c}{Available assignment options}&\tabincell{c}{Mixed function of delay and PLR}\\
			\cline{2-6}
			&DDPG&\cite{9448736}&\tabincell{c}{The selected beamformer indexes,\\ the achievable rate and signal power at all users,\\ interferer links, and interfered links, respectively}&\tabincell{c}{Codeword in the beamformer codebook}&\tabincell{c}{Achievable rate minus the sum of \\the achievable rate losses of the interfered links}\\
			\bottomrule
	\end{tabular}}
	\label{ta:FL_DRL}
\end{table*}
For example, the authors in \cite{9155494} proposed an experience-based scheduling framework for client selection in each communication round to cope with the non-i.i.d. data distribution, where DQN was adopted to learn the optimal client selection policy aiming for higher test accuracy and fewer communication rounds under dynamic wireless conditions.  
DQN-based algorithm was also adopted in \cite{9500603} to optimize the user access in open radio access network for long-term throughput maximization and efficient FL.  
Further, the authors in \cite{9478892} developed a double DQN-based algorithm to optimize the amount of data, energy and computational resources allocated to each end device for FL training by minimizing the energy consumption and system delay. 
The joint optimization of radio resource allocation and device scheduling were also investigated in \cite{9348225}, where an actor-critic based DRL approach was proposed to optimize the transmit power at the BS and the computing frequencies at the local devices when participating in the FL training, 
such that the FL learning performance and the fairness of users can be maximized while reducing the energy and time consumption.
In \cite{9205252}, a value-based DRL method across computing, communication and caching was proposed for FL optimization in a 5G ultra-dense edge computing networks, where DQN was adopted to solve the complex optimization objectives integrating the QoS metric and communication delays. 
To reduce the back-haul traffic congestion in the large-scale IoT network, the authors in \cite{9772619} adopted a policy-based algorithm to allocate the back-haul data flow by maximizing the network utility.
In \cite{9772081}, the authors exploited the DRL technique to optimally allocate the available energy and data units at each device and design the block generation rate at miner in a blockchain-assisted FL system. 
To summarize, by transforming the implicit optimization target related to the FL learning accuracy and efficiency into a numerical reward, such DRL-based algorithms show better fitness in wireless FL for dynamic resource allocation compared with the COAs under dynamic environment.

\subsection{Application 2: Cross-Silo Federated Learning}
In contrast with cross-device FL, where a large number of IoT devices participate in FL to complete same learning task, the clients in cross-silo setting are silo servers, including organizations (e.g. medical or financial), geo-distributed data-centers, etc., each of which has an identity or name that allows the system to access it specifically \cite{kairouz2021advances}.
In cross-silo setting, a number of organizations can share incentive (e.g. the payoff-sharing scheme in financial FL system \cite{yu20fl}) rather than data directly to train an ML model due to confidentiality or law issues.
In cross-device FL, the training data are usually partitioned by examples, while in cross-silo FL, in addition to be partitioned by examples, the training data can be partitioned by features, e.g., different organizations keep the data of different features corresponding to same batch of customers \cite{kairouz2021advances}. 
The feature partitioned FL requires multi-clients to train the model collaboratively by exchanging intermediate parameters rather than model parameters to deal with the missing features. 
Lots of works have been done to address the security and privacy challenges \cite{liu2020backdoor} and communication bottlenecks \cite{liu2019communication} in feature-partitioned FL. 
Another source of heterogeneity in cross-silo FL is the non-i.i.d. and non-stationary siloed data. 
Typically in large-scale wireless networks, the clients in cross-silo setting can be clusters of cellular or D2D networks that contain non-i.i.d. and non-stationary heterogeneous data.

To overcome the model divergence and staleness in the training stage and poor accuracy in the inference stage induced by non-i.i.d. and non-stationary training data of different clients in cross-silo FL, an adaptive federated multi-task learning (FMTL) framework was proposed by \cite{hongwei_ICCfl}, which preserves multiple models at the server and the clients through adaptive model updating and cluster splitting to deal with non-stationary environments and multi-task learning.
Specifically, in model update scheme, model dichotomy \cite{savaresi2002cluster} was adopted to find the geometric centers of local models in two virtual sub-clusters, whose model update directions were compounded to determine the update direction of global model. 
The recursive cluster splitting, through decreasing the
sub-clusters' distances of updating directions, was also proposed to avoid the poor performance induced by the mutation of data distributions. 
Furthermore, a binary tree-based low-complexity model selection scheme was proposed to choose the best model for fitting the current data in both training and testing stages.
Such adaptive FMTL has been shown to accelerate the model training convergence and reduce the computation complexity while ensuring model accuracy when it is applied to solve the GNN-based power control problem in cross-silo FL system consisting of D2D networks.

\subsection{Advantages and Disadvantages}
FL can not only embed the training capabilities of DNN for hyper-scale wireless networks, but also build a unified multi-source data application system with privacy preserving among multiple devices. Besides, FL can realize data sharing and integration across silo servers for supporting  high-precision model construction \cite{kairouz2021advances}.
However, FL still faces many challenges in terms of privacy protection, theoretical analysis and wireless deployment. For example, the privacy preserving techniques usually sacrifice the learning performance and induce additional computational cost, which are undesirable for efficient FL. The convergence analysis of FL considering the trade-off between learning performance and resource consumption is difficult due to the highly non-convexity and intractability of optimization problem.  Moreover, the deployment of FL in large-scale wireless networks should jointly consider the issues of dynamic fading channel, communication overhead, low power constraint of end devices, as well as the availability and the willingness of participants, which can be challenging for efficient algorithm design.


\section{Discussions and Future Research Directions}\label{sec:8}
In light of the appealing benefit of ML for large-scale optimization in 6G wireless networks, significant efforts are still needed to upgrade the existing ML techniques or develop new ones considering the constraints of practical wireless communication systems. 
To further pave the path for its more comprehensive applications in future wireless communication systems, in this section, we summarize the existing DNN design principles to accommodate to different kinds of optimization problems in wireless networks, after which, we summarize the existing theoretical tools to characterize the performance of MOAs. 
Subsequently, we discuss the software platforms and implementation issues for MOAs in 6G wireless networks and some potential research directions. 

\subsection{Neural Network Design for Wireless Communications}
The design of NNs (e.g. loss function design, network architecture design and training algorithm design) for solving complex optimization problems in various wireless communication applications requires careful consideration. 
In the following, we summarize the design principles of DNNs from different aspects when applied to solve large-scale optimization problems in wireless networks. 

\subsubsection{Loss Functions}
The design of loss function is generally dependent on the communication problem being solved and the availability of training labels. When training labels are available, the DNN can be effectively trained in a supervised manner by constructing the loss functions using the training labels. 
For example, regression-based losses (e.g., MSE or weighted MSE) are usually optimized for estimation problems, such as channel estimation \cite{solt19DL4ce} and MIMO detection \cite{hengtao2020MIMOdetection}, and for resource allocation problems, such as power allocation \cite{shen2019graph} and beamforming design \cite{Elbir2022TwentyFiveYO}, when the labels of targeting signals are available.  
Cross-entropy losses are adopted for classification problems, such as codebook-based precoder design \cite{elbir2019hybrid}, user scheduling \cite{lee2019learning}, and sub-channel selection \cite{ahmed2019deep}, when the class labels are available. 
In some specific applications, information theory-based losses can help to fulfill the goals of optimization task, which can be effectively calculated using the empirical data (including training labels), such as the mutual information-based loss in semantic communication \cite{beck2022semantic}, min-max generative adversarial net (GAN) loss in channel estimation \cite{dong2020channel}, information-bottleneck-based loss in edge inference system \cite{shao20edgeinfer}, etc.  
Alternatively, when training labels are unavailable, the average performance measurements are adopted to formulate the loss function for model training in an unsupervised manner \cite{mark19l2opt}. 
For example, in resource management problem, the average performance measurement $\E[\bm{f}(\bm{p}(\bm{h}),\bm{h})]$, e.g., (weighted) sum-rate \cite{Shen_JSAC21,Jiang2021Learning}, energy efficiency \cite{huang2020deep}, communication delay \cite{she2021tutorial}, secrecy capacity \cite{fritschek2019deep}, etc., are utilized to guide the model training.

\subsubsection{Network Architectures}
For network architecture design, MLP is usually adopted as a benchmark algorithm for
comparison \cite{hengtao2020MIMOdetection,Shen2022GraphNN}, while for the problems with special structures (e.g., data structure, algorithm structure and problem structure), specialized NN architectures are preferred to better serve the underlying purpose of the task. 
For data with graph structure (e.g., network data \cite{Shen_JSAC21}), GNN is more suitable for exploiting the inherent graph structure of the problem. 
For data with spatio-temporal correlations (e.g. traffic prediction \cite{wang2017end}), CNN or RNN is specialized in capturing the correlations within data. 
For data with low dimensional structure, the generative model can be utilized to capture the fundamental sparse structure within data (e.g., the underlying probability distribution of spatial channel can be learned by generative network \cite{balevi2020high}). 
For algorithms with iterative nature, NN can be designed to imitate the forward operation of iterative algorithms (e.g., deep unrolling NN inherits the structure of original iterative algorithm \cite{monga2021algorithm}). 
For problems with the distributed data, FL framework can be exploited to meet the distributed requirement \cite{yuanming22edgeai}. 
For problems with complex dynamic environment, DRL constitutes a viable technology to address the stochastic optimization problem through agents and environment interactions. 

To cope with the more stringent requirements in 6G system and support diverse applications for future wireless networks, hybrid DL-based optimizations have attracted increasing attention to fully integrate the intrinsic diverse features of different tasks into the design of customized NNs and make the most of the advantages of various DL techniques to improve the algorithm efficiency. 
The core of hybrid DL is to match different features of the problem with appropriate learning technology. 
For example, DRL-enabled FL system has been discussed in Section VII to solve dynamic resource allocation problems in FL.  
Besides, the algorithm unrolling approaches can be easily combined with not only conventional optimization \cite{mingmin21twcwmmse,an21tspdeepunfolding} but also some DL techniques such as GNN \cite{arinadm21wmmsegnn} to speed up the iterative algorithms. 
GNN integrated with DRL has been discussed in \cite{Munikoti2022ChallengesAO} for algorithmic and methodological improvements, where DRL and GNN complement each other for better utility or application-specific enhancements. 
In particular, the versatility of DRL and the flexible encoding capability of GNNs can be combined to address challenging optimization problems in different applications. 
Moreover, contrastive learning, a self-supervised learning approach, has been leveraged in GNN to address the challenge of data heterogeneity in graphs \cite{NEURIPS2020_3fe23034, Zhu:2020vf}, where the node features are learned in an unsupervised way.

\subsubsection{Training Algorithms}
When the network structure is fixed with reasonable loss function, Adam \cite{kingma2014adam}, the most popular back propagation algorithm, is usually applied for stable training. 
To further depict the recurrence relation of gradients between two adjacent layers, the generalized chain rule was proposed in \cite{Hu2020unrolling} to perform back propagation in unrolling-based DNN algorithms. 
To obtain better learning performance for algorithm unrolling methods, the layer-wise training approach is widely adopted \cite{shi2022twc4ma}.
End-to-end training of DNNs based on a large number of channel samples can bypass the explicit channel modeling procedure and potentially provide system-level performance gains compared to the COAs when solving the optimization problems in wireless communication systems \cite{Yu2022RoleOD}. 
Furthermore, to accommodate the NN to the dynamic changing environment in wireless communication systems, several techniques can be applied. 
For example, transfer learning has been adopted in \cite{yifei20lorm} to tackle the task mismatch issue (e.g., the network setting in training is different from that in testing) in LBB algorithm via self-imitation. 
Similarly, transfer learning in semantic communication in \cite{xie2021deep,weng2021semantic} enables the trained NN to adapt to the dynamic communication environment quickly with reduced number of training data. 
Meta learning is another technique to improve the generalizability of DNN in dynamic settings. 
For instance, model-agnostic meta-learning is used in \cite{Zeng2021DownlinkCF} to learn a good model initialization by alternating inner-task and across-task updates, such that the learned model parameters can adapt to a new environment with a small number of labeled data. 

Even though transfer learning and meta learning can significantly speed up the learning process of NN in new environment, they still require batch-sized training data to be available before the learning. 
In many scenarios of wireless networks, the training data arrives sequentially in a stream whose inherent features can be drifted due to the dynamic environment.  
In this case, a promising solution is to learn the models on the fly, which can improve the generalizability and scalability of model sufficiently and save the memory of system. 
Online deep learning \cite{anderson2008theory}, to learn the DNNs from the sequentially received data in an online manner, has been investigated in the various contexts of wireless networks. 
For example, in \cite{gao2022online}, an online DNN framework was proposed to solve the general optimization problems in wireless communication, where the self-defined layers rather than convolutional or full-connected layers were adopted to estimate optimization variables for each data sample. 
The CNN-based \cite{hong2015online} and GNN-based \cite{gao2022wide} online learning algorithms were further proposed, where the online module is retrained based on the observed testing samples to overcome the mismatches between training and testing data induced by the dynamic environment. 
Besides the network generalization issues, the complex and unpredictable environment may also lead to implicit system performance functions in resource allocation problem, which hinders the gradient calculation in the training process.  
To tackle this, a model-free approximation approach was proposed in \cite{kalogerias2020zeroth}, in which the gradients are approximated by their zeroth-ordered updates through sampling the model functions.   

\subsubsection{Optimization Constraints} 
Optimizations in wireless communications usually need to deal with various complex constraints to meet specific requirements, which brings additional difficulties to the design of ML algorithm. Normalization layer as the output layer of DNN provides a simple and effective way to deal with modulus constraint (e.g. power constraint \cite{Jiang2021Learning}). 
The proper activation functions can help to restrain the network output within a feasible region. 
For example, sigmoid can keep the output between $0$ and $1$. 
For discrete constraints, e.g., quantization, various smooth functions can be constructed to approximate the discontinuous function, or we can directly set the gradient to be $1$ at the back propagation to avoid the gradient vanishing and explosion \cite{zheng2021going}. 
For other more complex constraints, primal-dual learning \cite{markTSPL2o,eisen2020large} plays an important role, where DNNs are designed to solve the corresponding unconstrained Lagrangian dual problem. 
The authors in \cite{naderializadeh2022learning} showed that the duality gap of radio resource management optimization problem in wireless networks is negligible if the parameterized learning through NN is near universal. 

\subsection{Theoretical Tools}
ML technology, represented by DNN, has made great achievements in the fields of computer vision, natural language processing and communications in recent years with the great improvement of computing power and the great enrichment of data. 
However, due to the ``black box" nature of DNN, the lack of interpretability and theoretical guarantee of the DL-based framework is a critical issue that needs to be addressed for applications requiring highly transparent and reliable technologies (e.g. wireless communications, healthcare and automatic driving). 
DNN is a model function characterizing complex relationships among data, and its ``black box" nature is mainly manifested in the fact that there is a huge unknown gap between the design of the model and its final performance on specific tasks. 
In other words, it is impossible to accurately predict and control DNN performance when designing the model, to clearly understand the reasons for its good or bad performance, and to systematically improve its performance, but only to rely on some lucky model design and training tricks.
From ML perspective, this is due to the fact that the ML task (including training data) and DNN theory aspects (e.g., loss quantities and the generalization performance of the model) have not been thoroughly understood.
Especially when dealing with high-dimensional real data (e.g. images with the millions of dimensions), many of the existing statistical quantities and information-theoretic concepts such as entropy, mutual information, maximum likelihood and KLD suffer from the curse of dimensionality for computation, the ill-posedness for degenerate distributions, as well as the lack of guarantees for finite samples \cite{wright2022high}. 
To avoid these issues, the principled formulations are replaced with approximate bounds, simplified assumptions, heuristics and special tricks in practice, resulting in a serious performance gap between theory and practice.

In wireless communications, the specialized MOAs potentially enable rigorous analytical results within certain performance limits \cite{hengtao19DL4plc} due to their task-specific features and the model-inspired property.
In the following, we summarize the existing research results and progresses on the theoretical aspects of the MOAs. 
\subsubsection {Algorithm Unrolling} 
Inherited from the traditional iterative algorithm, the behavior of each layer of unrolled NN is interpretable.
The unrolled iterative hard threshold (IHT), used to solve $\ell_0$ norm constrained sparse recovery problem, has been theoretically analyzed in \cite{Blumensath2008IterativeHT}, which provided the optimality condition for the exact sparse recovery of the unrolled NN and proved the linear convergence rate of the unrolled IHT. 
The theoretical studies for the unrolled ISTA can be found in \cite{chen2018theoretical,ablin2019learning}, where the linear convergence rate of unrolled ISTA has been established and the structure of optimal network parameters for unrolled ISTA has been analyzed as well to guide the network structure design and the training parameters downsizing. 
The extension of unrolled ISTA to solve the group-sparse matrix estimation problem has been studied in \cite{shi2022twc4ma} and applied to JADCE problem in wireless networks, which established the linear convergence rate of unrolled ISTA with group sparsity structure. 
While some progresses have been achieved to establish the performance guarantees of algorithm unrolling, the underlying mechanism and the impact of learned parameters on the convergence and learning accuracy are still to be further discovered. 
\subsubsection {Learning to Branch-and-Bound}
The complexity of LBB for solving MINLPs is analyzed in \cite{yifei20lorm} following the common assumptions and analysis of imitation learning \cite{ross2011reduction}, which shows the expected number of nodes explored and the number of relaxed problems solved are $\mathcal{O}(L^2)$ and $\mathcal{O}(L)$ with $L$ integer variables under proper parameter settings. 
Therefore, the computational complexity of LBB is much lower than that of the traditional BB algorithm especially for large-scale network with large $L$. 
However, the theoretical analysis of learning performance is still missing.

\subsubsection {Graph Neural Network for Structured Optimization}
The graph optimization problem for large-scale wireless networks exhibits the property of permutation invariance or permutation equivariance depending on the output features. 
The classic GNN frameworks enjoy the same properties, i.e., permutation invariance or equivariance, which guarantees advantageous performance of GNN when applied to solve the graph-structured optimization problems \cite{yifei21gnn}.
Furthermore, Shen \textit{et al.} \cite{Shen2022GraphNN} firstly provided the generalization analysis of GNNs to theoretically verify the advantages of GNNs over MLPs in solving wireless communication problems in terms of the generalization error and the required number of training samples.
Based on the probably approximately correct (PAC) learning framework, they showed that the GNNs’
generalization error and required number of training samples are $\mathcal{O}(n)$ and $\mathcal{O}(n^2)$ lower than those of MLPs, where $n$ is the number of nodes in the graph.
Therefore, the GNNs can be theoretically proved to solve the graph optimization problem with near-optimal performance and much fewer training samples than generic NNs. 

\subsubsection {Deep Reinforcement Learning for Stochastic Optimization}
The existing theoretical framework of DRL is established based on the classic control theory and the theoretical results of conventional RL. Besides the optimization performance in terms of the expected accumulated reward, another main concern of RL community is the convergence performance in terms of the sample efficiency over collected experience data.
Xie \textit{et al.} \cite{NEURIPS2021_34f98c7c} proposed a pessimism-based approach that guarantees the convergence with $\mathcal{O}(d)$ sample complexity while only requiring Bellman closedness as standard in the exploratory setting. 
Furthermore, Zhan \textit{et al.} \cite{pmlr-v178-zhan22a} proposed a novel theoretical analysis framework for DRL according to the primal-dual formulation of MDPs. 
By relaxing the stringent requirement on the all-policy concentrability and Bellman-completeness, the framework proposed therein enables polynomial sample complexity under single-policy concentrability. 

\subsubsection {End-to-End Learning for Semantic Optimization}
The theory of semantic communication has caught extensive attentions
in the past several years with some preliminary research results \cite{wen2022challenges, Gndz2022BeyondTB}.
In \cite{beck2022semantic}, the IB theory was used to design the semantic communication systems by taking the meaning of semantic information and the compression ratio into consideration.
However, the IB formulation is task and label dependent, that is, the measurement quantity changes as tasks and labels change.
Besides, the IB provides an information-theoretic guidance for semantic information extraction and transfer, where the implementation of each functionality relies on the DNNs. 
Accordingly, a theoretical understanding of DNN can facilitate the theoretical analysis of semantic communication system, which hinges on the development of ML theories. 

\subsubsection{Federated Leaning for Distributed Optimization}
The theoretical development of FL depends on the research progress of the underlying DL theories. 
The existing theoretical research of FL focused on the convergence performance analysis with simple ML models, such as convex loss functions in \cite{Tian_SPM20}. 
The generic performance analysis for DL model in FL is still missing.
Furthermore, the theoretical analysis of FL shall be developed in view of various challenging practical issues including expensive communication, system and data heterogeneity as well as privacy and security concerns \cite{Tian_SPM20}.

\subsubsection{End-to-End Learning for General Non-convex Optimization}
The DL theories also enable the theoretical analysis of deep generative networks \cite{balevi2020high}, ReLU-based DNNs \cite{hu2020deep}, continual learning \cite{sun2022learning}, etc., for end-to-end learning frameworks.
The error bound of generative model for CS was analyzed in \cite{bora2017compressed} for ReLU-based generative NN and $L$-Lipschitz generative models, which can guide the high dimensional channel estimation applications in wireless communications \cite{balevi2020high}.
The ReLU-based DNN was proved mathematically equivalent to a piecewise linear function under some mild conditions, which can be applied to theoretically prove that the end-to-end DNNs can be utilized as a universal approximator of the MMSE channel estimator to supply theoretical support for DNN-based channel estimation algorithm design \cite{hu2020deep}. 
Moreover, the convergence analysis of continual learning framework \cite{sun2022learning} and model-free online DNNs \cite{zheng2021online} were further established for end-to-end wireless applications.
As the development of theoretical understandings on various learning techniques, more solid and in-depth description of theoretical analysis of MOA designs can be obtained to guide their usage in practical 6G wireless networks. 

\subsection{Implementation Issues and Software Platforms}
\subsubsection{Implementation Issues}
In academic, most existing \textit{learning to optimize} methods are trained and tested on an offline simulator with designed dataset (e.g. the DeepMIMO \cite{Alkhateeb2019} and Raymobtime \cite{klautau20185g} for collecting realistic training data) or generated dataset from simulator (e.g., the data generated from BB algorithm for training LBB models).
The software platforms for offline training simulator of ML-based model shall be discussed next. 
For distributed ML to be implemented on massive low-power end devices \cite{yuanming22edgeai}, FL, decentralized learning, model-split learning, distributed RL as well as trustworthy learning are considered as promising edge learning algorithms which are amenable to distributed implementations in large-scale networks. The edge inference implementation issues can be categorized as horizontal edge inference and vertical edge inference based on different collaborative computing mechanisms, which are also well discussed in literatures \cite{jia20edgeinference, yang20iot, hua21serverinference} to realize real-time inference in practical distributed systems.

The standardizations of \textit{AI/ML for communication} project have just been established and discussed in December 2021. 
The first technical standard for AI/ML was approved in the 3rd generation partnership project (3GPP) Release 18 \cite{RAN94} to investigate the implementation-related issues of AI/ML in physical layer. 
In order to ensure that the AI/ML model can be stably applied to the communication system, the general AI/ML architecture, collaboration between user and network, life cycle management of AI/ML models, model activation/deactivation, model monitoring, and model switching/updating shall be further discussed and designed in 3GPP \cite{CATT,google,huawei}, which can evaluate whether to update the AI model or go back to the traditional algorithm according to the monitoring results.

\subsubsection{Software Platforms}
For offline training of ML models, there are a rapidly growing body of software platforms for simulations and productization of ML algorithms and models, which can greatly simplify the construction of the complicated NNs, including the forward operation, gradient back propagation as well as the parallel computing, 
and provide potentially feasible platforms for the implementation of MOAs 
in practical wireless communication systems.
MATLAB Neural Network Toolbox, TensorFlow \cite{tf} and PyTorch \cite{pytorch} have provided excellent open-source software frameworks, which are compatible with common operating systems and can be installed in various communication devices, e.g., cloud server, BS, AP or a terminal with certain computational power, for real-time/offline data collection, model training, updating and inference.  
For example, Pytorch can be used to build a DNN easily in network environment and the library is well optimized for graphic processing units (GPU).  
TensorFlow is more suitable for building advanced and large-scale NNs.  
As a production-oriented DL framework, Caffe2 \cite{Caffe2} has been developed in Facebook to train NNs on multiple GPUs in distributed setting for supporting mobile operating systems (e.g. iOS and Android).
In addition, there are many other software platforms such as Blocks \cite{Blocks}, CNTK \cite{cntk} and Lasagne \cite{Lasagne} that can also support mobile systems for commercial-grade distributed DL implementation.
From the perspective of promoting scientific research, Open-L2O \cite{chen2021learning}, a research-oriented open software package, has recently been established to support both model-free and model-based ``learning to optimize” approaches, which facilitates a fair performance comparison of different algorithms in a simulated environment and a fully automatic algorithm design of various kinds of optimization problems.  
The advances in computing capacities and data storage techniques further fertilize the development of more sophisticated and advanced MOAs.  
For example, the GPU can be utilized to execute the DL algorithms much faster than traditional processors. 
Besides general-purpose GPU, customizable field-programmable gate array (FPGA) and dedicated application specific integrated circuit (ASIC) also encourage the research progress of DL in big data processing.

\subsection{Challenges and Future Research Directions}
Even though great progresses have been made in the field of MOA designs, a large amount of data are still required to train the DNNs to achieve near-optimal performance, which leads to several challenges for the practical deployment of MOAs. 
The acquisition of training data in practice can be difficult due to the hard-measurable environment, high storage cost and dynamic nature of wireless networks. 
To address this, the task-oriented MOAs as introduced in this paper can significantly improve the sample efficiency and the generalizability of NNs by incorporating the prior knowledge and task-specific features into the DL design. 
To avoid the transmission cost of data acquisition, local dataset can be exploited to train the DL models locally.
However, how to address the heterogeneity of the distributed computing nodes and guarantee a satisfactory overall performance is another issue to be carefully looked at.  
Besides the quantify of training data, the quality of training data can also greatly affect the learning performance. 
Robust MOA design, which is robust to data errors, measurement noises, hardware faults and mismatched training/testing conditions, is critical to generate reliable and trustworthy results in the practical applications. In addition to the challenges related to the data acquisition, we suggest following possible directions for future research in this area.

\subsubsection{Theoretical Analysis of MOAs}
A huge amount of parameters need to be optimized when using ML-based algorithm to solve large-scale wireless optimization problems.
Hence, the research progress of formal and rigorous ML theories shall help us to understand the optimal parameters to be learned, which can significantly reduce the training time of ML-based algorithm and guide the design of MOAs.
For instance, the good model parameters were analyzed in \cite{shi2022twc4ma} and can be obtained by solving a simple convex optimization problem rather than through computational intensive back propagation algorithm, 
which significantly accelerates the training process of large-scale DL models for JADCE problem.
However, the theoretical understandings of MOAs for wireless communication applications are still in the initial stage.

\subsubsection{Ultra-Lite Neural Network Design}
Most of the existing MOAs are highly demanding on computational power and storage space, which renders their deployment on small size and low computational power end devices. As motivated by the edge computing, the DL should be implemented in a distributed manner based on the local datasets. 
The straightforward network sparsification or pruning can alleviate the storage and computational burden, while it can deteriorate the learning performance significantly. 
Therefore, a light-weight and low-complexity MOA achieving on-par performance with the traditional algorithm is attracting increasing attention in edge computing systems, which allows on-device model training and computing with high accuracy, small model size and low computational complexity.

\subsubsection{Advanced Methodologies and Extended Applications of MOAs}
A trend of MOA is to exploit more sophisticated underlying features of specific optimization problems and design more advanced and task-specific MOAs to embed expert prior knowledge into DL techniques to speed up the convergence. 
In particular, the model-based approaches can be utilized to inspire/assist the design of MOAs and provide theoretical insight for the designed networks. 
In addition to the development of the methodologies, another trend is to explore new applications in wireless networks. 
Besides the optimization problems mentioned before, \textit{learning to optimize} techniques are expected to solve other problems involved in wireless communication for future research directions, including multi-object optimization problems \cite{shi2021capacity}, bi-level optimization problems \cite{beykal2020domino}, conic programming \cite{venkataraman2021neural}, maximum-likelihood estimation problems \cite{jialin22mra}, etc, in various emerging applications.

\section{Conclusion}
Integrating high-performance intelligent algorithm into the wireless networks has been an inevitable trend and disruptive shift for supporting highly transparent, reliable and large-scale 6G communication systems.
In this paper, we investigated some of the most groundbreaking ML technologies applied for solving challenging large-scale optimization problems in 6G wireless networks, including algorithm unrolling, learning to branch-and-bound, graph neural network for structured optimization, deep reinforcement learning for stochastic optimization, end-to-end learning for semantic-aware optimization as well as the 
federated wireless learning for distributed optimization.
In each section, the general algorithm was first introduced, followed by its case studies of the formulated optimization problems arising from wireless applications as well as the summary of its advantages and disadvantages.
In the last part, the neural network design for wireless communications, theoretical tools, implementation issues together with the future research directions were also discussed to implement ML algorithms in wireless communications from theory to practice.
Note that the research contributions discussed are representative but not complete, we hope that this article will serve as a valuable reference and guideline for ML-based optimization algorithm design in wireless networks across algorithmic regulation, theoretical understanding, systematic design and the practical deployment to support 6G intelligent communications.

\bibliographystyle{ieeetr}
\bibliography{reference}
\end{document}